\documentclass[twocolumn,showpacs,preprintnumbers,amsmath,amssymb]{revtex4-1}
\usepackage{graphicx}
\usepackage{bm}
\usepackage{amsmath}
\usepackage{dsfont}
\usepackage{multirow}

\begin{document}

\title{Classification of topological crystalline insulators based on representation theory}

\author{Xiao-Yu Dong$^{1}$ and Chao-Xing Liu$^{2}$}

\affiliation{$^1$Department of Physics and State Key Laboratory of Low-Dimensional Quantum Physics, Tsinghua University, Beijing 100084, P.R.China;}
\affiliation{$^{2}$Department of Physics, The Pennsylvania State University, University Park, Pennsylvania 16802-6300, USA;}

\begin{abstract}
	Topological crystalline insulators define a new class of topological insulator phases with gapless surface states protected by crystalline symmetries. In this work, we present a general theory to classify topological crystalline insulator phases based on the representation theory of space groups. Our approach is to directly identify possible nontrivial surface states in a semi-infinite system with a specific surface, of which the symmetry property can be described by 17 two-dimensional space groups. We reproduce the existing results of topological crystalline insulators, such as mirror Chern insulators in the $pm$ or $pmm$ groups, $C_{nv}$ topological insulators in the $p4m$, $p31m$ and $p6m$ groups, and topological nonsymmorphic crystalline insulators in the $pg$ and $pmg$ groups. Aside from these existing results, we also obtain the following new results: (1) there are two integer mirror Chern numbers ($\mathbb{Z}^2$) in the $pm$ group but only one ($\mathbb{Z}$) in the $cm$ or $p3m1$ group for both the spinless and spinful cases; (2) for the $pmm$ ($cmm$) groups, there is no topological classification in the spinless case but $\mathbb{Z}^4$ ($\mathbb{Z}^2$) classifications in the spinful case; (3) we show how topological crystalline insulator phase in the $pg$ group is related to that in the $pm$ group; (4) we identify topological classification of the $p4m$, $p31m$, and $p6m$ for the spinful case; (5) we find topological non-symmorphic crystalline insulators also existing in $pgg$ and $p4g$ groups, which exhibit new features compared to those in $pg$ and $pmg$ groups. We emphasize the importance of the irreducible representations for the states at some specific high-symmetry momenta in the classification of topological crystalline phases. Our theory can serve as a guide for the search of topological crystalline insulator phases in realistic materials.
\end{abstract}


\maketitle

\section{Introduction}
According to the ability to conduct electric currents, materials can usually be classified into insulators and metals. Recently, researchers discovered a new class of materials, of which their interior cannot conduct electric currents, similar to insulators, but their surfaces have conducting channels. Although the surface states can also exist in ordinary insulators, the uniqueness of this class of materials lies in the fact that the existence of surface states is related to its bulk property, which can be described by the mathematical theory, topology. Thus, these materials are dubbed ``topological insulators (TIs)'' \cite{qi2011,hasan2010,moore2010,qi2010phystoday}. The quantum Hall effect \cite{klitzing1980}, discovered in 1980, can be viewed as the first example of topologically non-trivial states. Under a strong magnetic field, two-dimensional (2D) electron gases form Landau levels and get localized, while at the edge, there are gapless conducting modes propagating along one direction, known as the ``chiral edge state.'' A recent progress in this field is the discovery of time-reversal (TR) invariant TIs \cite{kane2005a,kane2005b,bernevig2006a,bernevig2006c,fu2007b,fu2007a,moore2007}, of which the gapless edge/surface states are protected by TR symmetry. Different from the chiral edge states in the quantum Hall effect, the edge/surface states of TR invariant TIs contain two counter-propagating modes with opposite spins, thus dubbed ``helical edge/surface states.'' Two branches of helical edge/surface states are related to each other by TR symmetry, forming the Kramers' pairs. According to the Kramers' theorem, the energy of one branch at the momentum ${\mathbf{k}}$ must be the same as that of the other branch at $-{\mathbf{k}}$. Therefore, at the momenta invariant under TR, such as ${\mathbf{k}}=0$, two branches of surface states are degenerate and form a Dirac-cone-like energy dispersion. TR invariant TIs have been confirmed in various materials, such as HgTe quantum wells \cite{bernevig2006c,koenig2007}, BiSb \cite{hsieh2008}, Bi$_2$Se$_3$ family of materials \cite{hjzhang2009,xia2009,chen2009}, etc.\cite{yan2012,ando2013topological}, by different experimental methods, including angular-resolved photon emission spectroscopy (ARPES) \cite{hsieh2008,hsieh2009,xia2009,chen2009}, scanning tunneling microscopy (STM) \cite{roushan2009,tzhang2009,alpichshev2010}, and transport measurements \cite{koenig2007,brune2011b,qu2010,analytis2010}. Since the topologically non-trivial edge/surface states require the protection from TR symmetry, these systems are also regarded as symmetry-protected topological states \cite{chen2012a} in free-fermions systems.

Based on the above discussion, we can see that degeneracy due to symmetry plays an essential role in topological states of free-fermion systems that are protected by symmetry. It is well known that in crystals, there are different types of degeneracy, including accidental degeneracy and essential degeneracy. The essential degeneracy arises from the presence of certain symmetries and is related to the representation theory of symmetry groups. Therefore, one may ask the following questions: (1) whether or not the concept of TR invariant TIs can be generalized to other types of symmetries, leading to new classes of topological states; and (2) if one can use representation theory to classify topological states that are protected by other types of symmetries. It was recently realized that crystalline symmetry can indeed protect new types of topological states, which are dubbed ``topological crystalline insulators (TCIs)'' \cite{fu2011a,hsieh2012,slager2012,fang2012a,jadaun2013topological, fang2012multi,kargarian2013,ueno2013symmetry,tanaka2013,liu2013two, fang2013theory,fulga2014statistical,ando2015topological, chiu2013classification, morimoto2013topological, shiozaki2014topology, fang2013entanglement,wrasse2014prediction,chiu2015classification}. In particular, SnTe systems are found to be a mirror-symmetry-protected topological state, which has recently been confirmed by the ARPES measurement \cite{dziawa2012,tanaka2012,xu2012a}. Various physical properties, including mass acquisition \cite{okada2013observation}, unconventional orbital texture \cite{zeljkovic2014mapping}, spin-filtered edge states \cite{liu2014spin}, quasi-particle interference \cite{gyenis2013quasiparticle,fang2013theory}, and interface superconductivity \cite{tang2014strain}, have been considered.

In this paper, we will systematically study different types of TCIs based on the representation theory of crystalline symmetry groups. Instead of directly studying three-dimensional (3D) bulk system, we consider a 3D semi-infinite system with one surface and investigate non-trivial gapless surface states which cannot be gapped once the crystalline symmetry is preserved. Since a 3D semi-infinite crystal is classified by 2D space group, this approach allows us to classify different types of non-trivial surface states based on the representation theory of 2D space group. After classifying 2D topologically non-trivial surface states, we can further investigate topological property of the 3D bulk system and identify the corresponding 3D bulk topological invariants. The details of our approach will be discussed in the following, but we would like first to illustrate our main results.

\subsection{Summary of main results}
\begin{figure} 
\includegraphics[width=8.6 cm]{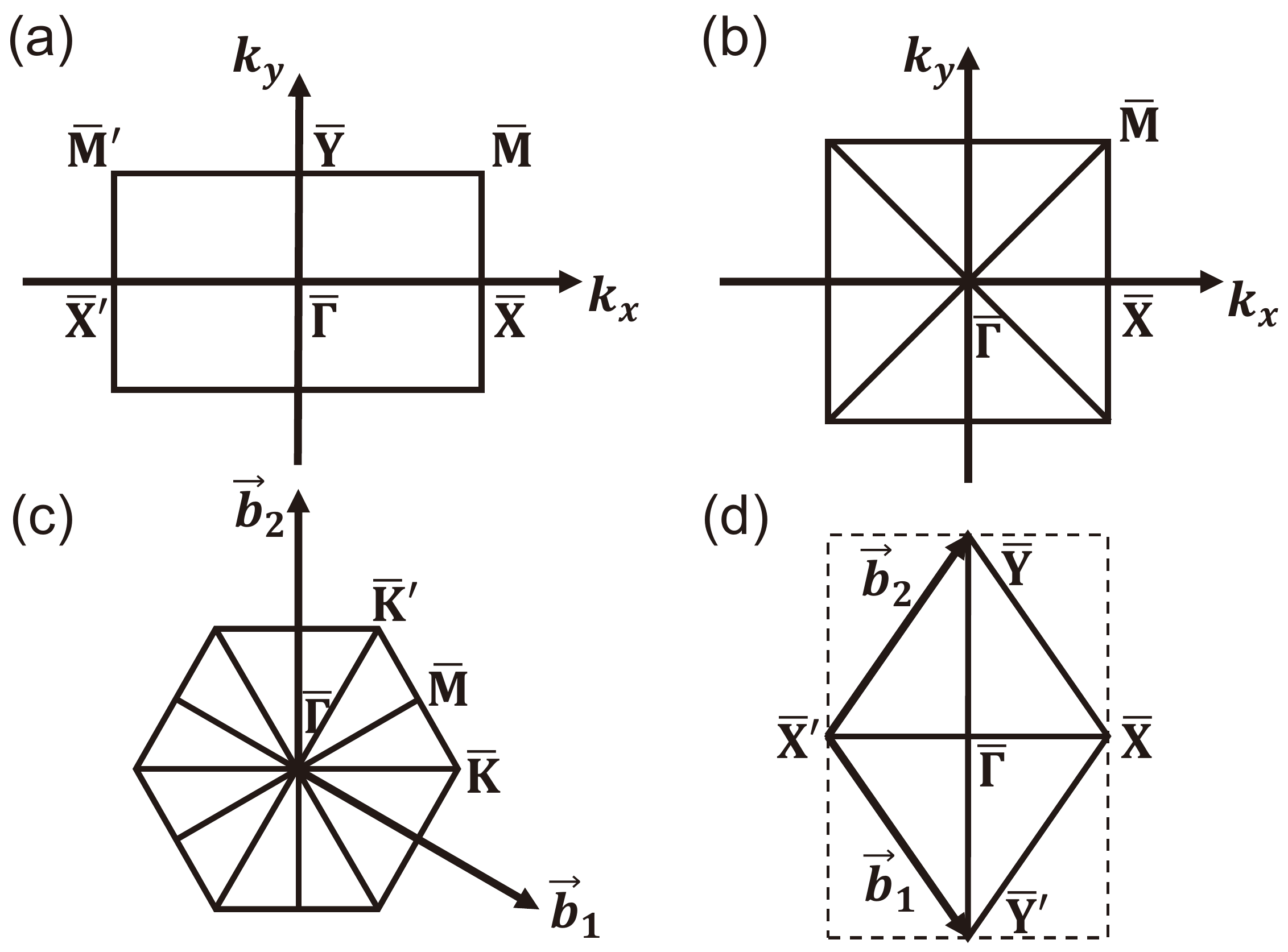}
\caption{Schematic plot of the surface BZ of rectangular, square, hexagonal, and rhombic groups in (a), (b), (c), and (d), respectively. }\label{Fig:1}
\end{figure}

Our main results are summarized in Table~\ref{tab1} for 17 2D space groups. The first column of the table is the group name and the second column gives the shape of 2D surface Brillouin zone (BZ), in which ``re'' is for rectanglar, ``h'' for hexagonal, ``rh'' for rhombic, and ``s'' for square. Different 2D surface BZs are shown in Fig.~\ref{Fig:1}. The next two columns list high symmetry momentum points (HSPs) and high symmetry momentum lines (HSLs), which are essential for determining topologically non-trivial surface states. Topological classifications of TCIs for the spinless and spinful cases are shown in the last two columns. Here the symbols $\mathbb{Z}$ and $\mathbb{Z}_2$ indicate whether the space of ground states for the insulating phases are partitioned into topological sectors labeled by an integer or a $\mathbb{Z}_2$ quantity, respectively. $\mathbb{Z}^n$ ($n=2,3,4$) means that we can define $n$ different integer topological invariants. We would like to emphasize that for several space groups, including $p4m$, $p31m$, $p6m$, $pgg$, $pmg$ and $p4g$, the classification depends on which irreducible representations (Irreps) of the states near the band gap belong to. Therefore, for these cases, we list the required condition of Irreps in the bracket after the classification. Below, we will discuss our results of TCIs for 17 2D space groups in several categories.

(1) For the groups $pm$, $p3m1$ and $cm$ (Sec.~\ref{Sec:pm}), we have integer number of copies of topologically non-trivial surface states, which are characterized by mirror Chern numbers (MCNs), belonging to a $\mathbb{Z}$ classification, for both the spinless and spinful cases. Two independent MCNs ($\mathbb{Z}^2$) can be defined for the $pm$ group, but only one can be defined for the $p3m1$ and $cm$ groups.

(2) For the groups $pmm$ and $cmm$ (Sec.~\ref{Sec:pmm}), MCNs are not allowed for the spinless case. In contrast, four (two) independent MCNs can be defined for the spinful case of $pmm$ ($cmm$) group, leading to a $\mathbb{Z}^4$ ($\mathbb{Z}^2$) classification.

(3) The $pg$ group (Sec.~\ref{Sec:pg}) is classified by one $\mathbb{Z}_2$ topological invariant.

(4) For the $p4m$, $p31m$ and $p6m$ (Sec.~\ref{Sec:p4m}), the classification depends on the Irreps of the states near the Fermi energy, and we need to illustrate the classification for each group separately. For the spinless case of the $p4m$ group, when all the states at $\bar{\Gamma}$ and $\bar{\mathrm{M}}$ belong to the doublet (or 2D) Irrep, we can define two integer topological invariants (halved mirror chirality (HMC)), giving rise to $\mathbb{Z}^2$ classification. However, if any state at $\bar{\Gamma}$ or $\bar{\mathrm{M}}$ belongs to one-dimensional (1D) Irreps, no topological invariant can be defined. For the spinful case of the $p4m$ group, three topological invariants of HMC type ($\mathbb{Z}^3$) can be defined, irrespective of any condition. For the $p31m$ group, there are three HSPs $\bar{\Gamma}$, $\bar{\mathrm{K}}$, and $\bar{\mathrm{K}}'$ with the point group $C_{3v}$ and three HSLs with mirror symmetry. When the states at $n$ HSPs ($n=2,3$) belong to doublet Irreps, the topological classification is $\mathbb{Z}^n$ with $n$ distinct HMCs for both the spinful and spinless cases. If there is only one HSP or no HSP with the states in doublet Irreps, one MCN ($\mathbb{Z}$) can be defined. For the $p6m$ group, we also have three HSPs, in which $\bar{\Gamma}$ belongs to the $C_{6v}$ group, $\bar{\mathrm{K}}$ to the $C_{3v}$ group, and $\bar{\mathrm{M}}$ to the $C_{2v}$ group, and also three HSLs with mirror symmetry. For the spinless case, there are two doublet Irreps at $\bar{\Gamma}$ and one doublet Irrep at $\bar{\mathrm{K}}$. One HMC ($\mathbb{Z}$ classification) can be defined when all the states at $\bar{\Gamma}$ and $\bar{\mathrm{K}}$ belong to doublet Irreps. The $C_{2v}$ group at $\bar{\mathrm{M}}$ has four 1D Irreps $A_i$ and $B_i$ ($i=1,2$). Please refer to Table.~\ref{tab:2} for the definition of $A_i$ and $B_i$. The characters of the mirror symmetry $m_x$ and $m_y$ are the same in Irreps $A_i$, while opposite in Irreps $B_i$. The topological classification on $\bar{\Gamma}$-$\bar{\mathrm{M}}$-$\bar{\mathrm{K}}$ depends on the Irreps of the bands near the Fermi surface at $\bar{\mathrm{M}}$. If all the states at $\bar{\mathrm{M}}$ belong to $A_i$ (or $B_i$), another HMC can be defined on $\bar{\Gamma}$-$\bar{\mathrm{M}}$-$\bar{\mathrm{K}}$. A detailed discussion about representation dependent classification is given in Section.~\ref{Sec:pmg}. For the spinful case, all the states at $\bar{\Gamma}$ and $\bar{\mathrm{M}}$ belong to the doublet Irrep, while both doublet and singlet Irreps are possible for the states at $\bar{\mathrm{K}}$. Thus, three independent HMCs ($\mathbb{Z}^3$) can be defined on $\bar{\Gamma}$-$\bar{\mathrm{K}}$, $\bar{\mathrm{K}}$-$\bar{\mathrm{M}}$, and $\bar{\mathrm{M}}$-$\bar{\Gamma}$ if the states at $\bar{\mathrm{K}}$ belong to doublet Irreps. Otherwise, there are two HMCs that can be defined on $\bar{\Gamma}$-$\bar{\mathrm{M}}$ and $\bar{\Gamma}$-$\bar{\mathrm{K}}$-$\bar{\mathrm{M}}$.

(5) For three non-symmorphic symmetry groups $pmg$, $pgg$, and $p4g$ (Sec.~\ref{Sec:pmg}), the classification also depends on Irreps at the HSPs $\bar{\Gamma}$, $\bar{\mathrm{X}}$, $\bar{\mathrm{Y}}$ and $\bar{\mathrm{M}}$. For the spinless case of the $pgg$ group, all the states at $\bar{\mathrm{X}}$ and $\bar{\mathrm{Y}}$ are doubly degenerate (doublet Irreps). The classification depends on 1D Irreps at $\bar{\Gamma}$ and $\bar{\mathrm{M}}$. There are four types of different 1D Irreps at $\bar{\Gamma}$ ($\bar{\mathrm{M}}$), denoted as $A_i$ and $B_i$ ($i=1,2$). If all the states at $\bar{\Gamma}$ ($\bar{\mathrm{M}}$) belong to the $A_i$ or $B_i$ Irreps, one integer topological invariant ($\mathbb{Z}$ classification), dubbed halved glide chirality (HGC), in analogy to HMC, can be defined along the HSL $\bar{\mathrm{X}}$-$\bar{\Gamma}$-$\bar{\mathrm{Y}}$ ($\bar{\mathrm{X}}$-$\bar{\mathrm{M}}$-$\bar{\mathrm{Y}}$). If both the $A_i$ and $B_i$ states exist at $\bar{\Gamma}$ and $\bar{\mathrm{M}}$ near the band gap, one can no longer define HGC along these momentum lines. However, a $\mathbb{Z}_2$ topological invariant can still be defined. For the spinful case, the classification is similar to that of the spinless case. The only difference is that now all the states at $\bar{\Gamma}$ and $\bar{\mathrm{M}}$ are doubly degenerate and it depends on the 1D Irreps at $\bar{\mathrm{X}}$ and $\bar{\mathrm{Y}}$ if a $\mathbb{Z}_2$ or $\mathbb{Z}$ topological invariant can be defined along the momentum line $\bar{\Gamma}$-$\bar{\mathrm{X}}$-$\bar{\mathrm{M}}$ ($\bar{\Gamma}$-$\bar{\mathrm{Y}}$-$\bar{\mathrm{M}}$). For the $pmg$ group, all the states at $\bar{\mathrm{X}}$ and $\bar{\mathrm{M}}$ are doubly degenerate, and thus a HMC can always be defined along the line $\bar{\mathrm{X}}$-$\bar{\mathrm{M}}$ ($\mathbb{Z}$ classification). Another independent $\mathbb{Z}$ topological invariant can be defined along the line $\bar{\mathrm{X}}$-$\bar{\Gamma}$-$\bar{\mathrm{Y}}$-$\bar{\mathrm{M}}$ if all the states at $\bar{\Gamma}$ and $\bar{\mathrm{Y}}$ belong to either $A_i$ or $B_i$ Irreps. For the spinful case of the $pmg$ group, the classification is similar and only the role of $\bar{\mathrm{X}}$, $\bar{\mathrm{M}}$ and $\bar{\mathrm{Y}}$, $\bar{\Gamma}$ is switched. For the $p4g$ group, in the spinless case, when the states at $\bar{\Gamma}$ and $\bar{\mathrm{M}}$ belong to doublet Irreps, three independent HGCs can be defined on $\bar{\Gamma}$-$\bar{\mathrm{X}}$, $\bar{\mathrm{X}}$-$\bar{\mathrm{M}}$ and $\bar{\mathrm{M}}$-$\bar{\Gamma}$. If only the states at $\bar{\mathrm{M}}$ ($\bar{\Gamma}$) belong to doublet Irrep, and all the states at $\bar{\Gamma}$ ($\bar{\mathrm{M}}$) belong to the 1D Irreps $A_i$ or $B_i$ of point group $C_{4v}$, two HGCs can be defined on $\bar{\mathrm{X}}$-$\bar{\mathrm{M}}$ ($\bar{\mathrm{X}}$-$\bar{\Gamma}$) and $\bar{\mathrm{X}}$-$\bar{\Gamma}$-$\bar{\mathrm{M}}$ ($\bar{\mathrm{X}}$-$\bar{\mathrm{M}}$-$\bar{\Gamma}$). If both the $A_i$ and $B_i$ states exist at $\bar{\Gamma}$ ($\bar{\mathrm{M}}$), the HGC can only be defined on $\bar{\mathrm{X}}$-$\bar{\mathrm{M}}$ ($\bar{\mathrm{X}}$-$\bar{\Gamma}$). If the states at both $\bar{\Gamma}$ and $\bar{\mathrm{M}}$ belong to singlet Irreps, no topologically non-trivial phase can be realized. In the spinful case, all the states are doubly degenerate at $\bar{\Gamma}$ and $\bar{\mathrm{M}}$, and thus one HGC can be defined on $\bar{\Gamma}$-$\bar{\mathrm{M}}$. When the states at $\bar{\mathrm{X}}$ belong to $A_i$ (or $B_i$) of the group $C_{2v}$, another one HGC can be defined on $\bar{\Gamma}$-$\bar{\mathrm{X}}$-$\bar{\mathrm{M}}$.

\begin{table*}
\centering
\caption{Summary of the symmetry and topological properties of 2D space groups. SBZ: surface BZ; HSP: high symmetry point; HSL: high symmetry line; re: rectangular; h: hexagonal; rh: rhombic; s: square. The surface BZ of rectangular, square, hexagonal, and rhombic group are shown in Fig.~\ref{Fig:1}s(a), (b), (c), and (d), respectively. For HSL, $m$ is for mirror symmetry and $g$ for glide symmetry. $A_i$ and $B_i$ denote 1D Irreps at the corresponding HSP while $E$ denotes 2D Irrep.   }
\begin{tabular}{|c|c|c|c|c|c|}
    \hline
    \hline
    \multirow{2}{*}{Group}
    & \multirow{2}{*}{SBZ}
    & \multirow{2}{*}{HSP}
    & \multirow{2}{*}{HSL}
    & \multicolumn{2}{|c|}{Topological Classification}
    \\
    \cline{5-6}
    & & &
    & spinless & spinful
    \\
    \hline
    $pm$
    & re
    &
    & $\bar{\Gamma}$-$\bar{\mathrm{X}}$,$\bar{\mathrm{Y}}$-$\bar{\mathrm{M}}$:$m_y$
    & $\mathbb{Z}^2\equiv\mathbb{Z}\times\mathbb{Z}$
    & $\mathbb{Z}^2$
    \\
    \hline
    $p3m1$
    & h
    & $\bar{\Gamma}$: $C_{3v}$
    & $\bar{\Gamma}$-$\bar{\mathrm{M}}$:$m$
    & $\mathbb{Z}$
    & $\mathbb{Z}$
    \\
    \hline
    $cm$
    & rh
    &
    & $\bar{\Gamma}$-$\bar{\mathrm{X}}$:$m_y$
    & $\mathbb{Z}$
    & $\mathbb{Z}$
    \\
    \hline
    \multirow{2}{*}{$pmm$}
    & \multirow{2}{*}{re}
    & \multirow{2}{*}{$\bar{\Gamma}$,$\bar{\mathrm{X}}$,$\bar{\mathrm{M}}$,$\bar{\mathrm{Y}}$:$C_{2v}$}
    & $\bar{\Gamma}$-$\bar{\mathrm{X}}$,$\bar{\mathrm{Y}}$-$\bar{\mathrm{M}}$:$m_y$
    & \multirow{2}{*}{None}
    & \multirow{2}{*}{$\mathbb{Z}^4$}
    \\
     &  &  &$\bar{\Gamma}$-$\bar{\mathrm{Y}}$,$\bar{\mathrm{X}}$-$\bar{\mathrm{M}}$:$m_x$ & &
    \\
    \hline
    \multirow{2}{*}{$cmm$}
    & \multirow{2}{*}{rh}
    & \multirow{2}{*}{$\bar{\Gamma}$,$\bar{\mathrm{X}}$,$\bar{\mathrm{Y}}$:$C_{2v}$}
    & $\bar{\Gamma}$-$\bar{\mathrm{X}}$:$m_y$;
    & \multirow{2}{*}{None}
    & \multirow{2}{*}{$\mathbb{Z}^2$}
     \\
     &  &  &$\bar{\Gamma}$-$\bar{\mathrm{Y}}$:$m_x$ & &
    \\
    \hline
    $pg$
    & re
    &
    & $\bar{\Gamma}$-$\bar{\mathrm{X}}$,$\bar{\mathrm{Y}}$-$\bar{\mathrm{M}}$:$g_y$
    & $\mathbb{Z}_2$
    & $\mathbb{Z}_2$
    \\
    \hline
    \multirow{3}{*}{$p4m$}
    & \multirow{3}{*}{s}
    & $\bar{\Gamma}$,$\bar{\mathrm{M}}$:$C_{4v}$
    & $\bar{\Gamma}$-$\bar{\mathrm{X}}$:$m_y$;
    & $\mathbb{Z}^2(\bar{\Gamma},\bar{\mathrm{M}}\in E)$
    & $\mathbb{Z}^3$
    \\
    &  &$\bar{\mathrm{X}}$:$C_{2v}$  & $\bar{\mathrm{X}}$-$\bar{\mathrm{M}}$:$m_x$; & None ($\bar{\Gamma}$ or $\bar{\mathrm{M}}\notin E$) &
     \\
    &  & & $\bar{\Gamma}$-$\bar{\mathrm{M}}$:$m_{d}$ &  &
    \\
    \hline
    \multirow{3}{*}{$p31m$}
    & \multirow{3}{*}{h}
    & \multirow{3}{*}{ $\bar{\Gamma}$,$\bar{\mathrm{K}}$,$\bar{\mathrm{K}}'$:$C_{3v}$}
    & $\bar{\Gamma}$-$\bar{\mathrm{K}}$:$m_1$
    & $\mathbb{Z}^3$(3 HSPs $\in E$)
    & $\mathbb{Z}^3$(3 HSPs $\in E$)
    \\
     &  &  &  $\bar{\mathrm{K}}$-$\bar{\mathrm{K}}'$:$m_2$ &$\mathbb{Z}^2$(2 HSPs $\in E$) & $\mathbb{Z}^2$(2 HSPs $\in E$)
    \\
    &  &  & $\bar{\mathrm{K}}'$-$\bar{\Gamma}$:$m_3$ &$\mathbb{Z}$(general case) & $\mathbb{Z}$(general case)
    \\
    \hline
    \multirow{3}{*}{$p6m$}
    & \multirow{3}{*}{h}
    & $\bar{\Gamma}$:$C_{6v}$
    & $\bar{\Gamma}$-$\bar{\mathrm{K}}$:$m_1$
    & $\mathbb{Z}^2$($\bar{\Gamma}\in E_i(i=1,2$),$\bar{\mathrm{K}}\in E$,$\bar{\mathrm{M}}\in A_i(B_i)$)
    & $\mathbb{Z}^3$($\bar{\mathrm{K}}\in E$)
    \\
    &  &$\bar{\mathrm{K}}$:$C_{3v}$  &$\bar{\Gamma}$-$\bar{\mathrm{M}}$:$m_2$ & $\mathbb{Z}$($\bar{\Gamma}\in E_i(i=1,2$),$\bar{\mathrm{K}}\in E$,$\bar{\mathrm{M}}$ general) &$\mathbb{Z}^2$($\bar{\mathrm{K}}\notin E$)
    \\
    &  &$\bar{\mathrm{M}}$:$C_{2v}$  &$\bar{\mathrm{K}}$-$\bar{\mathrm{M}}$:$m_3$ & None ($\bar{\Gamma}\notin E_i(i=1,2$) or $\bar{\mathrm{K}}\notin E$)   &
    \\
    \hline
    \multirow{3}{*}{$pgg$}
    & \multirow{3}{*}{re}
    & \multirow{3}{*}{$\bar{\Gamma}$,$\bar{\mathrm{X}}$,$\bar{\mathrm{Y}}$,$\bar{\mathrm{M}}$:$C_{2v}$}
    & $\bar{\Gamma}$-$\bar{\mathrm{X}}$,$\bar{\mathrm{Y}}$-$\bar{\mathrm{M}}$:$g_y$;
    & $\mathbb{Z}^2(\bar{\Gamma},\bar{\mathrm{M}}\in A_i(B_i)$; \qquad\qquad\quad
    & $\mathbb{Z}^2(\bar{\mathrm{X}},\bar{\mathrm{Y}}\in A_i(B_i)$; \qquad\qquad\quad
    \\
    &  &
    & $\bar{\Gamma}$-$\bar{\mathrm{Y}}$, $\bar{\mathrm{X}}$-$\bar{\mathrm{M}}$:$g_x$
    & \quad~$\bar{\Gamma}\in A_i(B_i), \bar{\mathrm{M}}\in B_i(A_i))$
    & \quad~$\bar{\mathrm{X}}\in A_i(B_i), \bar{\mathrm{Y}}\in B_i(A_i))$
    \\
    &  &  &  & $\mathbb{Z}(\bar{\Gamma}$ or $\bar{\mathrm{M}}\in A_i; \bar{\Gamma}$ or $\bar{\mathrm{M}}\in B_i)$
    & $\mathbb{Z}(\bar{\mathrm{X}}$ or $\bar{\mathrm{Y}}\in A_i; \bar{\mathrm{X}}$ or $\bar{\mathrm{Y}}\in B_i)$
    \\
    &  &  &  & $\mathbb{Z}_2$(general case) & $\mathbb{Z}_2$(general case)
    \\
    \hline
   \multirow{2}{*}{$pmg$}
    & \multirow{2}{*}{re}
    &  \multirow{2}{*}{ $\bar{\Gamma}$,$\bar{\mathrm{X}}$,$\bar{\mathrm{Y}}$,$\bar{\mathrm{M}}$:$C_{2v}$}
    & $\bar{\Gamma}$-$\bar{\mathrm{X}}$,$\bar{\mathrm{Y}}$-$\bar{\mathrm{M}}$:$g_y$;
    &$\mathbb{Z}^2$($\bar{\mathrm{Y}},\bar{\Gamma}\in A_i$;  $\bar{\mathrm{Y}},\bar{\Gamma}\in B_i$;
    & $\mathbb{Z}^2$($\bar{\mathrm{X}},\bar{\mathrm{M}}\in A_i$;  $\bar{\mathrm{X}},\bar{\mathrm{M}}\in B_i$;
    \\
     &  &  & $\bar{\Gamma}$-$\bar{\mathrm{Y}}$,$\bar{\mathrm{X}}$-$\bar{\mathrm{M}}$:$m_x$  &  $\qquad\quad$$\bar{\mathrm{Y}}\in A_i(B_i),\bar{\Gamma}\in B_i(A_i)$)
     & $\qquad\quad$$\bar{\mathrm{X}}\in A_i(B_i),\bar{\mathrm{M}}\in B_i(A_i)$)
    \\
    &  &  &   & $\mathbb{Z}$(general case) & $\mathbb{Z}$(general case)
    \\
    \hline
    \multirow{4}{*}{$p4g$}
    & \multirow{4}{*}{s}
    & $\bar{\Gamma}$,$\bar{\mathrm{M}}$:$C_{4v}$
    & $\bar{\Gamma}$-$\bar{\mathrm{X}}$:$g_y$;
    & $\mathbb{Z}^3$($\bar{\Gamma},\bar{\mathrm{M}}\in E$)
    & $\mathbb{Z}^2$($\bar{\mathrm{X}}\in A_i(B_i)$)
    \\
    &  & $\bar{\mathrm{X}}$:$C_{2v}$  & $\bar{\mathrm{X}}$-$\bar{\mathrm{M}}$:$g_x$; &$\mathbb{Z}^2$($\bar{\mathrm{M}}\in E$, $\bar{\Gamma}\in A_i(B_i)$; $\bar{\Gamma}\leftrightarrow\bar{\mathrm{M}}$) & $\mathbb{Z}$(general case)
    \\
    &  &   & $\bar{\Gamma}$-$\bar{\mathrm{M}}$:$g_{d}$ & $\mathbb{Z}$($\bar{\mathrm{M}}\in E, \bar{\Gamma}$ general; $\bar{\Gamma}\leftrightarrow\bar{\mathrm{M}}$) &
    \\
    &  &   &  & None($\bar{\Gamma},\bar{\mathrm{M}}\notin E$) &
    \\
    \hline
    \hline
  \end{tabular}\label{tab1}
\end{table*}

\subsection{Outline}
This paper is organized as follows. Section~\ref{Sec:Method} is devoted to the description of our technique approach. We first describe the relationship between symmetry and degeneracy and then discuss our strategy to classify topologically non-trivial surface states for a 3D semi-infinite crystal. In Sec.~\ref{Sec:Classify}, we perform a complete analysis of 17 2D space groups to identify possible crystal structures that are able to host non-trivial surface states. The topological invariant is also identified for each topologically non-trivial case. A conclusion is drawn in Sec.~\ref{Sec:Conclusion}.

\section{Methodology}\label{Sec:Method}
\subsection{Symmetry and degeneracy}
In quantum mechanics, if two different eigenstates share the same eigenenergy, we say these two states are degenerate. The degeneracy is usually due to the presence of symmetry, the invariance of the Hamiltonian under a certain operation. The relation between degeneracy and symmetry has been well established based on the representation theory of symmetry groups. Here we focus on the degeneracies induced by crystalline symmetries. According to different types of crystalline symmetries, we can further classify the corresponding degeneracies into three types, which are discussed in details following.

The first type of degeneracy (type-I) is due to the non-commutation relation between symmetry operations, which leads to the presence of high dimensional Irreps of the symmetry group. From the group theory, the eigenstates of a system can form the basis to construct Irreps of the symmetry group for this system and the degeneracy is equal to the dimension of the corresponding Irreps\cite{dresselhaus2008}. In an Abelian group, in which all the symmetry operators commute with each other, only one 1D Irreps exist. Thus, the existence of high-dimensional Irreps requires that at least two of the group elements do not commute, $AB\neq BA$ ($A, B\in G$, $G$ denotes the symmetry group of the system with the Hamiltonian $H$). We may consider a special case, of which the system has two symmetry operators $A$ and $B$ anti-commuting with each other ($[H,A]=0$, $[H,B]=0$ and $\{A,B\}=0$).
Since $[H,A]=0$, we can take $|\psi\rangle$ as a common eigenstate of $H$ and $A$, i.e., $H|\psi\rangle=E|\psi\rangle$ and $A|\psi\rangle=a|\psi\rangle$. $B|\psi\rangle$ is also an eigenstate of the Hamiltonian with the same eigenenergy $E$ since $H(B|\psi\rangle)=BH|\psi\rangle=E(B|\psi\rangle)$. At the same time, we have $A(B|\psi\rangle)=-BA|\psi\rangle=-a(B|\psi\rangle)$, which indicates that $B|\psi\rangle$ and $|\psi\rangle$ are two orthogonal  and degenerate eigenstates of $H$.

The second type of degeneracy (type-II) occurs for the eigenstates belonging to different Irreps of a symmetry group. As mentioned above, all eigenstates of a system can form the basis for the representations of the corresponding symmetry group. Let us consider two eigenstates $|\psi_1(k)\rangle$ and $|\psi_2(k)\rangle$ belonging to two different Irreps of the symmetry group of the Hamiltonian, where $k$ is a tuning parameter and can be regarded as the momentum of the BZ for our case. Due to different Irreps, the coupling between these two states is forbidden by symmetry, indicating that the corresponding Hamiltonian must be block diagonal under this basis. The eigenenergies of these two states vary with $k$ and at a certain $k$, these two states may cross each other. Since the coupling between these two states must vanish due to symmetry, this accidental degeneracy at the crossing point cannot open a gap.

The third type of degeneracy (type-III) originates from the anti-unitary symmetry operators. For a unitary symmetry operation $A$, $\langle A\phi|A\psi\rangle=\langle\phi|\psi\rangle$, where $|\psi\rangle$ and $|\phi\rangle$ are two arbitrary wave functions. In contrast, TR operator $\Theta$ is anti-unitary, which means that $\langle \Theta\phi|\Theta\psi\rangle=(\langle\phi|\psi\rangle)^*$. For spinful fermions, the TR symmetry operator satisfies the relation $\Theta^2=-1$, which can lead to a double degeneracy in a TR invariant system according to the so-called Kramers' theorem\cite{dresselhaus2008}. As discussed above, this degeneracy plays an essential role in protecting the gapless nature of surface states of TR invariant TIs. Moreover, magnetic systems can also possess some types of anti-unitary operations, which have been discussed in details in Ref.~\cite{zhang2015topological}. This paper mainly focuses on the type-I and II degeneracies, which only concern unitary operators.

\subsection{Overview of classification principles}
After understanding the above three types of degeneracies induced by symmetry, we will next describe our approach for classification and show how these different types of degeneracies can lead to non-trivial surface states.
\begin{figure} 
\includegraphics[width=8.6 cm]{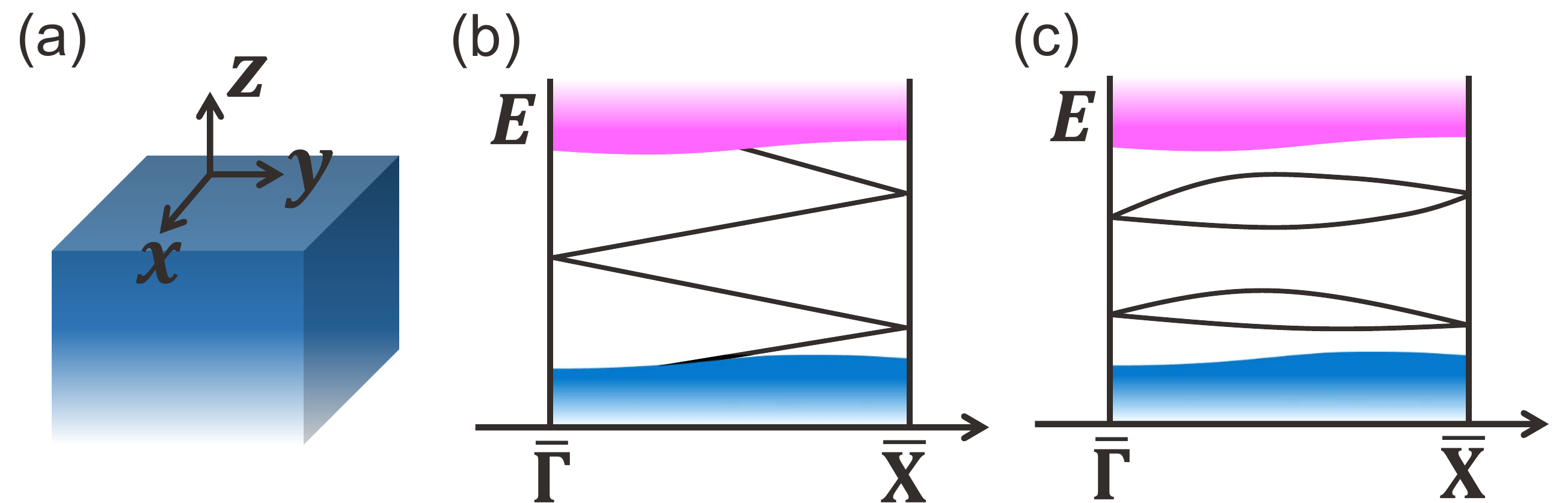}
\caption{(Color online) (a) Schematic plot of a semi-infinite system. (b) Schematic plot of the surface states of TR invariant TIs. $\bar{\Gamma}=(0,0)$ and $\bar{\mathrm{X}}=(\pi,0)$ are two TI invariant momenta in the surface BZ. (c) Schematic plot of the surface states of trivial insulators. }\label{Fig:2}
\end{figure}

Let us start from the discussion about how TR symmetry protects the gapless nature of surface states in TR invariant TIs and why topological surface states are essentially different from normal surface states, such as dangling bonds, when TR symmetry is preserved. Our discussion follows Ref.~[\onlinecite{fu2006}]. We consider a semi-infinite system with one surface along the $z$ direction, as shown in Fig.~\ref{Fig:2}(a). The difference for the surface states between TR invariant TIs and trivial insulators are shown in Figs.~\ref{Fig:2}(b) and (c). One can see that the surface states for TIs connect the conduction bands to valence bands while those for trivial insulators are isolated from either conduction or valence bands. This difference relies on the fact that all the states at $\bar{\Gamma}(0,0)$ and $\bar{\mathrm{X}}(\pi,0)$ in Figs.~\ref{Fig:2}(b) and (c) are doubly degenerate due to TR symmetry. For non-trivial surface states in Fig.~\ref{Fig:2}(b), the degeneracies at $\bar{\Gamma}$ and $\bar{\mathrm{X}}$ point come from different surface bands (switching partners) while for trivial surface states in Fig.~\ref{Fig:2}(c), degenerate states at $\bar{\Gamma}$ and $\bar{\mathrm{X}}$ are from the same surface bands. Thus, the surface states in Figs. \ref{Fig:2}(b) and (c) are essentially different. If these degeneracies are removed by the breaking of TR symmetry, there will be no difference between the surface bands in Figs.~\ref{Fig:2}(b) and (c).

From this discussion, one can clearly see that the degeneracies at $\bar{\Gamma}$ and $\bar{\mathrm{X}}$ play an essential role in protecting topological surface states. This physical picture can be naturally generalized. If one can find a similar surface dispersion as that in Fig.~\ref{Fig:2}(b) and the degeneracies at the $\bar{\Gamma}$ and $\bar{\mathrm{X}}$ points are protected by other types of symmetry, surface states will be robust against any perturbation at the surface once the symmetry exists. In a crystal, degeneracy is usually induced by crystalline symmetry. Thus, we will next study how different types of degeneracy due to crystalline symmetry can help to protect non-trivial surface states.

We again consider a semi-infinite system in 3D with one surface as shown in Fig.~\ref{Fig:2}(a), which can be described by 2D space group instead of 3D space group since there is no translational symmetry along the direction normal to the surface. The 2D BZ is defined for this semi-infinite system, as shown in Fig.~\ref{Fig:3}(a), and can be viewed as the projection of 3D bulk BZ into the 2D surface. Before discussing surface states of this semi-infinite system, it is necessary to first review some basic concepts of space group\cite{dresselhaus2008}\cite{bir1974}. The space group $G$ of a crystal consists of the symmetry operators with the form $g=\{r|\mathbf{R}+\bm{\tau}\}$ that transform the crystal into itself, where $r$ is a point group operation, $\mathbf{R}$ is a primitive translation of the corresponding Bravais lattice, and $\bm{\tau}$ is a non-primitive translation vector. Non-zero $\bm{\tau}$ appears in the compound crystals with more than one atoms in each primitive cell. The translation symmetry operators form an Abelian subgroup $T$. The factor group $F=G/T$ of $G$ with respect to $T$ is isomorphic to a point group. If one can find a set of generators for the whole space group $G$, for which one can always choose $\bm{\tau}=0$, the group $G$ is called a symmorphic group; otherwise, it is a non-symmorphic group. For the subgroup $T$ of translational operators, we can define the BZ in the momentum space. In the BZ, a momentum $\mathbf{k}$ can either be preserved under an operator or be transformed to another one. All the symmetry operations that preserve $\mathbf{k}$ form a subgroup of the space group $G$, usually known as a wave-vector group or a little group at the momentum $\mathbf{k}$, denoted as $G_{\mathbf{k}}$. The little group $G_{\mathbf{k}}$ also consists of translational group $T$ and the corresponding factor group is denoted as $F_{\mathbf{k}}$. The dimension of the Irreps of the wave-vector group $G_{\mathbf{k}}$ or the corresponding factor group $F_{\mathbf{k}}$ determines the degeneracy of energy bands at the momentum $\mathbf{k}$. In Appendix \ref{App:rep}, we give a systematic review of the representation theory of space groups.

By identifying the wave-vector group for each momentum in the BZ, we can figure out how different types of degeneracy occur in the whole BZ. The detailed analysis for 17 2D space groups will be presented in the next section. Here, we will first illustrate how the occurrence of degeneracy can lead to non-trivial surface states. We first look at the type-I degeneracy in a crystal. This type of degeneracy only occurs at some HSPs, where the corresponding wave-vector group consists of at least two symmetry operations that do not commute with each other. To have a non-trivial surface state as that in Fig.~\ref{Fig:2}(b), we require two separate HSPs with degenerate states, denoted as $K_1$ and $K_2$. This degeneracy should be split in the line $K_1$-$K_2$ except $K_1$ and $K_2$. The TCI due to type-I degeneracy, dubbed type-I TCI, can exist when the above conditions are satisfied. This can occur in both symmorphic and non-symmorphic crystals for the spinless case. For symmorphic crystals, since the wave vector group is identical to a certain point group, the corresponding Irreps can be directly read out from the character tables of the point group. By searching for point groups with high-dimensional Irreps, we can easily identify which types of crystal structures can host type-I TCIs. It turns out that among 17 2D space groups, the symmorphic groups $p4m$, $p31m$, and $p6m$ are allowed for type-I TCIs.
Here, we emphasize that for a symmorphic crystal, aside from high-dimensional Irreps, 1D Irreps also exist. Thus, if some states belonging to these 1D Irreps coexist with non-trivial surface states near the Fermi energy, the system becomes trivial.
For non-symmorphic crystals, the situation is different and the eigenstates are no longer related to the representations of the point group. Instead, the so-called projective representations \cite{bir1974} of point groups are required to describe eigenstates in a non-symmorphic system. It turns out that at certain HSPs, the projective representations only possess high-dimensional Irreps. Correspondingly, all states at these momenta are at least doubly degenerate, in contrast to the symmorphic case. Therefore, the type-I TCI is more robust for non-symmorphic crystals. The non-symmorphic groups $pmg$, $pgg$, and $p4g$ belong to this case. When spin is taken into account, we need to consider the spinor representations (or double representations) of point groups. The spinor representation can also be included into the theory of projective representations, which is discussed in details in Appendix \ref{App:rep}. For the spinful case, the type-I TCI can exist in $pmm$, $pmg$, $pgg$, $cmm$, $p4m$, $p4g$, $p31m$ and $p6m$ groups.

In constrast to the type-I TCIs, the type-II degeneracy occurs because of multiple different Irreps. As discussed above, the type-II degeneracy requires a tuning parameter, which can be taken as the momentum $k$ in the BZ. Thus, it is required that multiple Irreps should exist at least on a momentum line in the BZ. In the 2D space group, only mirror and glide symmetry can exist on a momentum line and allow for multiple Irreps. Thus, the type-II degeneracy is possible for the $pm$, $pg$, $cm$, $pmm$, $pmg$, $pgg$, $cmm$, $p4m$, $p4g$, $p3m1$, $p31m$ and $p6m$ groups. The TCI due to the type-II degeneracy is dubbed the type-II TCI. One example of the type-II TCI is the mirror Chern insulator\cite{hsieh2012}, for which gapless surface states are protected by mirror symmetry. It should be emphasized that for the groups with type-II degeneracy listed above, gapless surface states are not always allowed. One example is that mirror Chern insulators do not exist in the $pmm$ for spinless fermions. We will analyze these different situations separately below.

Before considering concrete 2D space groups, we summarize our major steps of identifying crystal space groups with the required symmetry for TCIs. For a given crystal, the wave-vector group for each momentum in the surface BZ of a semi-infinite system should be first identified. Then, based on the representation theory of wave-vector groups, degeneracy can be extracted for each momentum. If we can find two or more than two momenta with high-dimensional Irreps, the type-I TCI is possible. If we can find a line of momenta in the BZ with more than one Irreps, the type-II TCI can exist. It should be emphasized that type-I and -II TCIs can co-exist in one space group. Our analysis of the relation between non-trivial surface states and degeneracy can be applied to each 2D space group, which will be discussed in details following. Although the discussion on non-trivial surface states can indicate topologically nontrivial phase in the bulk system based on the surface-bulk correspondence\cite{hatsugai1993chern}, this approach does not directly give us topological nature of bulk systems, as well as bulk topological invariants. It also does not tell us which kinds of models can be topologically nontrivial. Thus, it is necessary to construct toy models for TCIs and compute topologically non-trivial surface states explicitly. We will also identify topological invariants for each class of TCIs.

\section{Classification of TCIs based on 2D space groups}\label{Sec:Classify}
Next, we will reveal our results for the classification of 17 2D space groups for TCI phases based on the approach described above. It turns out that there are no TCIs in the $p1$, $p2$, $p3$, $p4$, and $p$6 groups. The type-II TCI can exist in the $pm$, $pg$, $cm$, and $p3m1$ groups for both single (spinless fermions) and double groups (spinful fermions). For the $pmm$ and $cmm$ groups, there is no TCI for the single-group case but for the double group case, both type-I and -II degeneracies are possible, leading to a mixed type--I-II TCI phase. For the $pmg$, $pgg$, $p4m$, $p4g$, $p31m$ and $p6m$ groups, both type-I and -II TCIs can exist for both single and double groups. It is interesting to notice that type-I degeneracy is always accompanied with type-II degeneracy for a 2D space group. This is because all the HSPs can be connected by momentum lines that are invariant under mirror or glide operations.

\subsection{$pm$, $p3m1$ and $cm$ groups}\label{Sec:pm}

\begin{figure} 
\includegraphics[width=8.6 cm]{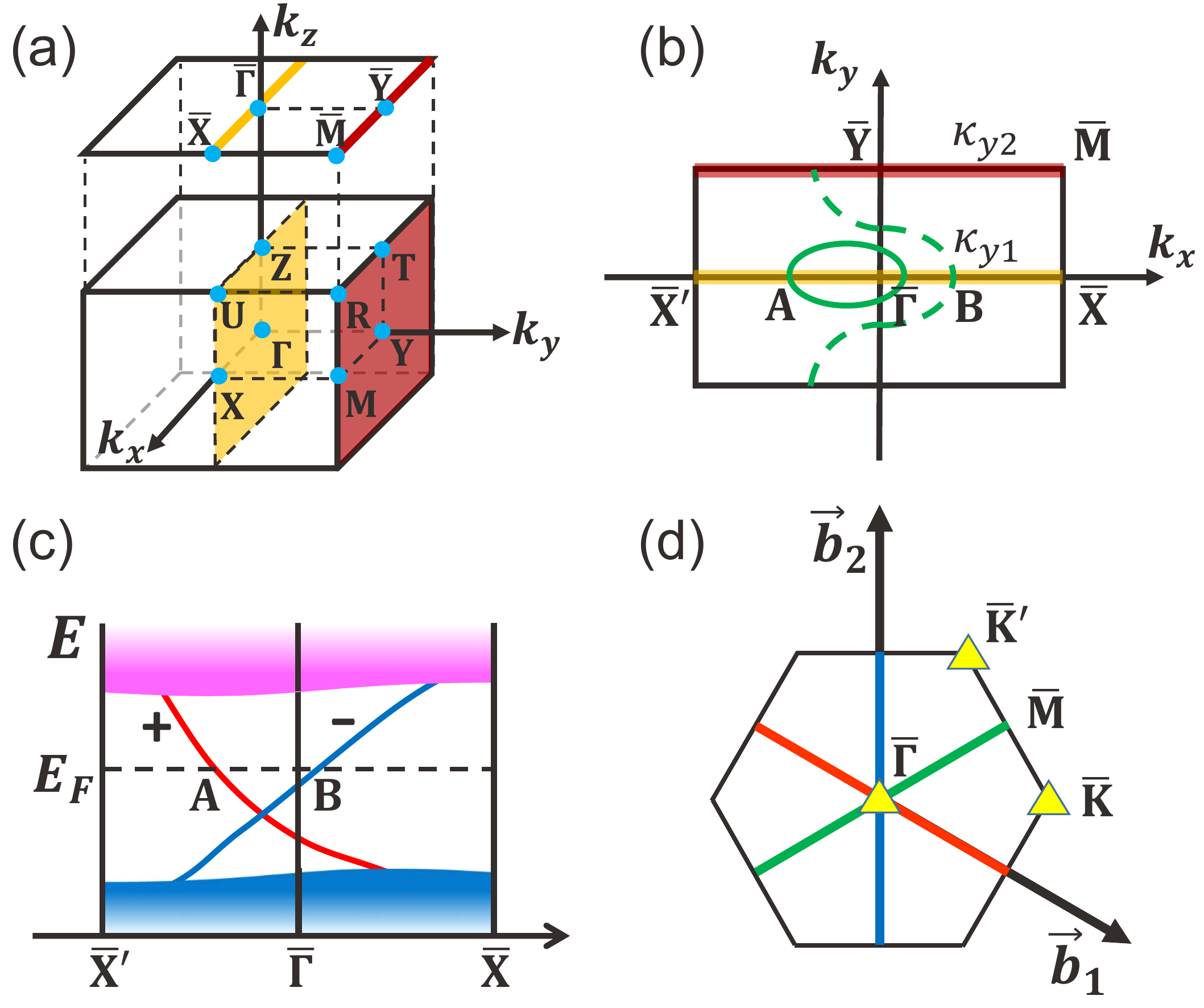}
\caption{(Color online) (a) Schematic plot of the $z$-projection relation between the 3D bulk BZ and the surface BZ. The points in the surface BZ are marked with bar over them. (b) Schematic plot of the surface BZ of $pm$ group with two $m_y$ invariant lines high-lighted and denoted by $\kappa_{y1}$ and $\kappa_{y2}$. (c) Schematic plot of surface states of TCI in $pm$ group on $\bar{\mathrm{X}}'$-$\bar{\Gamma}$-$\bar{\mathrm{X}}$ with $C_{o1}=-C_{e1}=1$. The red and blue lines denote the bands in even and odd mirror subspace, respectively. (d) Schematic plot of the BZ of $p3m1$ group. The MILs are denoted by the red, blue and green lines. The three-fold invariant points are marked by the yellow triangles.}\label{Fig:3}
\end{figure}

We start our discussion from the systems with $pm$, $p3m1$ and $cm$ groups, for which only type-II degeneracy exists in both the spinless and spinful cases. As stated above, a symmetry invariant line is needed to realize the type-II TCI. In a 2D space group, only the mirror reflection and glide symmetry operations can exist along a line. The most simple case is the $pm$ group, which is a symmorphic group with parallel mirror reflection axes. Let us assume the mirror operation is along the $y$-direction, described by the notation $m_y:(x,y)\rightarrow(x,-y)$. As shown in Fig.~\ref{Fig:3}(b), there are two mirror symmetry invariant momentum lines (MILs), denoted as $\kappa_{y1}=(k_x,0)$ and $\kappa_{y2}=(k_x,\pi)$, in the surface BZ. Along these momentum lines, the wave vector group is given by $C_{1h}$ and has two 1D Irreps, which can be distinguished by the parity $m_{\pm}=\pm i^f$ of the mirror operator $m_y$, where $f=0$ for the spinless case and $f=1$ for the spinful case. Consequently, on each line, the Hamiltonian can be diagonalized into two blocks, which are distinguished by mirror parities. Thus, each block belongs to the subspace of the states with a definite mirror parity $m_+$ or $m_-$, which is dubbed the mirror even or odd subspace. If two bands belonging to two different mirror subspaces cross each other along these momentum lines, the crossing point is protected by the mirror symmetry.

After understanding the wave vector group of the $pm$ group, we next ask what types of Fermi surfaces and energy dispersions are possible for non-trivial surface states. Since the Fermi surface must be a closed contour on the surface BZ (a torus) of a bulk insulator, the Fermi surface always crosses the HSLs $\kappa_{y1}$ and $\kappa_{y2}$ with even number of times. There are two possibilities: (i) the Fermi surface intersects with each of the HSLs an even number of times; or (ii) it crosses both $\kappa_{y1}$ and $\kappa_{y2}$ with an odd number of times. For the case (i), we may consider an example with the Fermi surface crossing the line $\kappa_{y1}$ twice, as depicted by the points $A$ and $B$ in Fig.~\ref{Fig:3}(b), in which the solid green circle is the contour of the Fermi surface for surface states. The energy dispersions around these two points have opposite velocities for non-chiral surface states. We further require that the states at these two crossings points have opposite mirror parities so that the surface states can be gapless. An example of this type of surface energy dispersion is shown along the $\bar{\mathrm{X}}'$-$\bar{\Gamma}$-$\bar{\mathrm{X}}$ ($\kappa_{y1}$) line with mirror symmetry $m_y$ in Fig.~\ref{Fig:3}(c), where the red and blue lines denote surface bands with even and odd mirror parities, respectively. Because of opposite velocities, these two surface bands must cross at a certain momentum along the $\bar{\mathrm{X}}'$-$\bar{\Gamma}$-$\bar{\mathrm{X}}$ line. Due to opposite mirror parities, the crossing point is protected by mirror symmetry and cannot open a gap. Thus, type-II TCI exists in the $pm$ group. For case (ii), we may consider an example of the Fermi surface crossing both $\kappa_{y1}$ and $\kappa_{y2}$ once, forming an open Fermi surface, as shown by the dashed green line in Fig.~\ref{Fig:3}(b). In this case, the surface energy dispersion must be chiral along the lines $\kappa_{y1}$ and $\kappa_{y2}$. This corresponds to either a layered quantum Hall system \cite{balents1996chiral,taherinejad2014wannier} or a Weyl semimetal \cite{wan2011topological,xu2011chern}. In this paper, we only focus on the insulating systems with non-chiral surface states in the case (i).

Next, we discuss bulk topological invariants for the TCI of the $pm$ group. As shown in Fig.~\ref{Fig:3}(a), we can always find a mirror invariant plane (MIP), denoted as $M_1$ and $M_2$ in the 3D BZ, which are projected into the MILs $\kappa_{y1}$ and $\kappa_{y2}$ in the surface BZ. We can treat a MIP as a 2D BZ and define the Chern number for the occupied states in the even (odd) subspace for a MIP as
\begin{eqnarray}
  C_{e(o)}=\frac{1}{2\pi}\int_{\text{MIP}}d\bm{\Omega} \cdot\mathbf{F}_{e(o)}(\mathbf{k}),
\end{eqnarray}
where $\mathbf{F}_{e(o)}(\mathbf{k})$ is the Berry curvature of the occupied energy bands in the even (odd) subspace
\begin{eqnarray}
  \mathbf{F}_{e(o)}(\mathbf{k})=i\sum_{n\in occ,e(o)}\nabla_{\mathbf{k}}\times\langle u_{n,\mathbf{k}}^{e(o)}| \nabla_{\mathbf{k}}|u_{n,\mathbf{k}}^{e(o)}\rangle.
\end{eqnarray}
Here, $|u^{e(o)}_{n,\mathbf{k}}\rangle$ is eigenfunction of the Hamiltonian $H(\mathbf{k})$ for the $n$-th occupied energy band, and $d\bm{\Omega}$ is an infinitesimal, directed area of the MIP. Since there is no coupling between two mirror subspaces, the Chern numbers $C_{e}$ and $C_{o}$ are independent. Alternatively, one can also define the mirror Chern number \cite{teo2008surface}
\begin{eqnarray}
  C_{M}=\frac{1}{2}(C_e-C_o)
\end{eqnarray}
aside from the total Chern number $C=\frac{1}{2}(C_e+C_o)$.

The surface states for the $pm$ group can be directly related to two Chern numbers $C_i$ and $C_{Mi}$ (or $C_{oi}$ and $C_{ei}$) defined above, where $i=1,2$ denotes the MIPs $M_1$ and $M_2$, respectively. In the case (i), surface energy dispersion in Fig.~\ref{Fig:3}(c), since there is a chiral edge mode along the 1D line $\kappa_{y1}$ in each mirror subspace (even or odd), this suggests that Chern number in each mirror subspace is non-zero for the MIP $M_1$ and we have $C_{o1}=-C_{e1}=1$ according to the bulk-edge correspondence \cite{hatsugai1993chern}. This corresponds to the non-chiral case with the total Chern number $C_1=C_2=0$, but the MCN $C_{M1}=-1$ and $C_{M2}=0$. This type of TCI is also called ``mirror Chern insulator'' due to its non-zero MCN\cite{teo2008surface,hsieh2012}. The case (i) of the $pm$ group can be characterized by two independent MCNs $C_{M1}$ and $C_{M2}$ \cite{kim2015layered}, leading to the $\mathbb{Z}^2\equiv\mathbb{Z}\times\mathbb{Z}$ classification. For the case (ii), not only MCNs, but also the total Chern numbers in the MIPs $M_1$ and $M_2$ are non-zero, leading to either the layered quantum Hall effect or Weyl semi-metals, as discussed above.

Similar discussion of type-II TCIs can also be applied to the $p3m1$ and $cm$ groups. For the $p3m1$ group, the surface BZ is shown in Fig.~\ref{Fig:3}(d), where $\vec{b}_1$ and $\vec{b}_2$ are the reciprocal lattice vectors. The blue, red, and green lines are three equivalent MILs, and the yellow triangles mark the positions of three-fold-rotation-invariant points. Since these MILs are related to each other by three-fold rotations, only one independent MCN can be defined on the MIPs that are projected to these MILs, giving rise to the $\mathbb{Z}$ classification of the $p3m1$ group \cite{alexandradinata2014spin}. The point group of $\bar{\Gamma}$ is $C_{3v}$ and allows for both 1D and 2D Irreps. If all the bands at $\bar{\Gamma}$ belong to the 2D Irreps, the crossing points between the two surface bands with opposite mirror parities will be pinned at the $\bar{\Gamma}$ point. Otherwise, the crossing points can locate at any place of the MILs.

For the $cm$ group, we consider the rhombic BZ shown in Fig.~\ref{Fig:1}(d), instead of the standard Wigner-Seitz cell. In this case, the mirror-symmetric line is $\bar{\mathrm{X}}'$-$\bar{\Gamma}$-$\bar{\mathrm{X}}$ for mirror reflection $m_y$. Different from the $pm$ group, there is only one MIL in the first BZ, leading to one MCN ($\mathbb{Z}$ classification) for the $cm$ group.

For the $pm$, $p3m1$ and $cm$ groups, the spinful case is similar to spinless case and the corresponding topological classifications are equivalent.

\subsection{$pmm$ and $cmm$ group}\label{Sec:pmm}
 For the $pmm$ group, the factor group $F$ is given by the point group $C_{2v}$, which is generated by two mirror symmetries with perpendicular mirror reflection axes (let's take them as $m_x$ and $m_y$). For the spinless case, the $pmm$ group does not have any non-trivial surface states, while a mixed type--I-II TCI can exist for the spinful case. The essential difference between these two cases is the commutation relation between two mirror operators $m_x$ and $m_y$. Direct calculation gives $m_xm_y=C_2(z)$ and $m_ym_x=Q(y)C_2(z)$, where $C_2(z)$ denotes the $\pi$ rotation around the $z$ axis and $Q(y)$ is for the $2\pi$ rotation around the $y$ axis. For the spinless case, two mirror operators commute with each other, $[m_x,m_y]=0$, since $Q(y)=1$. In contrast, for the spinful case, the $2\pi$ rotation $Q(y)$ gives $-1$, leading to $\{m_x,m_y\}=0$.

\begin{figure} 
\includegraphics[width=8.6 cm]{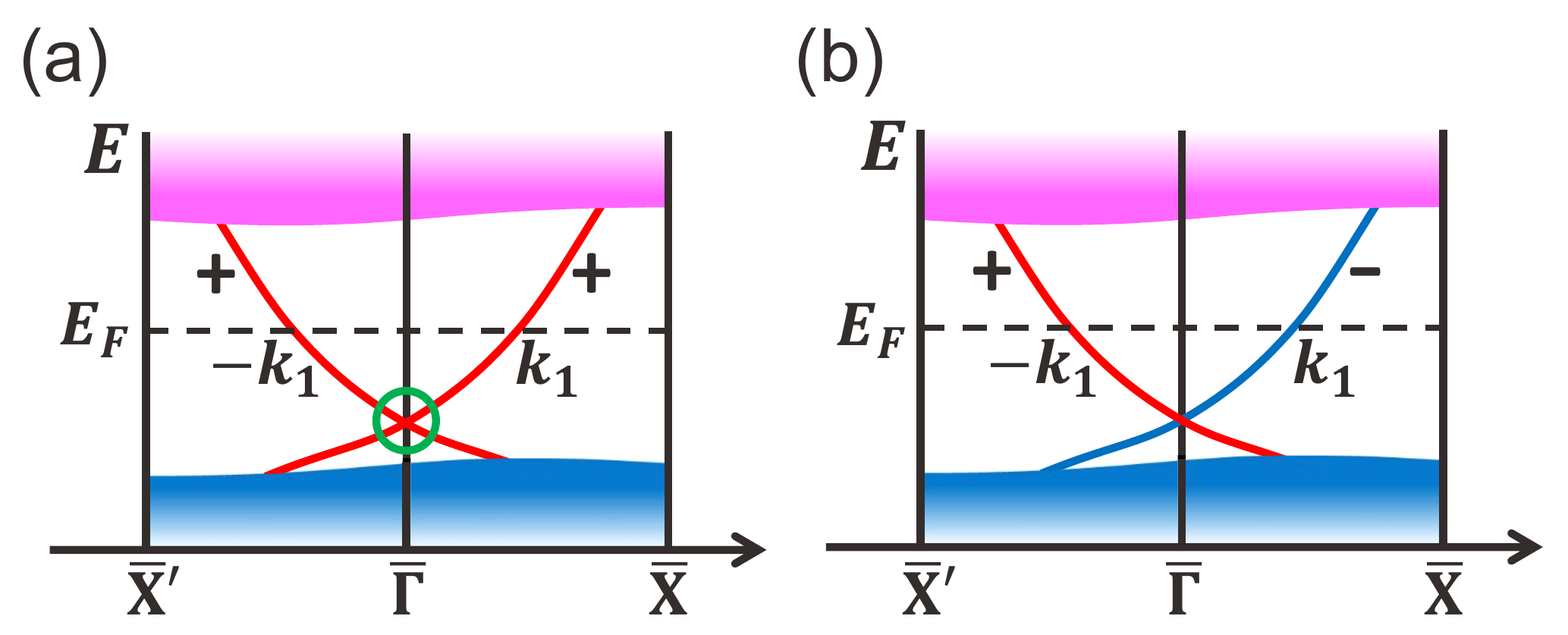}
\caption{(Color online) Schematic plot of surface states of TCIs in the $pmm$ group. The red and blue lines denote the bands with even and odd mirror parities, respectively. (a) The trivial phase in spinless case. The crossing marked by the green circle is not protected and can be gapped. (b) The non-trivial phase in the spinful case. }\label{Fig:4}
\end{figure}

Next let us check how the commutation relation between $m_x$ and $m_y$ makes the surface states trivial for the spinless case. Compared to the $pm$ group, the $pmm$ group has an additional mirror reflection symmetry. We consider surface bands in the line $\kappa_{y1}=(k_x,0)$ that is invariant under $m_y$, and assume that there is a surface state with the momentum $(k_1,0)$ at the Fermi energy along the momentum line $\kappa_{y1}$. The mirror operation $m_x$ requires that at the Fermi energy, another surface state must exist with the momentum $(-k_1,0)$, as shown in Fig.~\ref{Fig:4}. For spinless fermions, since $m_x$ and $m_y$ commute with each other, these two branches of surface states at $(\pm k_1,0)$ possess the same mirror parity under $m_y$ but opposite velocities, as depicted in Fig.~\ref{Fig:4}(a). Thus, the crossing between these two branches at the $\bar{\Gamma}$ point (marked by the green circles) will open a gap, resulting the trivial surface states.

In contrast, the spinful case of the $pmm$ group is dramatically different from the spinless case since $m_x$ and $m_y$ anti-commute with each other. There are two consequences due to the anti-commutation relation. Firstly, at HSPs, such as $\bar{\Gamma}$, $\bar{\mathrm{X}}$, $\bar{\mathrm{Y}}$, and $\bar{\mathrm{M}}$, of which the wave vector group includes both $m_x$ and $m_y$, the anti-commutation relation yields type-I degeneracy. Secondly, the surface state at $( k_1,0)$ in the line $\kappa_{y1}$ has opposite mirror parities of $m_y$ compared to that at $(-k_1,0)$ for the spinful case, as shown in Fig.~\ref{Fig:4}(b), which is in sharp contrast to the same mirror parity for the spinless case. Since these two branches of surface bands in Fig.~\ref{Fig:4}(b) belong to different mirror parity subspaces, no coupling is allowed between these two branches. Thus, type-II degeneracy can also exist in this group.

\begin{figure} 
\includegraphics[width=8.6cm]{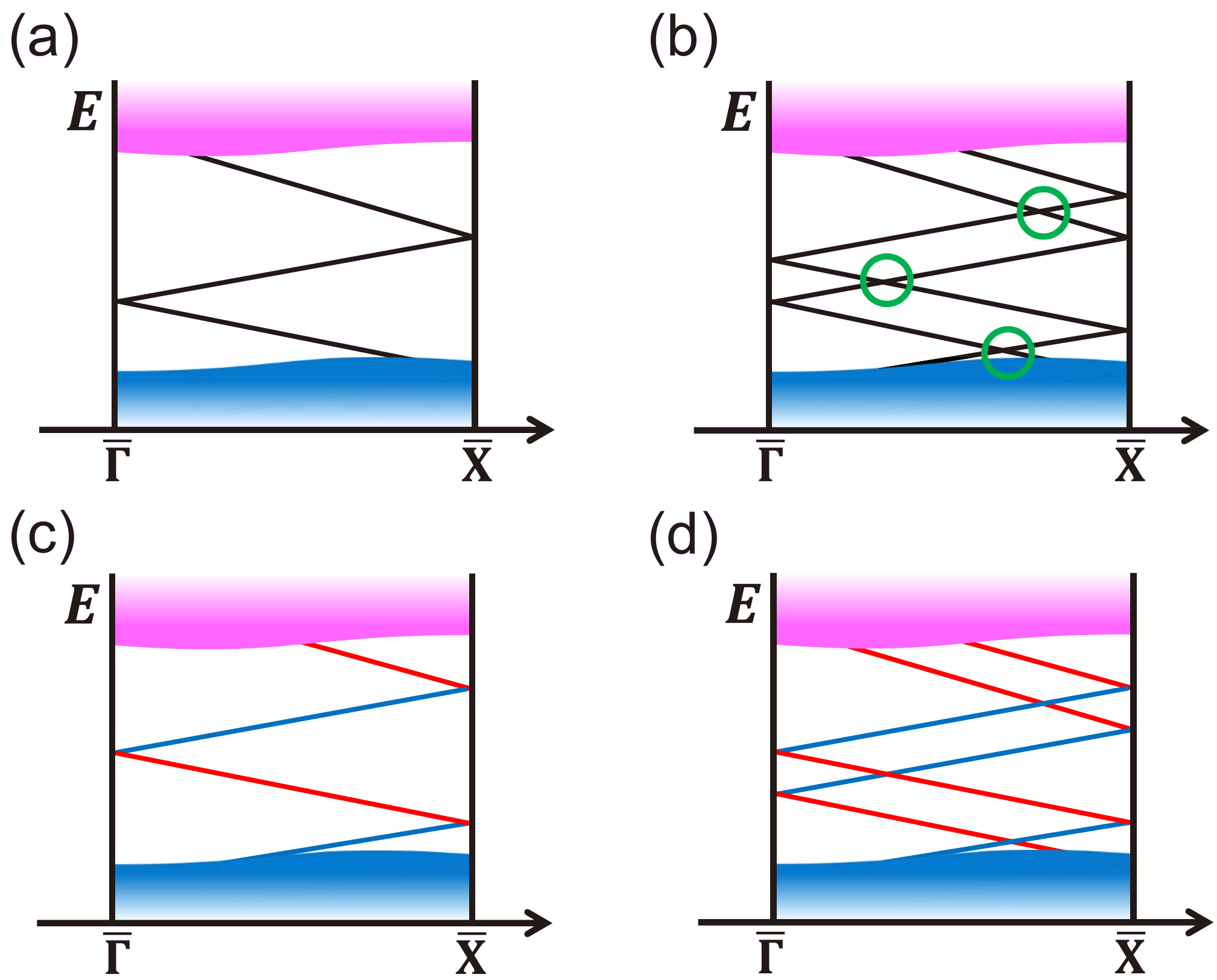}
\caption{(Color online) (a) Schematic plot of non-trivial surface states of TR invariant TIs. (b) Schematic plot of the trivial surface states which are two copies of that in (a). The crossings marked by the green circles are not protected and gap can be opened. (c) Schematic plot of the non-trivial surface states of the $pmm$ group in the spinful case with $C_M=1$. (d) Schematic plot of non-trivial surface states of the $pmm$ group in the spinful case with $C_M=2$. The red and blue lines in (c) and (d) denote the bands with even and odd mirror parities, respectively.}\label{Fig:5}
\end{figure}

The coexistence of type-I and -II degeneracies suggests a $\mathbb{Z}$ classification of the mixed type--I-II TCI phase in the $pmm$ group for the spinful case, which is different from the $\mathbb{Z}_2$ classification of TR invariant TIs. This difference can be illustrated in Fig.~\ref{Fig:5}, in which we compare surface states for TR invariant TIs and those of TCIs in the $pmm$ group. In Figs.~\ref{Fig:5}(a) and (b), we consider one and two copies of surface states for the TR invariant TIs, respectively. One surface state is stable while for two copies of surface states, the crossings (marked by the green circles) between them on the line connecting two TR invariant points [$\bar{\Gamma}$-$\bar{\mathrm{X}}$ in Figs.~\ref{Fig:5}(a) and (b)] are not protected, and thus a gap can be opened, leading to a trivial phase. In contrast, for the mixed type--I-II TCI phase of the $pmm$ group, the momentum line $\bar{\Gamma}$-$\bar{\mathrm{X}}$ is a MIL with $m_y$ symmetry in Fig.~\ref{Fig:5}(c). The doubly degenerate states at the HSPs $\bar{\Gamma}$ and $\bar{\mathrm{X}}$ always have opposite mirror parities of $m_y$, shown by red and blue lines in Fig.~\ref{Fig:5}(c), respectively. For the case of two copies of surface states of the mixed type--I-II TCI phase, the crossing points along the line $\bar{\Gamma}$-$\bar{\mathrm{X}}$ are between two states with opposite mirror parities [Fig.~\ref{Fig:5}(d)] and thus protected by the type-II degeneracy. Therefore, in the mixed type--I-II TCI phase, any integer number of branches of surface bands are stable.

All the above conclusions based on the analysis of surface states are consistent with the Chern number in the bulk systems. Detailed analysis of the total Chern number and MCN in this system can be found in the Appendix \ref{App:MCN}. It turns out that the MCN for a MIP must be zero for the spinless case, while it can be any integer for the spinful case. In the surface BZ, there are four independent MILs, and thus the topological classification of the system is $\mathbb{Z}^4\equiv\mathbb{Z}\times\mathbb{Z}\times\mathbb{Z}\times\mathbb{Z}$ for insulating systems. Further analysis shows that the system possesses gapless Weyl points in the 3D bulk if $\sum_{i=1}^4C_{Mi}$ is an odd number \cite{alexandradinata2014spin}.

Just like the similarity between the $pm$ and $cm$ groups, the discussion of the $cmm$ group is almost identical to that of the $pmm$ group. The only difference is the number of MILs in the BZ. For the $cmm$ group, there are only two MILs $\bar{\mathrm{X}}'$-$\bar{\Gamma}$-$\bar{\mathrm{X}}$ and $\bar{\mathrm{Y}}'$-$\bar{\Gamma}$-$\bar{\mathrm{Y}}$ in the rhombic BZ, as shown in Fig.~\ref{Fig:1}(d). Correspondingly, the classification of the $cmm$ group is $\mathbb{Z}^2$ in the spinful case, and no topologically non-trivial phase in the spinless case.

\subsection{$pg$ group}\label{Sec:pg}
The $pg$ group is generated by a glide symmetry $g_y=\{m_y|\bm{\tau}_x\}$ with $\bm{\tau}_x=(1/2,0)$, which is a combination of mirror operation and nonprimitive translation. The BZ of the $pg$ group is similar to that of the $pm$ group, and the glide symmetry exists along the lines $\kappa_{y1}$ and $\kappa_{y2}$ in Fig.~\ref{Fig:3}(b). The difference between the $pm$ and $pg$ group lies in the fact that the nonprimitive translation introduces an additional phase factor for the eigenvalues of glide operation, compared to that of mirror operation. We may take the line $\kappa_{y1}$ ($\bar{\mathrm{X}}'$-$\bar{\Gamma}$-$\bar{\mathrm{X}}$) as an example. Along this line, all the eigenstates can be characterized by the eigenvalues of glide operation (mirror operation) for the $pg$ ($pm$) group and the Hamiltonian can be block diagonal with two blocks. For the $pm$ group, these two blocks of the Hamiltonian can be labeled by the mirror eigenvalue $m_{\pm}=\pm i^f$ (even or odd subspace). In each subspace, the block Hamiltonian is periodic in the BZ, so $H_{+(-)}(\bar{\mathrm{X}}')=H_{+(-)}(\bar{\mathrm{X}})$, where $H_{+(-)}$ denotes the block Hamiltonian with the mirror parity $+i^f$ ($-i^f$). Thus, Chern number can be defined in each subspace. In contrast, for the $pg$ group, the eigenvalues of the glide operation $g_{\pm}(k_x)=\pm i^fe^{ik_x/2}$, which is dubbed ``glide parity,'' depends on the momentum $k_x$, where $f=0$ in the spinless case and $f=1$ in the spinful case. When $k_x$ changes by $2\pi$, there is an additional minus sign from the phase factor $e^{ik_x/2}$. As a consequence, $g_{\pm}(\bar{\mathrm{X}}')=g_{\mp}(\bar{\mathrm{X}})$ and the corresponding block Hamiltonian in each subspace is no longer periodic. Instead, we have $H_{+(-)}(\bar{\mathrm{X}}')=H_{-(+)}(\bar{\mathrm{X}})$. Therefore, for each subspace, the 2D BZ can not form a closed manifold to define a Chern number. However, the above property of the glide symmetry suggests that all the bands along the glide symmetry invariant lines must come in pairs with one band of $g_+(k_x)$ parity and the other of $g_-(k_x)$ parity. These properties have been shown to lead to a $\mathbb{Z}_2$ classification of TCIs in the $pg$ group\ cite{fang2015new}\cite{shiozaki2015z}.

\begin{figure} 
\includegraphics[width=8.6 cm]{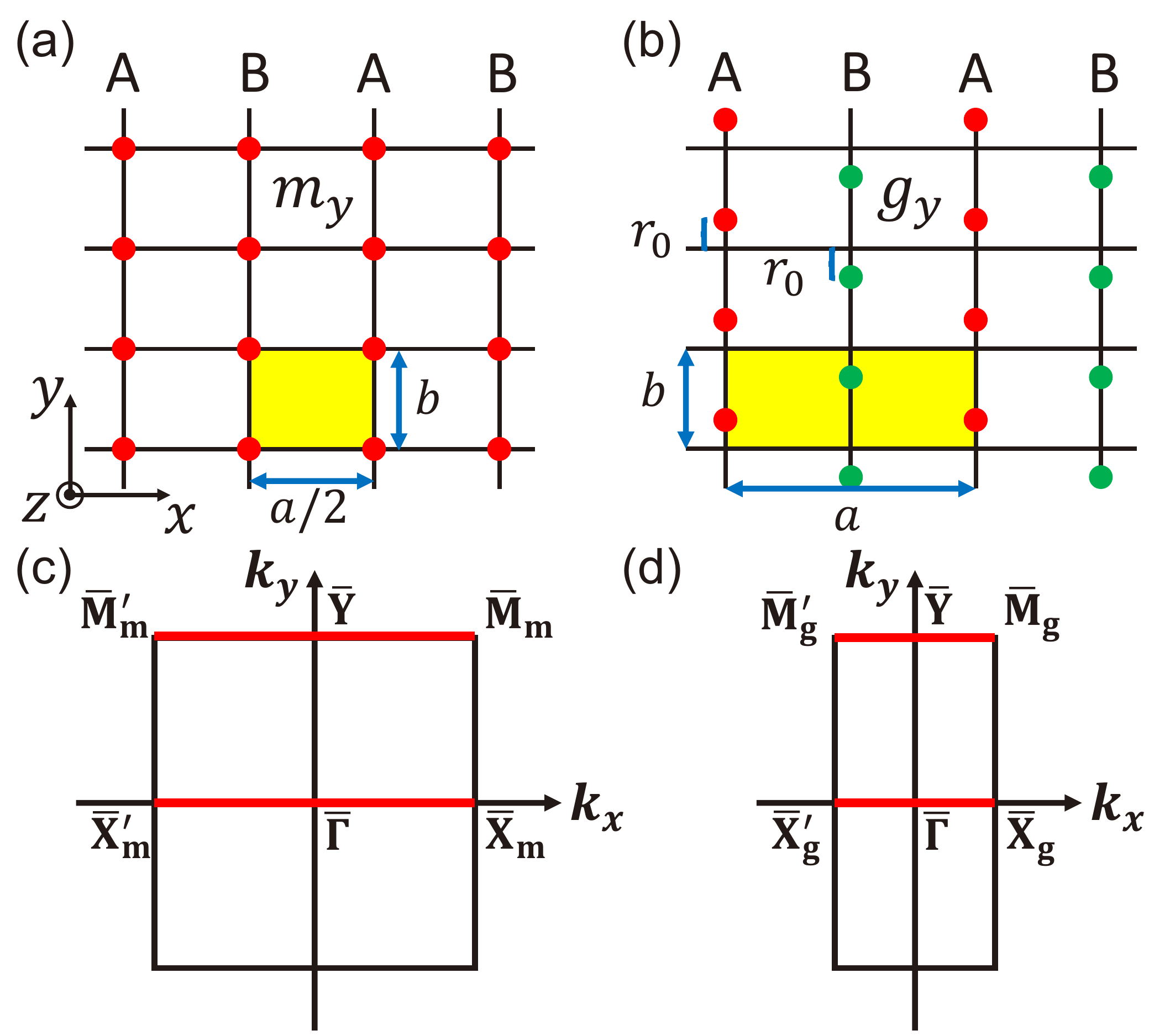}
\caption{(Color online)(a) Schematic plot of a lattice of the $pm$ group with mirror symmetry $m_y$. The lattice constants in $x$, $y$, and $z$ directions are $a/2$, $b$, and $c$, respectively. (b) Schematic plot of a lattice of the $pg$ group, which is obtained by a small distortion of the lattice in (a), with glide plane symmetry $g_y$. The atoms in layer A move in the $y$ direction by $r_0$, while the atoms in layer B move by $-r_0$. The yellow regions in (a) and (b) denote a primitive cell of the corresponding lattice. (c) Schematic plot of the surface BZ of the lattice in (a), where $\bar{\mathrm{X}}_m=(2\pi/a,0)$ and $\bar{\mathrm{X}}'_m=(-2\pi/a,0)$. (d) Schematic plot of the surface BZ of the lattice in (b), where $\bar{\mathrm{X}}_g=(\pi/a,0)$ and $\bar{\mathrm{X}}'_g=(-\pi/a,0)$. }\label{Fig:6}
\end{figure}

Since the $pg$ group is closely related to the $pm$ group, we will discuss how TCI phases in these two groups are related to each other below. Here, we consider a lattice of the $pm$ group, as shown in Fig.~\ref{Fig:6}(a). It is a layered structure stacked along the $x$ direction, of which each layer in the $y$-$z$ plane is a rectangular lattice. we consider the $pm$ group symmetry of this lattice with mirror symmetry $m_y:(y\rightarrow -y)$. The sites in the layers A and B are equivalent to each other and thus there is only one atom in each primitive cell for the $pm$ group. The corresponding lattice constants along the $x$, $y$, $z$ directions are $a/2$, $b$ and $c$, respectively. Now let us shift the layers A and B in the opposite directions along the $y$ axis with the same displacement $r_0$, as shown in Fig.~\ref{Fig:6}(b). The mirror symmetry is broken and changed to a glide plane symmetry $g_y=\{m_y|(\frac{a}{2},0,0)\}$. As a consequence, the primitive cell is doubled in the real space and the corresponding BZ is halved along the $x$ direction, as illustrated in Figs.~\ref{Fig:6}(c) and (d).

\begin{figure}
\includegraphics[width=0.5\textwidth]{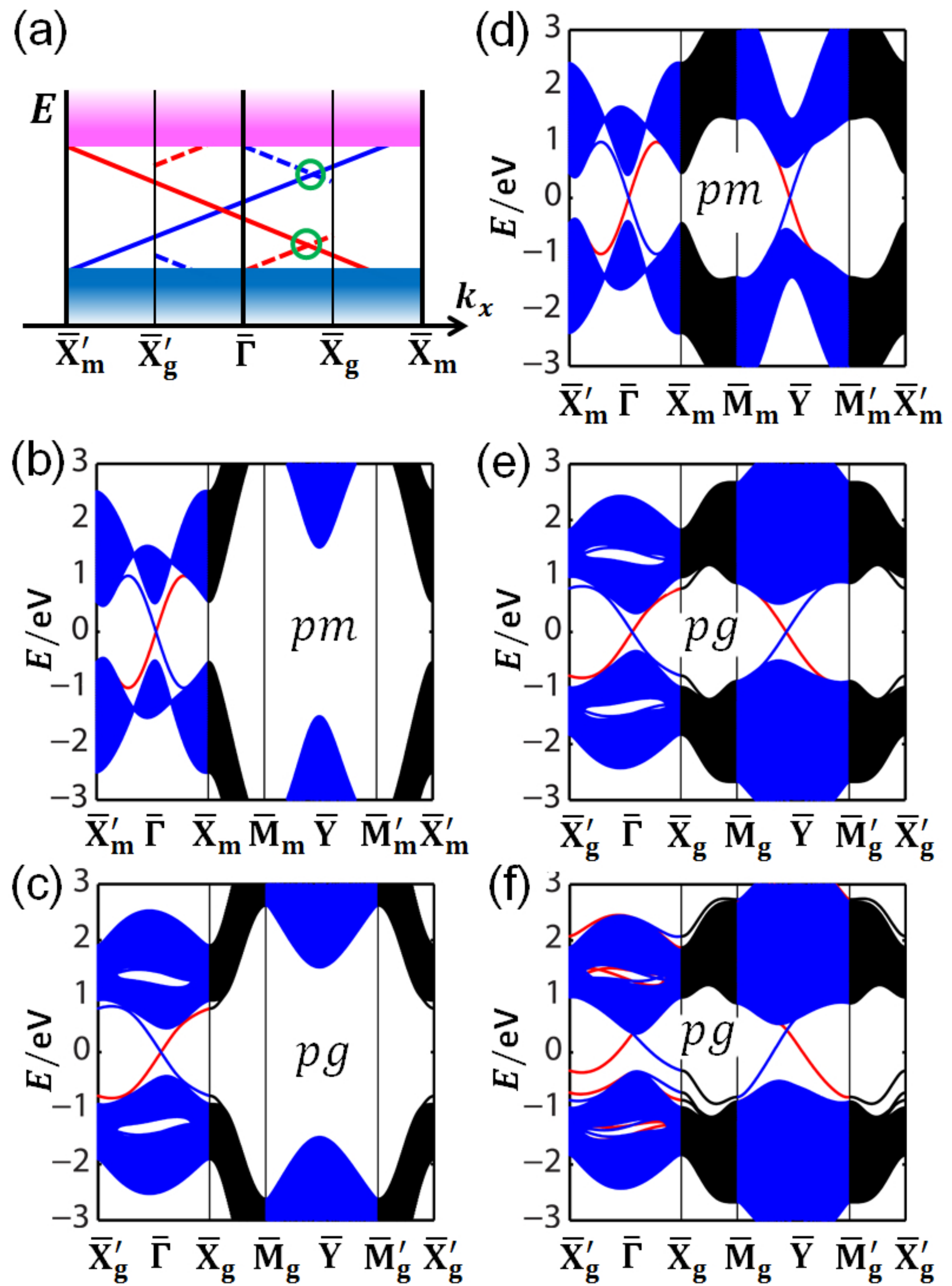}
\caption{(Color online)(a) Illustration of the BZ folding process from the lattice of $pm$ group to $pg$ group. The range of the first BZ of the lattice of $pm$ group in the $k_x$ direction is $[-2\pi/a,2\pi/a]$, and that of $pg$ group is $[-\pi/a,\pi/a]$. The solid lines are the bands of $pm$ group in the bulk gap, and the dashed lines are the bands that appear after the BZ folding. The band with glide even (odd) parity is drawn in red (blue) color. (b)The band dispersion of the model of the $pm$ group in a slab configuration with one Dirac cone on $\bar{\mathrm{X}}'_m$-$\bar{\Gamma}$-$\bar{\mathrm{X}}_m$. We only show the surface bands in the bulk gap on one surface of the slab. The parameters are $a=b=c=1$, $m_0=-2$, $t_0'=1.5$, $t_0=t_0''=t=t'=t''=1$, $\phi=0.4$. The number of layers in the $z$ direction is $N=60$. (c) The band dispersion of the model of the $pg$ group in a slab configuration. The parameters are the same with those in (b). The perturbation term $t_3=0.2$. (d)The band dispersion of the model of the $pm$ group with one Dirac cone on $\bar{\mathrm{X}}'_m$-$\bar{\Gamma}$-$\bar{\mathrm{X}}_m$, and one Dirac cone on $\bar{\mathrm{M}}_m$-$\bar{\mathrm{Y}}$-$\bar{\mathrm{M}}'_m$. The parameters, which are different with those in (b), are $m_0=-0.9$, $t_0'=0.5$. (e) The band dispersion of the model of the $pg$ group with the same parameters in (d) and $t_3=0.2$. (f) The band dispersion of the model of the $pg$ group got by adding a surface potential $V=0.5$ eV on the first layer of the slab. The other parameters are the same with those in (e).}\label{Fig:7}
\end{figure}

Next, we consider a tight-binding model with non-trivial surface states on this lattice and show how surface states evolve when the lattice is distorted from the $pm$ group to the $pg$ group. Let us consider the $pm$ group case and assume there are four orbitals on each site, and the basis are denoted as $\{$$|A,1\rangle$, $|A,2\rangle$, $|A,3\rangle$, $|A,4\rangle$$\}$, given by
\begin{eqnarray}
  |A,i\rangle
  =\frac{1}{\sqrt{N}} \sum_{\mathbf{R}}e^{i\mathbf{k}\cdot \mathbf{R}}\varphi_{A,i}(\mathbf{r}- \mathbf{R}-\mathbf{r}_{A,i}).
\end{eqnarray}
where $i=1,2,3,4$, $\mathbf{R}$ is the lattice vector, $\mathbf{r}_{A,i}$ is the relative position of the atoms in one primitive cell and $\varphi_{A,i}(\mathbf{r}- \mathbf{R}-\mathbf{r}_{A,i})$ is the corresponding atomic orbital wave function.
Furthermore, we assume the mirror parity of the first two orbitals $|A,1\rangle$ and $|A,2\rangle$ is $+1$, while that of the other two orbitals $|A,3\rangle$ and $|A,4\rangle$ is $-1$, in the spinless case. Thus, the matrix representation of the mirror operator in the above basis is $\tilde{U}(m_y)=\Gamma_{30}$, where $\Gamma_{ij}=\sigma_i\otimes\sigma_j(i,j=0,1,2,3)$, $\sigma_i(i=1,2,3)$ are Pauli matrices, and $\sigma_0$ is a two by two identity matrix. The Hamiltonian in the momentum space is given by
\begin{eqnarray}
  H_{pm}(\mathbf{k})&=&[m_0+t_0\cos(k_x a/2)+t_0'\cos(k_y b)\nonumber\\&&+t_0''\cos(k_z c))]\Gamma_{03} +t\sin(\frac{a}{2}(k_x-\phi))\Gamma_{31}\nonumber\\
  &&+t_1\sin(k_z c)\Gamma_{02}+t_2\sin(k_yb)\Gamma_{11}, \label{Eq:Hpmpg1}
\end{eqnarray}
which satisfies $\tilde{U}(m_y)H(\mathbf{k})\tilde{U}(m_y)^{\dag}=H(m_y\mathbf{k})$. By choosing the parameters in certain regime, the Hamiltonian can possess non-zero MCN. For example, one surface Dirac cone in the line $\bar{\mathrm{X}}'_m$-$\bar{\Gamma}$-$\bar{\mathrm{X}}_m$ is shown for the surface energy dispersion of this model in a slab configuration in Fig.~\ref{Fig:7}(b) and two surface Dirac cones, one in $\bar{\mathrm{X}}'_m$-$\bar{\Gamma}$-$\bar{\mathrm{X}}_m$ and the other in $\bar{\mathrm{M}}'_m$-$\bar{\mathrm{Y}}$-$\bar{\mathrm{M}}_m$, are shown in Fig.~\ref{Fig:7}(d).

After distorting the lattice from the $pm$ group symmetry to the $pg$ group symmetry, the bases become $\{$$|A,1\rangle$, $|A,2\rangle$, $|A,3\rangle$, $|A,4\rangle$, $|B,1\rangle$, $|B,2\rangle$, $|B,3\rangle$, $|B,4\rangle$$\}$, and orbital corresponding to $|B,i\rangle$ has the same mirror parity as that of $|A,i\rangle$. According to the matrix of mirror operator $\tilde{U}(m_y)$, one can show that the glide operation exchanges the basis $|A,i\rangle$ and $|B,i\rangle$ as
\begin{eqnarray}
  g_y|A,i\rangle&=&\pm |B,i\rangle,\nonumber\\
  g_y|B,i\rangle&=&\pm e^{-ik_x a}|A,i\rangle,
\end{eqnarray}
where the coefficient takes $+$ if $i=1,2$ and $-$ if $i=3,4$. Thus, the matrix of the glide operation is
\begin{eqnarray}
  \tilde{U}'(g_y)=e^{-i\frac{k_x a}{2}}(\cos(\frac{k_x a}{2})\sigma_1+\sin(\frac{k_x a}{2})\sigma_2)\otimes\Gamma_{30}.
\end{eqnarray}

Next we need to re-write the Hamiltonian (\ref{Eq:Hpmpg1}) into the new basis of the distorted lattice. We notice that the nearest neighbor hopping along the $x$ direction in the original lattice corresponds to the hopping between the adjacent A and B sublattices in the distorted lattice. Following this rule, one can show that the Hamiltonian in the $pg$ group is given by
\begin{eqnarray}
  H_{pg}(\mathbf{k})&=&\sigma_0\otimes\big\{[m_0+t_0'\cos(k_y b)+t_0''\cos(k_z c)]\Gamma_{03}\nonumber\\&&+t_1\sin(k_z c)\Gamma_{02}+t_2\sin(k_y b)\Gamma_{11}\big\}\nonumber\\&&
+\frac{t_0}{2}\left[(1+\cos(k_x a))\sigma_1+\sin(k_x a)\sigma_2\right]\otimes\Gamma_{03}\nonumber\\
&&+t\bigg[\sin\left(\frac{a}{2}(k_x-\phi)\right)\cos\left(\frac{ak_x}{2}\right)\sigma_1 \nonumber\\&&+\sin\left(\frac{a}{2}(k_x-\phi)\right)\sin\left(\frac{ak_x}{2}\right)\sigma_2\bigg]\otimes \Gamma_{31} \label{Eq:Hpmpg2}
\end{eqnarray}
in the $r_0\rightarrow 0$ limit, where $\sigma$ denotes the Pauli matrices in the A-B sublattice space. In this limit, the Hamiltonian has both the glide and mirror symmetries, so the energy dispersion of the Hamiltonian (\ref{Eq:Hpmpg2}) can be obtained from that of the Hamiltonian (\ref{Eq:Hpmpg1}) by BZ folding. For a non-zero $r_0$, additional terms, such as $t_3\sigma_3\otimes \Gamma_{11}$, can exist in the Hamiltonian (\ref{Eq:Hpmpg2}), and break mirror symmetry while preserving glide symmetry.

Next, we will consider how surface states of TCIs in the $pm$ group in Fig.~\ref{Fig:7}(b) evolve due to the BZ folding for the $pg$ group. This is illustrated schematically in Fig.~\ref{Fig:7}(a). The surface bands between $\bar{\mathrm{X}}'_m$-$\bar{\mathrm{X}}'_g$ ($\bar{\mathrm{X}}_g$-$\bar{\mathrm{X}}_m$) will be shifted to the $\bar{\Gamma}$-$\bar{\mathrm{X}}_g$ ($\bar{\mathrm{X}}'_g$-$\bar{\Gamma}$). Since this shift corresponds to a change of momentum $k_x$ by $\pm 2\pi/a$, the glide parity of the corresponding bands will be changed after this shift [$g_{\pm}(k_x+\frac{2\pi}{a})=g_{\pm}(k_x-\frac{2\pi}{a})=g_{\mp}(k_x)$]. For example, in Fig.~\ref{Fig:7}(a), the red solid line between $\bar{\mathrm{X}}'_m$-$\bar{\mathrm{X}}'_g$ is shifted to the blue dashed line between $\bar{\Gamma}$-$\bar{\mathrm{X}}_g$, where we have used the red and blue to represent glide parities $g_+(k_x)$ and $g_-(k_x)$, respectively. After the BZ folding, the surface bands have some more crossing points, as marked by green circles in Fig.~\ref{Fig:7}(a). Since two states at the crossing points have the same glide parity, a gap can be opened after we introduce additional terms to break the mirror symmetry down to the glide symmetry. For the case of one Dirac cone in Fig.~\ref{Fig:7}(b), we still have a single Dirac cone after BZ folding, which cannot be gapped, as discussed in Ref.~\cite{fang2015new}. However, if the original model (\ref{Eq:Hpmpg1}) of the $pm$ group has two Dirac cones [Fig.~\ref{Fig:7}(d)], two Dirac cones appear for the model of $pg$ group after the BZ folding in Fig.~\ref{Fig:7}(e). But, this is a trivial phase since if we add an energy potential on the surface, the surface bands can be pushed up, and a gap can be opened for the whole system, as shown in Fig.~\ref{Fig:7}(f). Therefore, the topological classification of the $pg$ group is $\mathbb{Z}_2$, which is consistent with results in Ref.~\cite{fang2015new}.

\subsection{$p4m$, $p31m$, and $p6m$ group}\label{Sec:p4m}
The type-I and -II degeneracies coexist in the $pmg$, $pgg$, $p4m$, $p4g$, $p31m$, $p6m$ group in both the single and double groups. The essential feature of these groups is that the HSPs with high dimensional Irreps are always connected by the mirror or glide symmetry invariant lines. Here we discuss three symmorphic symmetry groups $p4m$, $p31m$ and $p6m$ and in the next section, we will focus on three non-symmorphic groups $pmg$, $pgg$ and $p4g$.

For the spinless case of the groups $p31m$, $p4m$ and $p6m$, Alexandradinata \textit{et} \textit{al}.\cite{alexandradinata2014spin} have pointed out that a new type of topological invariant, the halved-mirror chirality, plays an essential role. The halved-mirror chirality is an integer topological invariant that characterizes the surface states on the half-mirror-lines (HMLs) of the surface BZ. The HMLs are defined as the MILs that connect two HSPs, at which the point groups have high dimensional Irreps. Correspondingly, we can also define half-mirror-planes (HMPs), which correspond to the planes in the bulk BZ that are projected onto the HMLs in the surface BZ. Here, we take the line $\bar{\Gamma}$-$\bar{\mathrm{M}}$ of the $p4m$ group as an example [see the Fig.~\ref{Fig:1}(b) for the BZ of the $p4m$ group]. The point group at the HSPs $\bar{\Gamma}$ and $\bar{\mathrm{M}}$ is $C_{4v}$, which possesses one 2D Irrep. Any state belonging to this 2D Irrep is doubly degenerate at these two momenta. Mirror symmetry exists along the momentum line $\bar{\Gamma}$-$\bar{\mathrm{M}}$, and thus this is a HML. The Hamiltonian on this line can be diagonalized into two blocks with each block labeled by mirror parity. Two degenerate states at $\bar{\Gamma}$ and $\bar{\mathrm{M}}$ have opposite mirror parities. Because of double degeneracy of two states with opposite mirror parities at two ends of a HML, it can be proved \cite{alexandradinata2014spin} that an integer topological invariant, the HMC, can be defined as the difference between the integral of the Berry curvature on the corresponding HMP in the even and odd mirror parity subspaces,
\begin{eqnarray}
  \chi=B_e-B_o,
\end{eqnarray}
where
\begin{eqnarray}
  B_{e(o)}=\frac{1}{2\pi}\int_{\mathrm{HMP}} dt dk_zF_{e(o)}(t,k_z),
\end{eqnarray}
where $e$ $(o)$ denote the mirror even (odd) subspace and $t$ denotes the momenta along the HML. This topological invariant determines the form of non-trivial surface states on the HML, e.g., $\bar{\Gamma}$-$\bar{\mathrm{M}}$ line. Aside from $\bar{\Gamma}$-$\bar{\mathrm{M}}$, there is another independent HML $\bar{\Gamma}$-$\bar{\mathrm{X}}$-$\bar{\mathrm{M}}$. Thus, for the spinless case, the classification of TCI in the $p4m$ group is $\mathbb{Z}^2$ if the low-energy bands at $\bar{\Gamma}$ and $\bar{\mathrm{M}}$ belong to the 2D Irrep of $C_{4v}$ group. We emphasize the importance of 2D Irreps at the $\bar{\Gamma}$ and $\bar{\mathrm{M}}$ here. If the states near the band gap at these two momenta belong to other 1D Irreps, there is no topological nontrivial phase.

For the spinful case of the $p4m$ group, all of the spinor representations of the $C_{4v}$ group (for $\bar{\Gamma}$ and $\bar{\mathrm{M}}$ point) and $C_{2v}$ group (for $\bar{\mathrm{X}}$ point) are 2D. Thus, all the bands at $\bar{\Gamma}$, $\bar{\mathrm{X}}$, and $\bar{\mathrm{M}}$ are at least doubly degenerate. Thus, we can identify three independent HMCs on three HMLs $\bar{\Gamma}$-$\bar{\mathrm{X}}$, $\bar{\mathrm{X}}$-$\bar{\mathrm{M}}$, and $\bar{\mathrm{M}}$-$\bar{\Gamma}$. The topological classification will change from $\mathbb{Z}^2$ in the spinless case to $\mathbb{Z}^3$ in the spinful case. We build a model of spinful $p4m$ group in Appendix \ref{App: p4m_model}.

The topological classification of TCIs in the $p31m$ and $p6m$ groups has also been discussed in Ref.~[\onlinecite{alexandradinata2014spin}] for the single-group case. One finds that the classification of TCI in $p31m$ group is $\mathbb{Z}^3$ and in $p6m$ group is $\mathbb{Z}$, assuming that the low-energy bands at all the HSPs belong to the doublet Irreps of the corresponding point group.
When the singlet Irreps are taken into consideration, the classifications will be different. For the $p31m$ group, of which the BZ are shown in Fig.~\ref{Fig:1}(c), the HSPs $\bar{\Gamma}$, $\bar{\mathrm{K}}$ and $\bar{\mathrm{K}}'$ (belonging to group $C_{3v}$) are connected by MILs. If only two of them belong to the 2D Irrep of $C_{3v}$ group, such as $\bar{\Gamma}$, and $\bar{\mathrm{K}}$, there are two MILs, $\bar{\Gamma}$-$\bar{\mathrm{K}}$ and $\bar{\Gamma}$-$\bar{\mathrm{K}}'$-$\bar{\mathrm{K}}$, connecting them. Here, we notice that since $C_3 m(\bar{\Gamma}$-$\bar{\mathrm{K}}')$=$m(\bar{\mathrm{K}}'$-$\bar{\mathrm{K}})$, where $m(\bar{\Gamma}$-$\bar{\mathrm{K}}')$ means $\bar{\Gamma}$-$\bar{\mathrm{K}}'$ is invariant under the mirror reflection operation, and the eigenvalue of $C_3$ rotation for all the 1D Irreps of $C_{3v}$ group are $1$, all the states along the line $\bar{\Gamma}$-$\bar{\mathrm{K}}'$ and the line $\bar{\mathrm{K}}'$-$\bar{\mathrm{K}}$ share the same mirror parity. Thus, the topological classification is $\mathbb{Z}^2$, characterized by two HMCs, in this case. If there are less than two momenta belonging to 2D Irrep, only one MCN can be defined in this system, leading to a $\mathbb{Z}$ classification. For the $p6m$ group, the BZ is the same with that of $p31m$, and there are only two independent points $\bar{\Gamma}$ (belonging to $C_{6v}$) and $\bar{\mathrm{K}}$ (belonging to $C_{3v}$) may have 2D Irreps. Thus, one HMC can be defined and give a $\mathbb{Z}$ classification. The momenta $\bar{\Gamma}$ and $\bar{\mathrm{K}}$ can be connected by another two independent MILs $\bar{\Gamma}$-$\bar{\mathrm{M}}$ and $\bar{\mathrm{M}}$-$\bar{\mathrm{K}}$. Whether a topological invariant with $\mathbb{Z}$ classification can be defined on the line $\bar{\Gamma}$-$\bar{\mathrm{M}}$-$\bar{\mathrm{K}}$ depends on the Irreps of the states at $\bar{\mathrm{M}}$. The point group of $\bar{\mathrm{M}}$ is $C_{2v}$, which has four 1D Irreps $A_i$ and $B_i$ ($i=1,2$). When all the states at $\bar{\mathrm{M}}$ belong to $A_i$ (or $B_i$), an independent HMC can be defined and the topological classification of the TCI will be $\mathbb{Z}^2$. We emphasize the importance of the 1D Irreps here, and the details of this argument will be discussed in the Sec.~\ref{Sec:pmg}. If the low-energy bands belong to 1D Irreps at either $\bar{\Gamma}$ or $\bar{\mathrm{K}}$, there will be no topological non-trivial phase.

For the spinful case, the classification of TCIs in the $p31m$ group is the same as those of the spinless case, because the dimensions of the Irreps of the spinor representations of the $C_{3v}$ group are the same with that of the Irreps of the single group (spinless case). For the $p6m$ group, the spinor representations at $\bar{\mathrm{M}}$ (belonging to $C_{2v}$) and $\bar{\Gamma}$ (belonging to $C_{6v}$) only contain 2D Irreps while those at $\bar{\mathrm{K}}$ (belong to $C_{3v}$) contain both the 1D and 2D Irreps. Therefore, if the low-energy bands at $\bar{\mathrm{K}}$ belong to the 2D Irreps, three independent HMCs can be defined along the HMLs $\bar{\Gamma}$-$\bar{\mathrm{M}}$, $\bar{\Gamma}$-$\bar{\mathrm{K}}$, and $\bar{\mathrm{K}}$-$\bar{\mathrm{M}}$ ($\mathbb{Z}^3$ classification). When the low-energy bands at $\bar{\mathrm{K}}$ belong to 1D Irreps, two HMCs can be defined along the HMLs $\bar{\Gamma}$-$\bar{\mathrm{M}}$ and $\bar{\Gamma}$-$\bar{\mathrm{K}}$-$\bar{\mathrm{M}}$ ($\mathbb{Z}^2$ classification). TCI models of $p31m$ and $p6m$ in the spinful case are given in Appendixes \ref{App: p31m_model} and \ref{App: p6m_model}, respectively.

\subsection{$pmg$, $pgg$, and $p4g$ group}\label{Sec:pmg}
In this section, we will consider three non-symmorphic groups: the $pmg$, $pgg$, and $p4g$ groups. The main difference between non-symmorphic groups and symmorphic groups lies in the fact that the degeneracy at certain HSPs in the BZ can not be described by the Irreps of the corresponding point groups. Instead, one needs to introduce the so-called projective representations, which will be illustrated in details following. One of the authors have studied the $pmg$ group in Ref.~\cite{liu2014topological}, and in this section, we will mainly focus on the example of the $pgg$ group below. After the discussion of the $pgg$ group, we will illustrate the results of the $pmg$ and $p4g$ groups for both the spinless and spinful cases.

\begin{figure} 
\includegraphics[width=8.6cm]{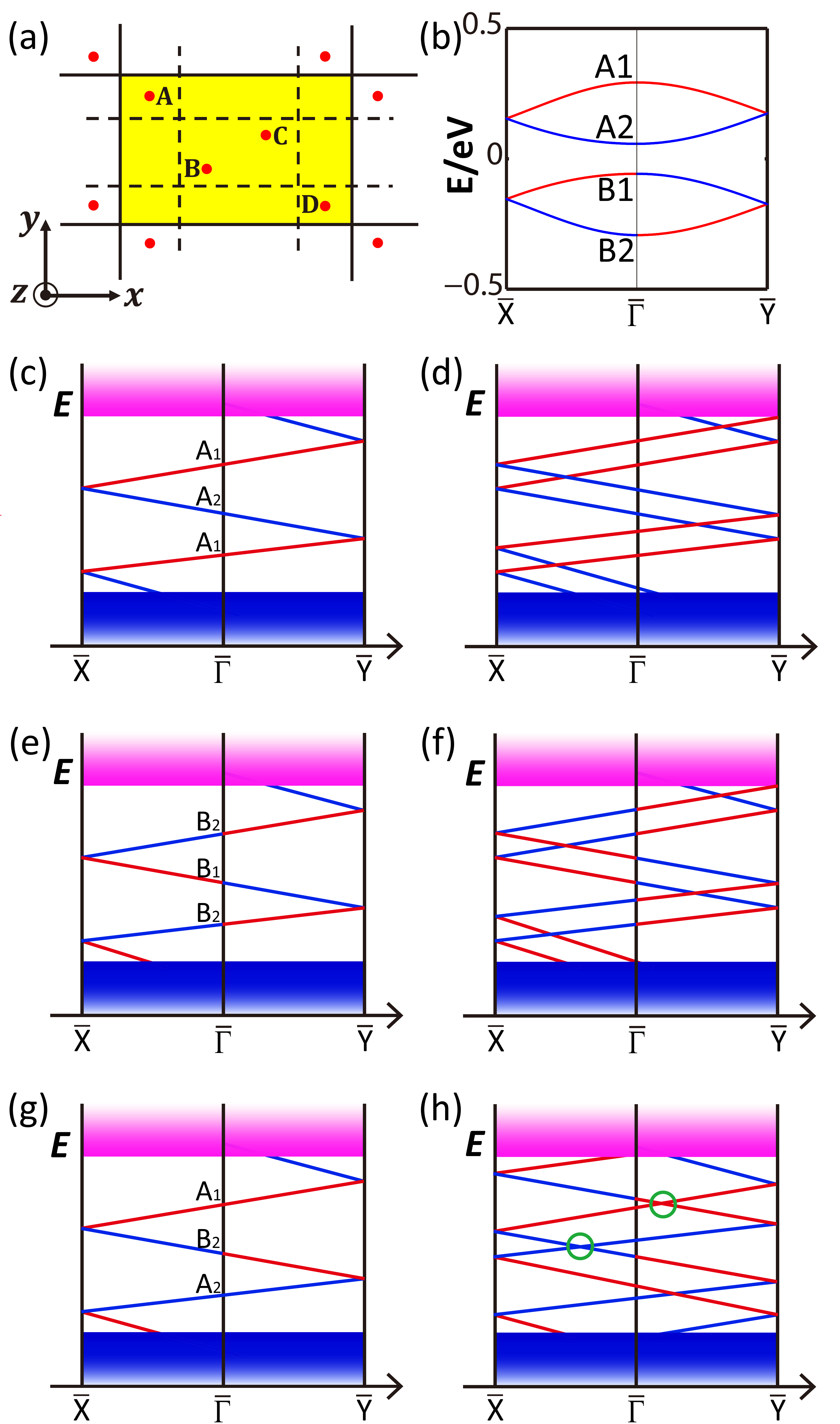}
\caption{(Color online)(a) Schematic plot of a lattice of the $pgg$ group. The yellow region denotes a primitive cell. (b) The energy bands of a 2D lattice of the $pgg$ group. $A_i$ and $B_i$ ($i=1,2$) denote the Irreps of the bands at $\bar{\Gamma}$. (c) One gapless surface states with all the bands belonging to Irreps $A_i$ at $\bar{\Gamma}$. (d) Two copies of gapless surface states in (c). (e )One gapless surface states with all the bands belonging to Irreps $B_i$ at $\bar{\Gamma}$. (f) Two copies of gapless surface states in (e). (g) One gapless surface states with some bands belonging to Irreps $A_i$ and some bands belonging to $B_i$ at $\bar{\Gamma}$. (h) Two copies of gapless surface states in (g). The crossings marked by the green circles can be gapped with perturbations that do not break the symmetry.}\label{Fig:8}
\end{figure}

The generators of the $pgg$ group include the translation operators, as well as two glide-symmetry operators $g_x=\{m_x|\bm{\tau}_y\}$ and $g_y=\{m_y|\bm{\tau}_x\}$, with $\bm{\tau}_y=(0,1/2)$ and $\bm{\tau}_x=(1/2,0)$. The eigenvalues of $g_x$ are $g_{\pm}^x=\pm i^f e^{ik_y/2}$ and of $g_y$ are $g_{\pm}^y=\pm i^f e^{ik_x/2}$, where $f=0$ for the spinless case and $f=1$ for the spinful case. A typical lattice for the $pgg$ group is shown in Fig.~\ref{Fig:8}(a). In the BZ of the $pgg$ group, $g_x$ symmetry exists on the momentum lines $\bar{\Gamma}$-$\bar{\mathrm{Y}}$ and $\bar{\mathrm{X}}$-$\bar{\mathrm{M}}$, while $g_y$ symmetry appears on the lines $\bar{\Gamma}$-$\bar{\mathrm{X}}$ and $\bar{\mathrm{Y}}$-$\bar{\mathrm{M}}$. Therefore, all the states in these momentum lines can be labeled by the eigenvalue of $g_x$ or $g_y$ operator and the type-I degeneracy is possible along these momentum lines. For four HSPs $\bar{\Gamma}=(0,0)$, $\bar{\mathrm{X}}=(\pi,0)$, $\bar{\mathrm{Y}}=(0,\pi)$, and $\bar{\mathrm{M}}=(\pi,\pi)$, both $g_x$ and $g_y$ exist, and thus the corresponding factor group $F_{\mathbf{k}}$ of the wave-vector group is isomorphic to $C_{2v}$, for which the character table is listed in Table~\ref{tab:2}. However, the states at these momenta cannot be directly described by the Irreps of the $C_{2v}$ in Table~\ref{tab:2}. For the momentum $\mathbf{k}=\bar{\Gamma}, \bar{\mathrm{X}}, \bar{\mathrm{Y}}, \bar{\mathrm{M}}$, the representation matrix for $g_{x(y)}$ can be denoted as $D^{\mathbf{k}}(g_{x(y)})=e^{-i\mathbf{k}\cdot \bm{\tau}_{y(x)}}D(m_{x(y)})$, where $e^{i\mathbf{k}\cdot \bm{\tau}_{y(x)}}$ comes from the translational part while $D(m_{x(y)})$ only depends on the point symmetry operation and satisfies the commutation relation
\begin{eqnarray}
	D(m_x)D(m_y)=\alpha(g_x,g_y)D(m_y)D(m_x)
\end{eqnarray}
with $\alpha(g_x,g_y)=e^{i(\mathbf{k}-m^{-1}_x\mathbf{k})\cdot\bm{\tau}_x} e^{-i(\mathbf{k}-m_y^{-1}\mathbf{k})\cdot\bm{\tau}_y}$ depending on $g_x$, $g_y$ and momentum $\mathbf{k}$. Since $[m_x,m_y]=0$, one can immediately see that $D(m_{x(y)})$ is not a representation of the point group $C_{2v}$ if $\alpha\neq1$. Instead, the so-called projective representations, which are discussed in details in Appendix \ref{App:rep}, are required to describe the degeneracy of the states. Depending on different values of $\alpha$, the projective representation belongs to different classes. Direct calculation shows $\alpha_{\bar{\Gamma},\bar{\mathrm{M}}}=1$ and $\alpha_{\bar{\mathrm{X}},\bar{\mathrm{Y}}}=-1$. This suggests that the states at $\bar{\Gamma}$ and $\bar{\mathrm{M}}$ are described by the conventional representations of $C_{2v}$ group, denoted as $K_0$ class, while those at $\bar{\mathrm{X}}$ and $\bar{\mathrm{Y}}$ are described by projective representations belonging to a non-trivial class, usually denoted as $K_1$ class. For the $C_{2v}$ group, all the Irreps in the $K_0$ class (conventional representations) are all 1D (see Table~\ref{tab:2}), and thus no degeneracy occurs at $\bar{\Gamma}$ and $\bar{\mathrm{M}}$. In contrast, all the Irreps in the $K_1$ class are 2D, leading to the type-II degeneracy at $\bar{\mathrm{X}}$ and $\bar{\mathrm{Y}}$. The degeneracy at these two momenta can also be understood from the anti-commutation relation between the representation matrix $D(g_x)$ and $D(g_y)$, $\{D(g_x),D(g_y)\}=0$, at $\bar{\mathrm{X}}$ and $\bar{\mathrm{Y}}$\cite{liu2014topological}.

\begin{table}
\centering
\caption{The character table of $C_{2v}$ single group. }
\begin{tabular}{c|cccc}
    \hline
    $C_{2v}$ & $E$ & $C_2$ & $m_x$ & $m_y$\\
    \hline
    $A_1$ & $1$ & $1$ & $1$ & $1$ \\
    $A_2$ & $1$ & $1$ & $-1$ & $-1$ \\
    $B_1$ & $1$ & $-1$ & $1$ & $-1$ \\
    $B_2$ & $1$ & $-1$ & $-1$ & $1$ \\
    \hline
  \end{tabular}\label{tab:2}%
\end{table}

The type-II degeneracy at $\bar{\mathrm{X}}$ and $\bar{\mathrm{Y}}$ suggests the possibility of type-II TCIs, which can be characterized by the $\mathbb{Z}_2$ topological invariant\cite{liu2014topological}. Furthermore, we also notice that type-I degeneracy exists along the momentum lines $\bar{\Gamma}$-$\bar{\mathrm{Y}}$, $\bar{\mathrm{X}}$-$\bar{\mathrm{M}}$, $\bar{\Gamma}$-$\bar{\mathrm{X}}$ and $\bar{\mathrm{Y}}$-$\bar{\mathrm{M}}$, which connect four HSPs $\bar{\Gamma}$, $\bar{\mathrm{X}}$, $\bar{\mathrm{Y}}$ and $\bar{\mathrm{M}}$. Therefore, one may ask if the mixed type-I-II TCIs with $\mathbb{Z}$ classification can exist in this system. Since $\bar{\mathrm{X}}$ and $\bar{\mathrm{Y}}$ are connected by the momenta $\bar{\Gamma}$ or $\bar{\mathrm{M}}$ with glide invariant lines, we need to analyze how the glide parity of a state evolves along the momentum lines $\bar{\mathrm{X}}$-$\bar{\Gamma}$-$\bar{\mathrm{Y}}$ and $\bar{\mathrm{X}}$-$\bar{\mathrm{M}}$-$\bar{\mathrm{Y}}$. Let us start from the momentum $\bar{\mathrm{X}}$, at which all the states are doubly degenerate. The double degenerate states will be split along $\bar{\mathrm{X}}$-$\bar{\Gamma}$. The two split states are the eigenstates of $g_y$ operators and always possess opposite glide parities $g_+^y$ and $g_-^y$, represented by red ($g_+^y$) and blue ($g_-^y$) in Fig.~\ref{Fig:8}(b), respectively. Along $\bar{\Gamma}$-$\bar{\mathrm{Y}}$, all the states are the eigenstates of $g_x$ with the eigenvalue $g^x_{\pm}$, which are still labeled by the red ($g^x_+$) and blue ($g^x_-$) in Fig.~\ref{Fig:8}(b). It should be emphasized that the glide operation is $g_y$ along $\bar{\mathrm{X}}$-$\bar{\Gamma}$ and $g_x$ along $\bar{\Gamma}$-$\bar{\mathrm{Y}}$ although we use the same colors (red and blue) to label glide parities. $\bar{\mathrm{X}}$-$\bar{\Gamma}$ and $\bar{\Gamma}$-$\bar{\mathrm{Y}}$ are connected at $\bar{\Gamma}$ and it turns out that depending on different Irreps at $\bar{\Gamma}$, the red (or blue) states along $\bar{\mathrm{X}}$-$\bar{\Gamma}$ can be connected to either red or blue states along $\bar{\Gamma}$-$\bar{\mathrm{Y}}$. To see this, we need to analyze the representations at $\bar{\Gamma}$. At the $\bar{\Gamma}$ point, the glide symmetry $g_x$ or $g_y$ is equivalent to the mirror symmetry $m_x$ or $m_y$, respectively. Thus, all the states are classified by the Irreps for the $C_{2v}$ group, as listed in the Table~\ref{tab:2}. There are four 1D Irreps for the $C_{2v}$ group, denoted as $A_i$ and $B_i$ ($i=1,2$). The eigenstates share the same parity for $m_x$ and $m_y$ if they belongs to the Irreps $A_i$, but opposite parities for the Irreps $B_i$. Therefore, for the Irreps $A_i$ at $\bar{\Gamma}$, the red (blue) states will be connected to the red (blue) states, while for the Irreps $B_i$, the red (blue) states will be connected to the blue (red) states.
If all the states at the $\bar{\Gamma}$ point near the Fermi energy belong to the Irreps $A_i$, any integer copies of surface states are possible to exist. For example, we show one gapless surface states in Fig.~\ref{Fig:8}(c), of which the gapless nature is guaranteed by type-II degeneracy at $\bar{\mathrm{X}}$ and $\bar{\mathrm{Y}}$.
In Fig.~\ref{Fig:8}(d), we consider two copies of gapless surface states and in this case, type-II degeneracy is not enough to guarantee the gapless nature since they may cross at a generic momentum in the line $\bar{\mathrm{X}}$-$\bar{\Gamma}$-$\bar{\mathrm{Y}}$. However, in this case, we find that all the crossings are always between surface bands with opposite glide parities and thus cannot lead to a gap opening. Therefore, the topological classification in this case is $\mathbb{Z}$.
If all the states near the Fermi energy at $\bar{\Gamma}$ belong to the Irreps $B_i$, as shown in Figs.~\ref{Fig:8}(e) and (f) for one and two copies of non-trivial surface states, respectively, we also find any integer copies of gapless surface states are stable, indicating the $\mathbb{Z}$ classification in this case.
When the $A_i$ states coexist with the $B_i$ states at $\bar{\Gamma}$ near the Fermi energy, multiple copies of surface states are no longer stable. We show an example of one and two copies of surface states in Figs.~\ref{Fig:8}(g) and (h), respectively. One can see that for two copies of surface states, some crossings are between bands with the same glide parity, as marked by the green circles in Fig.~\ref{Fig:8}(h). A gap can be opened at these points and drives the system into a topologically trivial phase. Thus, in this case, the classification is $\mathbb{Z}_2$.
Aside from the $\bar{\mathrm{X}}$-$\bar{\Gamma}$-$\bar{\mathrm{Y}}$ line, a similar analysis can be applied to the $\bar{\mathrm{X}}$-$\bar{\mathrm{M}}$-$\bar{\mathrm{Y}}$ line and lead to a similar topological classification at this momentum line. The above discussion suggests that the classification of gapless surface states in the $pgg$ group sensitively depends on the Irreps of the states at $\bar{\Gamma}$ and $\bar{\mathrm{M}}$. When the states at the $\bar{\Gamma}$ and $\bar{\mathrm{M}}$ all belong to the Irreps $A_i$ (or $B_i$), any integer copies of surface states are possible and the classification is $\mathbb{Z}\times\mathbb{Z}$, in which the first $\mathbb{Z}$ is for the number of surface states along $\bar{\mathrm{X}}$-$\bar{\Gamma}$-$\bar{\mathrm{Y}}$ while the second is for the states along $\bar{\mathrm{X}}$-$\bar{\mathrm{M}}$-$\bar{\mathrm{Y}}$ since these two momentum lines are independent.

After identifying the classification of surface states in the surface BZ, we next discuss topological invariants in the 3D bulk systems with the $pgg$ group. There are two types of topological invariants that can be defined here. Firstly, due to the type-II degeneracy at $\bar{\mathrm{X}}$ and $\bar{\mathrm{Y}}$, a $\mathbb{Z}_2$ topological invariant can be defined, which has been discussed in details for the $pmg$ group in Ref.~\cite{liu2014topological}.
The doubly degenerate bands at $\bar{\mathrm{X}}$ and $\bar{\mathrm{Y}}$ can be classified into two groups $I$ and $II$ based on the eigenvalue under $g_x$ (or $g_y$). Let us consider a state $|\phi_{\mathbf{k}}^I\rangle$ on the line $X$-$U$ ($Y$-$T$) in the 3D BZ which projects onto $\bar{\mathrm{X}}$ (or $\bar{\mathrm{Y}}$) [Fig.~\ref{Fig:3}(a)] satisfies $H(\mathbf{k})|\phi_{\mathbf{k}}^I\rangle=E_{\mathbf{k}}|\phi_{\mathbf{k}}^I\rangle$ and $g_x|\phi_{\mathbf{k}}^I\rangle=g^x_+|\phi_{\mathbf{k}}^I\rangle$. The state defined as $|\phi_{\mathbf{k}}^{II}\rangle=e^{i\chi_{\mathbf{k}}}g_y |\phi_{\mathbf{k}}^I\rangle$, where $\chi_{\mathbf{k}}$ is a phase factor, is degenerate with $|\phi_{\mathbf{k}}^I\rangle$ at $\bar{\mathrm{X}}$ (or $\bar{\mathrm{Y}}$) and has the eigenvalue $g^x_-=-g^x_+$ under $g_x$. The topological invariant can be defined as the difference between the ``doublet polarization" $P_d$ at $\bar{\mathrm{X}}$ and $\bar{\mathrm{Y}}$, i.e., $\Delta=P_d(\bar{\mathrm{X}})-P_d(\bar{\mathrm{Y}})$ mod $2$, where $P_d=P_I-P_{II}$, and $P_{\alpha}=\frac{1}{2\pi}\oint dk_z \langle \phi^{\alpha}_{\mathbf{k}}|i\partial_{k_z}|\phi^{\alpha}_{\mathbf{k}}\rangle$ $(\alpha=I,II)$. This is a $\mathbb{Z}_2$ topological invariant, similar to that of TR invariant TIs.
Secondly, when all the states at $\bar{\Gamma}$ or $\bar{\mathrm{M}}$ belong to the 1D Irreps $A_i$ ($B_i$), an integer topological invariant, in analogy to HMC, can be defined. Let us consider the line $\bar{\mathrm{X}}$-$\bar{\Gamma}$-$\bar{\mathrm{Y}}$ as an example. For a continuous band, if it belongs to $A_i$ ($B_i$) at $\bar{\Gamma}$, the $g_x$ glide parity of the band along $\bar{\Gamma}$-$\bar{\mathrm{Y}}$ is the same as (opposite to) the $g_y$ glide parity along $\bar{\Gamma}$-$\bar{\mathrm{X}}$. Therefore, the bands on the line $\bar{\mathrm{X}}$-$\bar{\Gamma}$-$\bar{\mathrm{Y}}$ can be labeled by a single operator $g_y$ in this case. The bands on $\bar{\mathrm{X}}$-$\bar{\mathrm{M}}$-$\bar{\mathrm{Y}}$ have the similar properties, and can be labeled by the glide parity of $g_y$ on $\bar{\mathrm{Y}}$-$\bar{\mathrm{M}}$. The topological invariant can be defined as
\begin{eqnarray}
  \chi_i=\frac{1}{2\pi}\int_{0}^1 dt_i\oint dk_z (\mathcal{F}_+-\mathcal{F}_-),\label{EqHGC}
\end{eqnarray}
where $i=1$ labels the line $\bar{\mathrm{X}}$-$\bar{\Gamma}$-$\bar{\mathrm{Y}}$, and $i=2$ labels $\bar{\mathrm{X}}$-$\bar{\mathrm{M}}$-$\bar{\mathrm{Y}}$, $t_i$ denotes the momentum along the line $i$ with $t_1=0$ at $\bar{\mathrm{X}}$, and $t_1=1$ at $\bar{\mathrm{Y}}$, while $t_2=0$ at $\bar{\mathrm{Y}}$ and $t_2=1$ at $\bar{\mathrm{X}}$. $\mathcal{F}_{+(-)}$ is the Berry curvature of the occupied bands in the glide parity $g_{+(-)}^y(k_x)$ subspace of $g_y$. This definition is similar to the HMC, and thus, we dub it ``halved glide chirality'' (HGC). Similarly, it can also be proved that when $\chi_1+\chi_2$ is an odd number, the bulk is gapless with Weyl points.

To confirm our classification, we construct a tight-binding model for a lattice with the $pgg$ group symmetry, as shown in Fig.~\ref{Fig:8}(a). There are four atoms in one primitive cell with the positions $\mathbf{r}_A=(r_x,r_y)$, $\mathbf{r}_B=(-\frac{1}{2}-r_x,r_y-\frac{1}{2})$, $\mathbf{r}_C=(\frac{1}{2}+r_x,\frac{1}{2}-r_y)$, $\mathbf{r}_D=(-r_x,r_y)$. The symmetry analysis below does not depend on the values of $r_x$ and $r_y$. We assume that there are two $s$ orbitals on each atom for simplicity and choose the basis in the momentum space as
\begin{eqnarray}
	|\alpha,i\rangle(\mathbf{r})=\frac{1}{\sqrt{N}} \sum_{\mathbf{R}}e^{i\mathbf{k}\cdot \mathbf{R}}\varphi_{\mathbf{R},\alpha,i}(\mathbf{r}- \mathbf{R}-\mathbf{r}_{\alpha}), \label{Eq:pggbasis}
\end{eqnarray}
where $\mathbf{R}$ denotes the vector of the Bravais lattice, $\alpha=A,B,C,D$ label the atoms, of which the relative position in a primitive cell is $\mathbf{r}_{\alpha}$, $i=a,b,$ label the two $s$ orbitals and $\varphi$ denotes the L\"{o}wdin orbitals.
In the above basis, the Hamiltonian is periodic in the momentum space with the reciprocal lattice vector, i.e., $H(\mathbf{k})=H(\mathbf{k}+\mathbf{G})$, where $\mathbf{G}$ is a reciprocal lattice vector.

Under the basis $\{$$|A,a\rangle$, $|B,a\rangle$, $|C,a\rangle$, $|D,a\rangle$, $|A,b\rangle$, $|B,b\rangle$, $|C,b\rangle$, $|D,b\rangle$$\}$ given by (\ref{Eq:pggbasis}), the explicit forms of the matrix $\tilde{U}$ for two glide plane operations are
\begin{eqnarray}
    &&\tilde{U}(\mathbf{k},g_x)=\sigma_{0}\nonumber\\
    &&  \otimes\left(
  \begin{array}{cccc}
    0 & e^{i(k_x+k_y)/2} & 0 & 0\\
    e^{i(k_x-k_y)/2} & 0 & 0 & 0\\
    0 & 0 & 0 & e^{i(-k_x+k_y)/2}\\
    0 & 0 & e^{i(-k_x-k_y)/2} & 0\\
  \end{array}\right)\nonumber\\
\end{eqnarray}
and
\begin{eqnarray}
    &&\tilde{U}(\mathbf{k},g_y)=\sigma_{0}\nonumber\\
    && \otimes\left(
  \begin{array}{cccc}
    0 & 0 & e^{i(-k_x-k_y)/2} & 0\\
    0 & 0 & 0 & e^{i(-k_x+k_y)/2}\\
    e^{i(k_x-k_y)/2} & 0 & 0 & 0\\
    0 & e^{i(k_x+k_y)/2} & 0 & 0\\
  \end{array}\right)\nonumber\\
\end{eqnarray}
where $\sigma_0$ is a two-by-two identity matrix.

\begin{figure}
\includegraphics[width=0.5\textwidth]{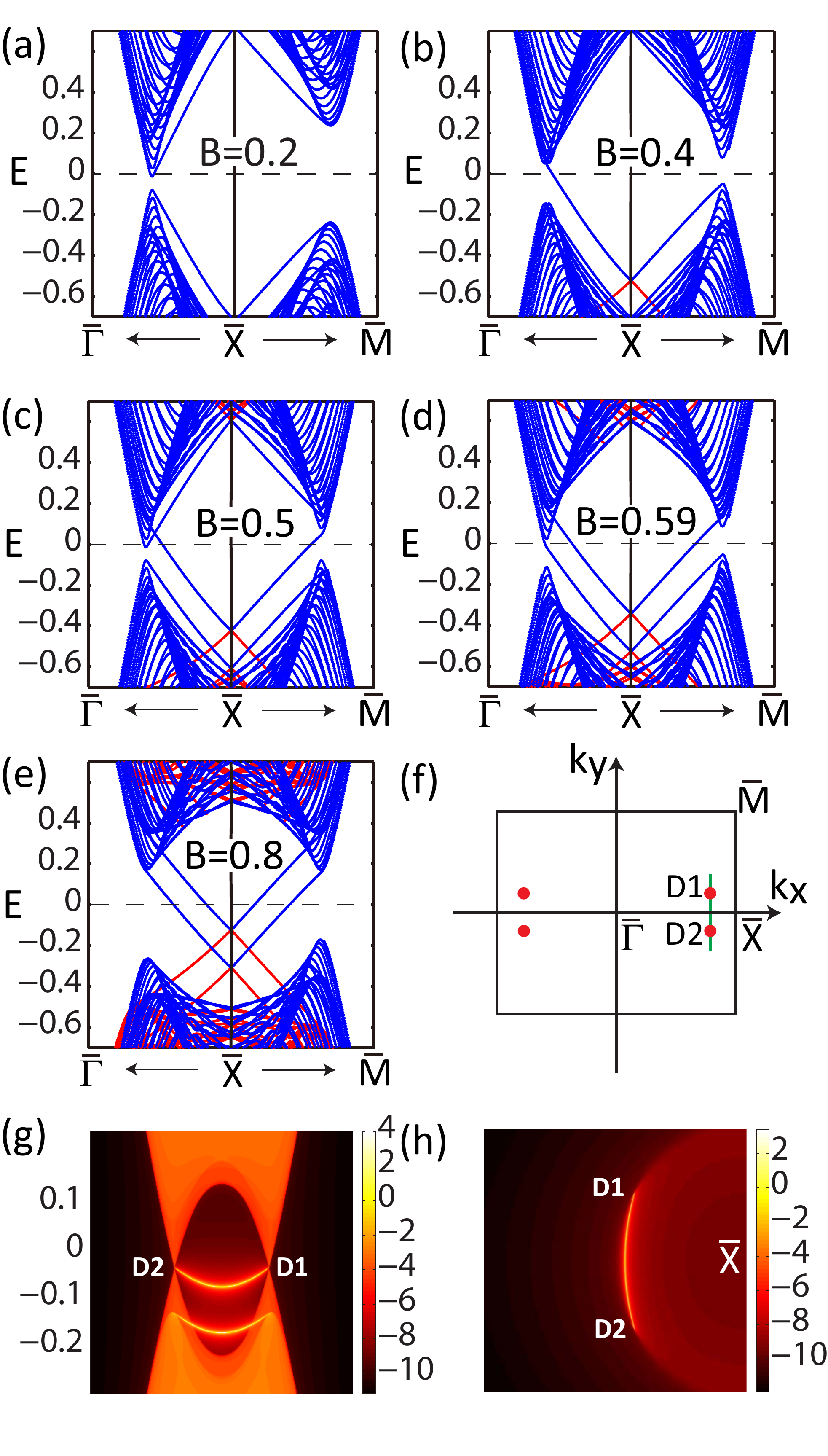}
\caption{(Color online) Band dispersions around $\bar{\mathrm{X}}$ of the model of TCI in the $pgg$ group in a slab configuration. $A=1.7$ is fixed and $B$ is tuned from $0.2$ to $0.8$ in (a)--(e). We only show the surface states on one surface of the slab. The red (blue) lines denote the bands with even (odd) glide parity. (f) Schematic plot of the positions of Weyl nodes (marked by the red dots) in the surface BZ when $B=0.4$. (g) The density of states calculated by iterative Green functions on the line connecting two Weyl nodes with $B=0.4$ in a semi-infinite configuration. (h) The density of states of the Fermi surface passing through the nodal points with $B=0.4$. The other parameters in all of the figures are $f_{1a}=f_{2a}=-f_{1b}=f_{2b}=0.1$, $f_{3a}=f_{3a}'=f_{3b}=f_{3b}'=0$, $\theta_{1a}=\theta_{2a}=\theta_{1b}=\theta_{2b}=0$, $f_1=f_2=1$, $f_4=0.1$, $\theta_1=\pi/2$, $\theta_2=0$ and $n=2$. }\label{Fig:9}
\end{figure}

The Hamiltonian in momentum space, which satisfies the constraints enforced by symmetries  $H(g_{x(y)}\mathbf{k})=\tilde{U}(\mathbf{k},g_{x(y)})H(\mathbf{k}) \tilde{U}(\mathbf{k},g_{x(y)})^{\dag}$, is
\begin{eqnarray}
  H=\left(
  \begin{array}{cc}
    H_{a} & H_{ab}\\
    H^{\dag}_{ab} & H_{b}\\
  \end{array}\right),
\end{eqnarray}
where
\begin{widetext}
\begin{eqnarray}
H_a(\mathbf{k})&=&\left(
  \begin{array}{cccc}
   0 & f_{1a}(e^{i\theta_{1a}}+e^{-i\theta_{1a}}e^{ik_y}) & f_{2a}(e^{i\theta_{2a}}+e^{-i\theta_{2a}}e^{-ik_x}) & f_{3a}e^{-i(k_x-k_y)} +f'_{3a}e^{ik_y}\\
   f_{1a}e^{-i\theta_{1a}}+e^{i\theta_{1a}}e^{-ik_y}) & 0 & f_{3a}+f'_{3a}e^{-ik_x} & f_{2a}(e^{-i\theta_{2a}}+e^{i\theta_{2a}}e^{-ik_x})\\
   f_{2a}(e^{-i\theta_{2a}}+e^{i\theta_{2a}}e^{ik_x}) & f_{3a}+f'_{3a}e^{ik_x} & 0 & f_{1a}(e^{-i\theta_{1a}}+e^{i\theta_{1a}}e^{ik_y})\\
   f_{3a}e^{i(k_x-k_y)} +f'_{3a}e^{-ik_y} & f_{2a}(e^{i\theta_{2a}}+e^{-i\theta_{2a}}e^{ik_x}) & f_{1a}(e^{i\theta_{1a}}+e^{-i\theta_{1a}}e^{-ik_y}) & 0
  \end{array}\right)\nonumber\\
  &&+[-1+8(3+\cos(k_x)-\cos(k_y)-\cos(n k_z))] \Gamma_{00},
\end{eqnarray}
\begin{eqnarray}
  H_b(\mathbf{k})&=&\left(
  \begin{array}{cccc}
   0 & f_{1b}(e^{i\theta_{1b}}+e^{-i\theta_{1b}}e^{ik_y}) & f_{2b}(e^{i\theta_{2b}}+e^{-i\theta_{2b}}e^{-ik_x}) & f_{3b}e^{-i(k_x-k_y)} +f'_{3b}e^{ik_y}\\
   f_{1b}e^{-i\theta_{1b}}+e^{i\theta_{1b}}e^{-ik_y}) & 0 & f_{3b}+f'_{3b}e^{-ik_x} & f_{2b}(e^{-i\theta_{2b}}+e^{i\theta_{2b}}e^{-ik_x})\\
   f_{2b}(e^{-i\theta_{2b}}+e^{i\theta_{2b}}e^{ik_x}) & f_{3b}+f'_{3b}e^{ik_x} & 0 & f_{1b}(e^{-i\theta_{1b}}+e^{i\theta_{1b}}e^{ik_y})\\
   f_{3b}e^{i(k_x-k_y)} +f'_{3b}e^{-ik_y} & f_{2b}(e^{i\theta_{2b}}+e^{-i\theta_{2b}}e^{ik_x}) & f_{1b}(e^{i\theta_{1b}}+e^{-i\theta_{1b}}e^{-ik_y}) & 0
  \end{array}\right)\nonumber\\
  &&-[-1+8(3+\cos(k_x)-\cos(k_y)-\cos(n k_z))] \Gamma_{00},
\end{eqnarray}
and
\begin{eqnarray}
  H_{ab}(\mathbf{k})&=&\left(
  \begin{array}{cccc}
   0 & f_1(e^{i\theta_1}+e^{-i\theta_1}e^{ik_y}) & f_2(e^{i\theta_2}+e^{-i\theta_2}e^{-ik_x}) & f_3e^{-i(k_x-k_y)} +f'_3e^{ik_y}\\
   f_1(e^{-i\theta_1}+e^{i\theta_1}e^{-ik_y}) & 0 & f_3+f'_3e^{-ik_x} & f_2(e^{-i\theta_2}+e^{i\theta_2}e^{-ik_x})\\
   f_2(e^{-i\theta_2}+e^{i\theta_2}e^{ik_x}) & f_3+f'_3e^{ik_x} & 0 & f_1(e^{-i\theta_1}+e^{i\theta_1}e^{ik_y})\\
   f_3e^{i(k_x-k_y)} +f'_3e^{-ik_y} & f_2(e^{i\theta_2}+e^{-i\theta_2}e^{ik_x}) & f_1(e^{i\theta_1}+e^{-i\theta_1}e^{-ik_y}) & 0
  \end{array}\right)\nonumber\\
  &&+(1+i\sin(n k_z)+f_4\cos(k_z))\Gamma_{00},
\end{eqnarray}
\end{widetext}
where $\Gamma_{00}=I_{4\times4}$ is a 4$\times$4 identity matrix, $f_{1a}$, $f_{2a}$, $f_{3a}$, $f_{3a}'$, $f_{1b}$, $f_{2b}$, $f_{3b}$, $f_{3b}'$, $f_1$, $f_2$, $f_3$, $f_3'$, and $f_4$ are real coefficients, $\theta_{1a}$, $\theta_{2a}$, $\theta_{1b}$, $\theta_{2b}$, $\theta_{1}$, and $\theta_{2}$ are real parameters which break the TR symmetry, and $n$ is a positive integer.
We further define $A=f_3+f'_3$ and $B=f_3-f'_3$. Different topological phases can be obtained by tuning the values of $A$ and $B$ in the above model. Surface energy dispersion can be calculated for a slab configuration and is shown along the momentum lines $\bar{\mathrm{X}}$-$\bar{\Gamma}$ and $\bar{\mathrm{X}}$-$\bar{\mathrm{M}}$ in Fig.~\ref{Fig:9} with a fixed $A=1.7$ and different values of $B$. When $B=0.2$, the system is a trivial insulator [Fig.~\ref{Fig:9}(a)]. With increasing $B$, the band gap closes on $\bar{\Gamma}$-$\bar{\mathrm{X}}$ between two bands with glide parity $g_-^y(k_x)$ (when $B\approx 0.27$) and drive the system into a phase with $\chi_1=-1$ and $\chi_2=0$. Energy dispersion for a slab configuration is shown along $\bar{\mathrm{X}}$-$\bar{\Gamma}$ and $\bar{\mathrm{X}}$-$\bar{\mathrm{M}}$ in Fig.~\ref{Fig:9}(b) with $B=0.4$. This is a semi-metal phase with a pair of Weyl nodes, emerging on $\bar{\Gamma}$-$\bar{\mathrm{X}}$ and moving in opposite direction perpendicular to $\bar{\Gamma}$-$\bar{\mathrm{X}}$, as schematically shown by red points in Fig.~\ref{Fig:9}(f). The band structure along the line connecting these two Weyl nodes is shown in Fig.~\ref{Fig:9}(g) (with $B=0.4$), calculated by iterative Green function method. Figure~\ref{Fig:9}(h) shows the density of states in the momentum space with the energy determined by the nodal points, and a Fermi arc connects these two Weyl nodes. By further increasing $B$, the band gap closes (when $B\approx 0.45$) and reopens on $\bar{\mathrm{X}}$-$\bar{\mathrm{M}}$. One Dirac cone appears at $\bar{\mathrm{X}}$ on the surface [see Fig.~\ref{Fig:9}(c) with $B=0.5$]. This is exactly the TCI phase with HGC $\chi_1=-1$ and $\chi_2=1$. Thus, the Weyl semi-metal phase separates the TCI phase and the trivial insulating phase in our model. Another topological phase transition occurs for an even larger $B$ and the system first evolves to another Weyl semi-metal phase with $\chi_1=-2$ and $\chi_2=1$ (when $0.55<B<0.64$) and then to a TCI phase with $\chi_1=-2$ and $\chi_2=2$ (when $0.64<B<1.4$). The corresponding surface bands from the calculations of a slab configuration are shown in Figs.~\ref{Fig:9}(d) and (e) for the Weyl semi-metal phase ($B=0.59$) and TCI phase ($B=0.8$), respectively.

\begin{figure} 
\includegraphics[width=8.6cm]{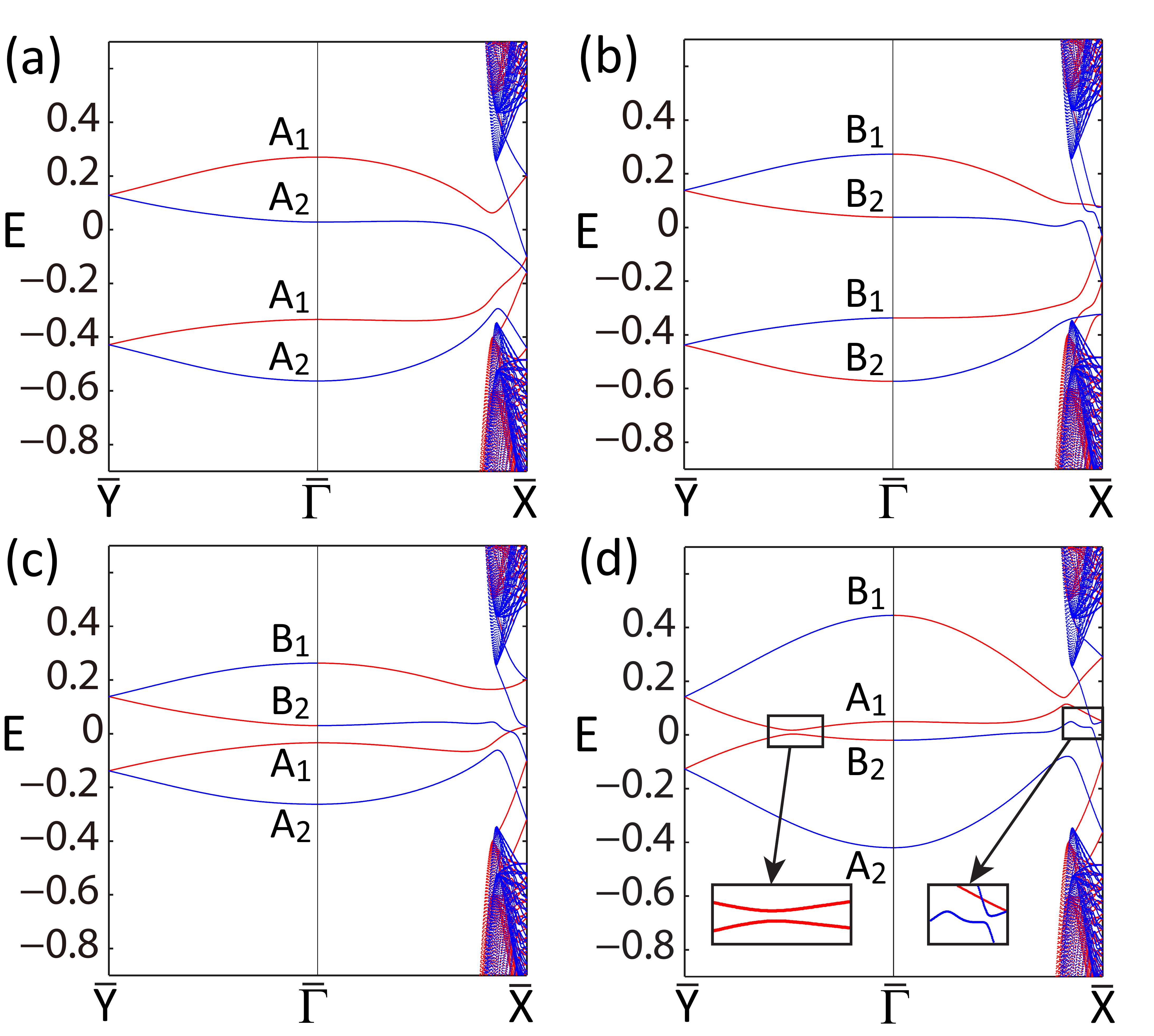}
\caption{(Color online) The band structure of the slab with a 2D layer on the top surface on $\bar{\mathrm{Y}}$-$\bar{\Gamma}$-$\bar{\mathrm{X}}$. The original slab is in $\chi_1=-2$ phase. (a) All the bands of the 2D layer belong to Irreps $A_i$ at $\bar{\Gamma}$. (b) All the bands of the 2D layer belong to Irreps $B_i$ at $\bar{\Gamma}$. (c) The bands of the 2D layer belonging to Irreps $A_i$ and $B_i$ coexist at $\bar{\Gamma}$. (d) The surface states are gapped by adiabatically tuning the parameters in (c).}\label{Fig:10}
\end{figure}

To check if these gapless Dirac surface states are stable or not, we add an additional 2D layer, which preserves the symmetry group $pgg$, on the top surface of the slab, and introduce the coupling between this 2D layer and the top layer of the slab. By tuning the parameters for the model of this additional 2D layer, we can move energy bands with the required Irreps in this 2D layer to the bulk band gap and couple them to topological surface bands. We explore the case with two Dirac cones ($B=0.9$), corresponding to the TCI phase with $\chi_1=-2$ on $\bar{\mathrm{X}}$-$\bar{\Gamma}$-$\bar{\mathrm{Y}}$. When the bands of the 2D layer belong to Irreps $A_i$ (or $B_i$) at $\bar{\Gamma}$, the gapless surface states remains stable, as shown in Figs.~\ref{Fig:10}(a) [or (b)]. However, if the bands in Irreps $A_i$ and $B_i$ coexist at $\bar{\Gamma}$, we find a local stability for gapless surface bands, as shown in Fig.~\ref{Fig:10}(c). The crossing on $\bar{\Gamma}$-$\bar{\mathrm{X}}$ between the bands in Irreps $A_1$ and $B_2$ at $\bar{\Gamma}$ are protected locally by glide plane symmetry. By tuning the parameters adiabatically, this crossing point can move from $\bar{\Gamma}$-$\bar{\mathrm{X}}$ to $\bar{\mathrm{Y}}$-$\bar{\Gamma}$, leading to an anti-crossing between these two bands due to the same glide parity, as shown in Fig.~\ref{Fig:10}(d). This leads to a trivial phase. Thus, in this case, only one surface Dirac cone is stable and the topological classification is $\mathbb{Z}_2$.
Therefore, our calculations based on an explicit tight-binding model indeed support our classification of TCIs in the $pgg$ group.

In the spinful case, the representation of $g_{x(y)}$ is $D^{\mathbf{k}}(g_{x(y)})\otimes D_{1/2}(g_{x(y)})=e^{-i\mathbf{k}\cdot\bm{\tau}_{y(x)}} D(m_{x(y)})\otimes D_{1/2}(m_{x(y)})=e^{-i\mathbf{k}\cdot\bm{\tau}_{y(x)}}D(m_{x(y)})'$, where $D_{1/2}$ is the transformation matrix of a spin-$\frac{1}{2}$ spinor and the non-primitive translation has no effect on spin. As discussed in details in the Appendix \ref{App:rep}, spin degree of freedom introduces an additional coefficient for the factor system of the projective representations. The representations satisfy
\begin{eqnarray}
  D(m_x)'D(m_y)'=\alpha_s D(m_y)'D(m_x)'
\end{eqnarray}
where $\alpha_s=\alpha\alpha_{1/2}$. Since $m_xm_y=C_2(z)$ and $m_ym_x=Q(y)C_2(z)$, where $C_2(z)$ denotes the $\pi$ rotation around the $z$ axis and $Q(y)$ is for the $2\pi$ rotation around the $y$ axis, $\alpha_{1/2}=-1$. Then $\alpha_s=\alpha\alpha_{1/2}=1$ at $\bar{\mathrm{X}}$ and $\bar{\mathrm{Y}}$, and $\alpha_s=-1$ at $\bar{\Gamma}$ and $\bar{\mathrm{M}}$. Thus, all the states at $\bar{\Gamma}$ and $\bar{\mathrm{M}}$, instead of $\bar{\mathrm{X}}$ and $\bar{\mathrm{Y}}$, are doubly degenerate. The topological classification is similar, but now depends on the Irreps at $\bar{\mathrm{X}}$ and $\bar{\mathrm{Y}}$.

Now, let us discuss the possible TCIs in the $pmg$ group. There is one mirror operation and one glide operation in the $pmg$ group, for which we can take $m_x$ and $g_y=\{m_y|\bm{\tau}_x\}$ with $\bm{\tau}_x=(1/2,0)$ as an example. The factor group of the $pmg$ group is the point group $C_{2v}$, which is generated by two mirror reflection symmetries with perpendicular mirror axes. The high-symmetry points $\bar{\mathrm{X}}=(\pi,0)$ and $\bar{\mathrm{M}}=(\pi,\pi)$ in the surface BZ possess 2D Irreps in the spinless case, due to the anti-commutation relation between $m_x$ and $g_y$ at these two points, as illustrated in Ref.~\cite{liu2014topological}. These two points are connected by a MIL. The coexistence of two double-degenerate points and the MIL which connects these two points leads to a mixed type--I-II TCI phase. The topological property of the $pmg$ group can be characterized by the HMC defined on the HMP which projects onto the $\bar{\mathrm{X}}$-$\bar{\mathrm{M}}$ line and the topological classification is $\mathbb{Z}$, more than the $\mathbb{Z}_2$ classification discussed in Ref.~\cite{liu2014topological}. The points $\bar{\mathrm{X}}$ and $\bar{\mathrm{M}}$ can also be connected by the HSL $\bar{\mathrm{X}}$-$\bar{\Gamma}$-$\bar{\mathrm{Y}}$-$\bar{\mathrm{M}}$, where $\bar{\Gamma}$-$\bar{\mathrm{Y}}$ is invariant under $m_x$, $\bar{\mathrm{X}}$-$\bar{\Gamma}$ and $\bar{\mathrm{Y}}$-$\bar{\mathrm{M}}$ are invariant under $g_y$. The states at $\bar{\Gamma}$ and $\bar{\mathrm{Y}}$ belong to 1D Irreps of the group $C_{2v}$. Similar to the case of the $pgg$ group, if all the states at $\bar{\Gamma}$ and $\bar{\mathrm{Y}}$ belong to $A_i$ (or $B_i$), the states in the line $\bar{\mathrm{X}}$-$\bar{\Gamma}$-$\bar{\mathrm{Y}}$-$\bar{\mathrm{M}}$ can be labeled by the parity on one segment, such as $\bar{\mathrm{X}}$-$\bar{\Gamma}$. This allows us to define an integer topological invariant with the same form as Eq.~(\ref{EqHGC}), which is a mixture of HMC and HGC and leads to a $\mathbb{Z}$ classification. Therefore, depending on the Irreps at $\bar{\Gamma}$ and $\bar{\mathrm{Y}}$, we can have either $\mathbb{Z}^2$ or $\mathbb{Z}$ classification for the spinless case of the $pmg$ group.

In the spinful case, similar to the analysis of the $pgg$ group, the spin degree of freedom gives $\alpha_{1/2}=-1$ on the four high-symmetry points. Thus, $\alpha_s=\alpha\alpha_{1/2}=1$ at $\bar{\mathrm{X}}$ and $\bar{\mathrm{M}}$, and $\alpha_s=-1$ at $\bar{\Gamma}$ and $\bar{\mathrm{Y}}$. In this case, the double degeneracy occurs at $\bar{\Gamma}$ and $\bar{\mathrm{Y}}$. Nevertheless, topological classification is still similar as that in the spinless case.

In the $p4g$ group, there are a $C_4$ rotation symmetry and four glide symmetries $g_x=\{m_x|\bm{\tau}\}$, $g_y=\{m_y|\bm{\tau}\}$, $g_d=\{m_d|\bm{\tau}\}$ and $g_d'=\{m_d'|\bm{\tau}\}$, where $\bm{\tau}=(1/2,1/2)$, $m_d$ transforms $(x,y)$ to $(y,x)$ and $m_d'$ transforms $(x,y)$ to $(-y,-x)$. In the surface BZ (see Fig.~\ref{Fig:1}(b)), the point group is $C_{4v}$ at $\bar{\Gamma}=(0,0)$ and $\bar{\mathrm{M}}=(\pi,\pi)$, and is $C_{2v}$ at $\bar{\mathrm{X}}=(\pi,0)$ ($\bar{\mathrm{Y}}=(0,\pi)$ is equivalent to $\bar{\mathrm{X}}$ in this group). There are three independent glide invariant lines $\bar{\Gamma}$-$\bar{\mathrm{X}}$, $\bar{\mathrm{X}}$-$\bar{\mathrm{M}}$, and $\bar{\mathrm{M}}$-$\bar{\Gamma}$. The projective representations of $m_x$ and $m_y$ satisfy
\begin{eqnarray}
 D(m_x)D(m_y)=\alpha(\mathbf{k}) D(m_y)D(m_x),
\end{eqnarray}
where $\alpha(\mathbf{k})=e^{i(\mathbf{k}-m_x\mathbf{k})\cdot{\bm{\tau}}} e^{-i(\mathbf{k}-m_y\mathbf{k})\cdot\bm{\tau}}$. At the $\bar{\mathrm{X}}$ point, $\alpha=-1$, while at $\bar{\Gamma}$ and $\bar{\mathrm{M}}$, $\alpha=1$. The glide plane symmetries guarantee the double degeneracy at the $\bar{\mathrm{X}}$ point. It can also be checked that for any two symmetry operations in the $C_{4v}$ group, denoted as $a$ and $b$, if $[a,b]=0$, we have $\alpha=\frac{D(a)D(b)}{D(b)D(a)}=1$ at $\bar{\Gamma}$ and $\bar{\mathrm{M}}$ for the $p4g$ group. Thus, the representations at these two points are conventional representations of the $C_{4v}$ group. A more rigorous proof is shown in Appendix \ref{App:rep}. The $C_{4v}$ group has both the 1D and 2D Irreps. If the surface bands belong to the 1D Irreps at both $\bar{\Gamma}$ and $\bar{\mathrm{M}}$, no TCIs can exist. If the surface bands belong to the 1D Irrep at $\bar{\Gamma}$, but belong to the 2D Irreps at $\bar{\mathrm{M}}$, the HGC can be defined along the line $\bar{\mathrm{X}}$-$\bar{\mathrm{M}}$, yielding the mixed type-I-II TCI phase. Another independent HGC is possible to exist along the line $\bar{\mathrm{X}}$-$\bar{\Gamma}$-$\bar{\mathrm{M}}$, depending on the 1D Irreps of the states at the momentum $\bar{\Gamma}$. Here, $\bar{\mathrm{X}}$-$\bar{\Gamma}$ is invariant under $g_y$ and $\bar{\Gamma}$-$\bar{\mathrm{M}}$ invariant under $g_d$. At $\bar{\Gamma}$, the glide symmetries $g_y$ and $g_d$ behave the same as $m_y$ and $m_d$, which satisfy the relation $m_d=C_4m_y$, where $C_4$ is the fourfold rotation in the anticlockwise direction and transforms $(x,y)$ to $(-y,x)$. The group $C_{4v}$ has four 1D Irreps $A_i$ and $B_i$ ($i=1,2$). In the Irreps $A_i$, the character of $C_4$ is $1$, and the mirror parities of $m_y$ and $m_d$ are the same, while in the Irreps $B_i$, the character of $C_4$ is $-1$, and the mirror parities of $m_y$ and $m_d$ are opposite. This is similar to the case of the $pgg$ group. If all the bands at $\bar{\Gamma}$ belong to the Irreps $A_i$ (or $B_i$), the HGC can be defined on $\bar{\mathrm{X}}$-$\bar{\Gamma}$-$\bar{\mathrm{M}}$, leading to the $\mathbb{Z}\times\mathbb{Z}$ classification. Similar analysis can be applied to the case with 2D Irrep at $\bar{\mathrm{M}}$ and 1D Irreps at $\bar{\Gamma}$. If the surface bands belong to the 2D Irreps at both $\bar{\Gamma}$ and $\bar{\mathrm{M}}$, three independent HGCs can be defined on the lines $\bar{\Gamma}$-$\bar{\mathrm{X}}$, $\bar{\Gamma}$-$\bar{\mathrm{M}}$, and $\bar{\mathrm{X}}$-$\bar{\mathrm{M}}$, giving rise to a $\mathbb{Z}^3$ classification. A model of the $p4g$ group in the spinless case is given in Appendix \ref{App: p4g_model}.

In the spinful case, spin gives an additional coefficient $\alpha_{1/2}=-1$ at $\bar{\Gamma}$, $\bar{\mathrm{X}}$, and $\bar{\mathrm{M}}$. For $\bar{\Gamma}$ and $\bar{\mathrm{M}}$, $\alpha_s=\alpha\alpha_{1/2}=-1$, the spinor representations belong to class $K_1$ and all the Irreps are 2D. Thus, all the bands are doubly degenerate at $\bar{\Gamma}$ and $\bar{\mathrm{M}}$. For $\bar{\mathrm{X}}$, we have $\alpha_s=1$, so the spinor representations at $\bar{\mathrm{X}}$ belong to the class $K_0$ (conventional representations of group $C_{2v}$). All the conventional Irreps of $C_{2v}$ are 1D, and thus no degeneracy occurs at $\bar{\mathrm{X}}$. One HGC can be defined on the glide invariant line $\bar{\Gamma}$-$\bar{\mathrm{M}}$. Another possible HGC on the line $\bar{\Gamma}$-$\bar{\mathrm{X}}$-$\bar{\mathrm{M}}$ depends on which Irreps the bands belong to at $\bar{\mathrm{X}}$, which is similar to the case of the $pgg$ group. If all the states at $\bar{\mathrm{X}}$ belong to the Irreps $A_i$ (or $B_i$) ($i=1,2$) of group $C_{2v}$, the HGC can be defined on the line $\bar{\Gamma}$-$\bar{\mathrm{X}}$-$\bar{\mathrm{M}}$, and leads to a $\mathbb{Z}\times\mathbb{Z}$ classification.

\section{Conclusion}\label{Sec:Conclusion}
In this paper, we have developed a theory to systematically classify TCI phases based on the representation theory of 2D space groups. We have shown that the classification sensitively depends on the Irreps of the states at certain HSPs and HSLs. Our theory provides a basis for the search of realistic materials for TCI phases. Since our theory is based on a semi-infinite system with a specific surface, one can first identify which type of 3D crystals can allow for the surface with the required 2D space group symmetry. This can be achieved with the help of the discussion of the layer groups and the scanning tables in the International Tables for Crystallography \cite{kopsky2006}, as discussed in details in Appendix B of Ref.~\cite{liu2014topological}. For example, the diamond and spinel structures are described by the 3D space group $Fd\bar{3}m$ (227), for which the $(110)$ surface possesses the symmetry of the $pmg$ group. Therefore, one can look at the specific surfaces of the corresponding materials with the help of some crystal databases \cite{curtarolo2012aflowlib,villars1997pearson}. In addition, topological phases have also been discussed recently in cold atom systems with optical lattices \cite{goldman2014p,grushin2014floquet,zheng2014floquet,jotzu2014experimental,aidelsburger2015measuring} and photonic crystal systems \cite{lu2013weyl,lu2014topological,wang2009observation,rechtsman2013photonic,yannopapas2011gapless,lu2015three}. Our classification, as well as the toy model, can also help to design crystal structures with the required symmetry in these systems. Disorder effect can break the crystalline symmetry of the system \cite{fu2012topology,fulga2014statistical}. However, as discussed in Refs.~\cite{hsieh2012,fu2012topology,diez2015extended, PhysRevB.91.235111}, crystalline topological phases will remain stable as long as the corresponding symmetry protection is preserved on average. Aside from the crystalline symmetry, one may also ask if time-reversal symmetry is compatible to the TCI phases discussed above and if the combination of time-reversal symmetry and crystalline symmetry can lead to new TCI phases. For the first question, the existence of TR symmetry may or may not change topological classification of TCIs. Take the $pm$ group as an example. In the spinless case, for any state $|\psi\rangle$ on the MIPs, the state $\Theta|\psi\rangle$ has the same mirror parity with $|\psi\rangle$, and thus the MCNs will always be zero, and no topologically non-trivial phase can exist. While, in the spinful case, $\Theta|\psi\rangle$ has the opposite mirror parity with $|\psi\rangle$, and as a consequence, MCNs can be non-zero and TR symmetry is compatible to TCI phases. The combination of time reversal symmetry and crystalline symmetry can also lead to the so-called magnetic crystalline symmetry group, in which new topological phases indeed exist, as discussed in Ref.~\cite{zhang2015topological}. The classification of TCIs with time-reversal symmetry is beyond the scope of this paper and will be left to future work. It should be noted that the classification approach in this paper is only suitable for TCI phases with nontrivial surface states. The classification based on the 2D space group is a subgroup of the 3D bulk topology. For a given 3D bulk, the surfaces along different directions will possess different 2D space groups with different topological class according to our results and the classification of 3D bulk topology in principle should include all possible topological invariants for different surfaces (of course many of them give the same bulk topological invariant). Our classification may miss some topological invariants since some symmetry cannot be preserved by any surface (e.g., inversion symmetric topological insulators \cite{hughes2011a}). From the practical view, our classification is useful because non-trivial surface states are the main physical consequence of bulk topology.


\section{acknowledgements}
We would like to acknowledge X. Dai, C. Fang, X.-L. Qi, C.-K. Xu, Q.-Z. Wang, R.-X. Zhang and B.-F. Zhu for helpful discussions. X.-Y. Dong acknowledges the support from the Program of Basic Research Development of China (Grant No. 2011CB921901) and National Natural Science Foundation of China (Grant No 11374173). C.-X.L. acknowledges the 1290
support from Office of Naval Research (Grant No. N00014- 129115-1-2675) and from the Penn State MRSEC, Center for 1292
Nanoscale Science, under the Award NSF DMR-1420620.

\begin{appendix}
\section{A review of representation theory of symmetry groups}\label{App:rep}

In this section, we will give a short review of the representation theory of symmetry groups in the description of electronic band structures.

Due to the periodic lattice structure, electronic states in a crystal form energy bands and are labeled by the crystal momentum $\mathbf{k}$, which form the BZ. From the view of group theory, the crystal momentum $\mathbf{k}$ also labels the Irrep of the translation subgroup of the space group. At each $\mathbf{k}$, we can define the wave vector group or little group $G_{\mathbf{k}}$, which contains all the elements of the space group that leave $\mathbf{k}$ unchanged or map it onto an equivalent vector $\mathbf{k}+\mathbf{G}$, where $\mathbf{G}$ is a reciprocal lattice vector. Electronic states at each $\mathbf{k}$ can be described by the Irreps of the corresponding wave-vector group. In particular, the dimension of the Irreps of the wave-vector group determines the degeneracy of electronic states at the corresponding momentum. The representations of a wave-vector group can be constructed from the representations of the corresponding point-group that contains all the point group operations of the wave-vector group. For a space-group operation $g=\{r|\mathbf{R}+\bm{\tau}\}$, the representation of the wave-vector group ar $\mathbf{k}$ is $D^{\mathbf{k}}(g)=e^{-i\mathbf{k}\cdot(\mathbf{R}+\bm{\tau})}D(r)$, where $D(r)$ is the representation of the corresponding point group.

Let $g_i=\{r_i|\mathbf{R}_i+\bm{\tau}_i\}$, we have
\begin{eqnarray}
  &&D(r_1)D(r_2)=e^{i\mathbf{k}\cdot(\mathbf{R}_1 +\bm{\tau}_1)}e^{i\mathbf{k}\cdot(\mathbf{R}_2 +\bm{\tau}_2)}D^{\mathbf{k}}(g_1)D^{\mathbf{k}}(g_2)\nonumber\\
  &&=e^{i\mathbf{k}\cdot(\mathbf{R}_1 +\bm{\tau}_1)}e^{i\mathbf{k}\cdot(\mathbf{R}_2 +\bm{\tau}_2)}D^{\mathbf{k}}(g_1g_2)\nonumber\\
  &&=e^{i\mathbf{k}\cdot(\mathbf{R}_2-r_1\mathbf{R}_2)} e^{i\mathbf{k}\cdot(\bm{\tau}_2-r_1\bm{\tau}_2)}D(r_1r_2)
  \nonumber\\&&=e^{i(\mathbf{k}-r^{-1}_1\mathbf{k})\cdot\mathbf{R}_2} e^{i(\mathbf{k}-r^{-1}_1\mathbf{k})\cdot\bm{\tau}_2}D(r_1r_2).
\end{eqnarray}
Since $\mathbf{k}-r^{-1}_1\mathbf{k}=\mathbf{G}$, $e^{i(\mathbf{k}-r^{-1}_1\mathbf{k})\cdot\mathbf{R}_2}=1$. However, if $\mathbf{G}\neq 0$ and $\bm{\tau}_2\neq 0$, the coefficient  $\omega(r_1,r_2)\equiv e^{i(\mathbf{k}-r^{-1}_1\mathbf{k})\cdot\bm{\tau}_2}\neq 1$, and $D(r_1)D(r_2)=\omega(r_1,r_2)D(r_1r_2)$.
The existence of the factor $\omega(r_1,r_2)$ shows that $D(r)$ is may not be the conventional representations of the point group which always satisfies $D(r_1)D(r_2)=D(r_1r_2)$. Thus, $D(r)$ is usually dubbed a projective representation of the point group belonging to the factor system $\omega(r_1,r_2)$ \cite{bir1974}.  It should be noted that $\omega(r_1,r_2)$ depends on the momentum $\mathbf{k}$. The factor system is specified by $h^2$ coefficients $\omega(r_1,r_2)$, where $h$ is the order of the point group. The factor systems can be classified into different classes. If $D(r)$ is a projective representation of the factor system $\omega(r_1,r_2)$, the representation $D'(r)=D(r)/u(r)$, where $u(r)$ is an arbitrary single-valued function and $|u(r)|=1$, is a projective representation belonging to the factor system $\omega'(r_1,r_2)=\frac{\omega(r_1,r_2)u(r_1r_2)}{u(r_1)u(r_2)}$. The factor system $\omega'(r_1,r_2)$ is said to be projectively equivalent with $\omega(r_1,r_2)$. The set of all projectively equivalent factor systems is called a class of factor systems.

For every pair of commuting element $a$ and $b$ in the point group, if $\frac{\omega'(a,b)}{\omega'(b,a)}=\frac{\omega(a,b)}{\omega(b,a)}$, the two factor systems $\omega'$ and $\omega$ belong to the same class; otherwise, they belong to different classes. There is a class called $K_0$ for every group, which contains a factor system with all the coefficients $\omega(r_1,r_2)=1$. The representations belonging to the class $K_0$ are the conventional representations of the point group. Any factor system with $\omega(a,b)=\omega(b,a)$ for commuting $a$ and $b$ also belongs to class $K_0$. It can be proved that for a point group that has more than one classes of factor systems, the 1D Irreps can only exist in class $K_0$, that is, there are no 1D Irreps in any other classes $K_p\neq K_0$.

When spin of the electrons is taken into account, the wave functions have the form $\psi_i={\psi_{i1} \choose \psi_{i2}}$, where $1$ and $2$ denote the two spin states of the electrons with the $z$ projection $\pm \frac{1}{2}$. The spinors with spin-$\frac{1}{2}$ transform according to $\mathcal{D}_{1/2}(r_\theta)=e^{i\bm{\sigma}\cdot\hat{\bm{n}}\theta/2}$, where $\bm{\sigma}=(\sigma_x,\sigma_y,\sigma_z)$ are the Pauli matrices, $\hat{\bm{n}}$ is the primitive vector along the rotation axis, and $\theta$ is the angle of rotation around $\hat{\bm{n}}$. It has the property that $D_{1/2}(r_{\theta+2\pi})=-D_{1/2}(r_{\theta})$. For two point-group operations $r_1$ and $r_2$, the production $r_1r_2$ can be expressed as a rotation around an axis with angle $\theta$ (maybe with an inversion). Then the spinor representations satisfy
\begin{eqnarray}
  D_{1/2}(r_1)D_{1/2}(r_2)=\omega_{1/2}(r_1r_2)D_{1/2}(r_1r_2),
\end{eqnarray}
where the factor $\omega_{1/2}(r_1,r_2)=1$ if $0<\theta <2\pi$, or $-1$ if $2\pi<\theta<4\pi$. Thus, the spinor representations of point groups can be viewed as projective representations.

Under the space-group symmetry operation, both the spatial and spin parts of the wave functions transform as
\begin{eqnarray}
  g\psi_{im}&=&\sum_{j}[D(g)]_{ji}\sum_{n(=1,2)} [D_{1/2}(g)]_{nm}\psi_{jn}\nonumber\\
  &=&\sum_{j,n}[D(g)']_{jn,im}\psi_{jn},
\end{eqnarray}
where $D(g)'=D(g)\otimes D_{1/2}(g)$ is the direct product of the representations of the spatial and spin parts. The direct product is usually reducible and can be decomposed into several Irreps.

Combining the spatial and spin parts, we have
\begin{eqnarray}
  &&D(r_1)'D(r_2)'
  =[D(r_1)\otimes D_{1/2}(r_1)][D(r_2)\otimes D_{1/2}(r_2)]\nonumber\\
  &&~~=[D(r_1)D(r_2)]\otimes[D_{1/2}(r_1)D_{1/2}(r_2)]\nonumber\\
  &&~~=[\omega(r_1,r_2)D(r_1r_2)]\otimes[\omega_{1/2}(r_1,r_2)D_{1/2}(r_1r_2)]\nonumber\\
  &&~~=\omega(r_1,r_2)\omega_{1/2}(r_1,r_2) D(r_1r_2)'\nonumber\\
  &&~~=\omega_s(r_1,r_2)D(r_1r_2)',
\end{eqnarray}
where $\omega_s(r_1,r_2)=\omega(r_1,r_2)\omega_{1/2}(r_1,r_2)$. Thus, the spinor representation $D'(r)$ is a projective representation of the point group with the factor system determined by $\omega_s$.

Here, we focus on the point groups $C_{2v}$, $C_{3v}$, $C_{4v}$ and $C_{6v}$, which are relevant to the 2D space group. There are two classes $K_0$ and $K_1$ for $C_{2v}$, $C_{4v}$ and $C_{6v}$, while only one class $K_0$ for $C_{3v}$. For the groups $C_{2v}$, $C_{4v}$ and $C_{6v}$, all the Irreps in the class $K_1$ are 2D. If $\alpha=\omega(a,b)/\omega(b,a)=1$ for every commuting pair of $a$ and $b$ ($[a,b]=0$), the representations belong to the class $K_0$. There is a systematic way to determine which class the representations belong to. It depends on the specific properties of the group. The complete discussions are presented in Ref.~\cite{bir1974}.

We take the $pgg$ group as an example. The factor group of the $pgg$ group is $C_{2v}$ at four momenta $\bar{\Gamma}$, $\bar{\mathrm{M}}$, $\bar{\mathrm{X}}$, and $\bar{\mathrm{Y}}$. At these momenta, we need to use the projective representations to describe the states. The generator of $C_{2v}$ group can chosen as $m_x$ and $m_y$ with the relations: $m_x^2=e$, $m_y^2=e$, and $m_xm_y=m_ym_x$, where $e$ is the identity operation. In this case, the relations between the representations of the little group $G_{\mathbf{k}}$ and the projective representations of the point group are $D^{k}(g_x)=e^{-i\mathbf{k}\cdot\bm{\tau}_y}D(m_x)$ and $D^k(g_y)=e^{-i\mathbf{k}\cdot\bm{\tau}_x}D(m_y)$. Thus,
\begin{eqnarray}
  &&D(m_x)^2=[e^{i\mathbf{k}\cdot\bm{\tau}_y}D^k(g_x)]^2 =e^{i\mathbf{k}\cdot2\bm{\tau}_y}D^k(g_x^2)\nonumber\\
  &&=e^{i\mathbf{k}\cdot2\bm{\tau}_y}e^{-i\mathbf{k}\cdot2\bm{\tau}_y}I=I,
\end{eqnarray}
where $I$ is an identity matrix, and similarly $D(m_y)^2=I$. The class of the representations i determined by $\alpha=\frac{D(m_x)D(m_y)}{D(m_y)D(m_x)}$. If $\alpha=1~(-1)$, the representations belong to class $K_0$ ($K_1$).

Another interesting group is the non-symmorphic group is $p4g$, of which the factor group is $C_{4v}$ at $\bar{\Gamma}$ and $\bar{\mathrm{M}}$. The generator of $C_{4v}$ can be chosen as the four fold rotation $c_4$ and the mirror symmetry $m_x$ with the relations: $c_4^4=e$, $m_x^2=e$ and $m_xc_4=c_4^3m_x$. The mirror symmetry $m_x$ corresponds to the glide symmetry $g_x=\{m_x|\bm{\tau}\}$. Since  $g_x^2=\{e|m_x\bm{\tau}+\bm{\tau}\}$, $g_xc_4=\{m_d|\bm{\tau}\}$ and $c_4^3g_x=\{m_d|c_4^3\bm{\tau}\}$, the relations between the projective representation and the representations of the space group are
\begin{eqnarray}
  D(c_4)^4&=&I,\nonumber\\
  D(m_x)^2&=&[e^{i\mathbf{k}\cdot\bm{\tau}}D^k(g_x)]^2=e^{i\mathbf{k}\cdot 2\bm{\tau}}D^k(g_x^2)\nonumber\\
  &=&e^{i\mathbf{k}\cdot 2\bm{\tau}}e^{-i\mathbf{k}\cdot (m_x\bm{\tau}+\bm{\tau})}=e^{ik_x},\nonumber\\
  D(m_x)D(c_4)&=&e^{i\mathbf{k}\cdot\bm{\tau}}D^k(g_x)D^k(c_4)\nonumber\\
  &=&e^{i\mathbf{k}\cdot\bm{\tau}}e^{-i\mathbf{k} \cdot\bm{\tau}}D(m_d)=D(m_d),\nonumber\\
  D(c_4)^3D(m_x)&=&D^k(c_4^3)D^k(g_x)e^{i\mathbf{k}\cdot\bm{\tau}}\nonumber\\
  &=&D(m_d)e^{-i\mathbf{k}\cdot(c^3_4\bm{\tau})}e^{i\mathbf{k} \cdot\bm{\tau}}\nonumber\\
  &=&e^{ik_y}D(m_d)=e^{ik_y}D(m_x)D(c_4).
\end{eqnarray}
Assuming $u(c_4)=e^{ik_y/2}$ and $u(m_x)=e^{ik_x/2}$, the representations $D(c_4)'=D(c_4)/u(c_4)$ and $D(m_x)'=D(m_x)/u(m_x)$ are projectively equivalent to $D(c_4)$ and $D(m_x)$ and satisfy
\begin{eqnarray}
  \left[D(c_4)'\right]^4&=&\alpha I,\nonumber\\
  \left[D(m_x)'\right]^2&=& I,\nonumber\\
  D(m_x)'D(c_4)'&=&\left[D(c_4)'\right]^3D(m_x)',
\end{eqnarray}
where $\alpha=e^{-i2k_y}$. At the point $\bar{\Gamma}$ and $\bar{\mathrm{M}}$, $\alpha=1$. Thus, we have proved that in the $p4g$ group, the representations at $\bar{\Gamma}$ and $\bar{\mathrm{M}}$ belong to class $K_0$.

In the spinful case, we need to further consider the spin part, and thus
$\alpha_s=\alpha\alpha_{1/2}$, where $\alpha$ is for the spatial part, and $\alpha_{1/2}=\omega_{1/2}(a,b)/\omega_{1/2}(b,a)$ is for the spin part.

\section{Mirror Chern number in the $pmm$ group}\label{App:MCN}
In this section, we will consider the direct calculation of the mirror Chern number in the $pmm$ group and show why mirror Chern number must be zero for the spinless case but it can be non-zero for the spinful case.

As shown in the main text, the Chern number of the mirror plane in the mirror even (odd) subspace is defined as the integral of the Berry curvature of the even (odd) subspace in the MIP. Let us take MIP of $k_y=0$, which is invariant under $m_y$, as an example. On this plane, the mirror Chern number in each subspace is given by
\begin{eqnarray}
  C_{e(o)}=\frac{1}{2\pi}\int_{\mathrm{MIP}}dk_x dk_z F_{e(o)}(k_x,k_z),
\end{eqnarray}
where $F_{e(o)}(k_x,k_z)$ is the $y$ component of the Berry curvature of the occupied energy bands in the even (odd) subspace
\begin{eqnarray}
  F_{e(o)}(k_x,k_z)=\partial_{k_z}A_{x,e(o)} (k_x,k_z)-\partial_{k_x} A_{z,e(o)}(k_x,k_z),\nonumber\\
\end{eqnarray}
with $A_{x(z),e(o)}$ is the $x~(z)$ component of Abelian Berry connection of of the occupied bands in even (odd) subspace.

Next, let us show how the mirror operator $m_x$ acts on $F_{e(o)}(k_x,k_z)$. For the spinless case, since $m_x$ commutates with $m_y$ and does not change the mirror parity of the eigenfunctions under $m_y$, two subspaces are separate under the $m_x$ operation. Thus, we can discuss two subspaces independently. For one subspace, we can introduce the sewing matrix, which is defined as
\begin{eqnarray}
	B_{i,\alpha\beta}(k_x,k_z)=\langle \psi_i(-k_x,k_z,\alpha)|m_x|\psi_i(k_x,k_z,\beta)\rangle
\end{eqnarray}
where $i=e,o$ for the even or odd subspace of $m_y$, $\psi(k_x,k_z,\alpha)$ is the $\alpha$-th eigenfunction of $H(k_x,k_z)$. Equivalently, we have
\begin{eqnarray}
  |\psi_i(-k_x,k_z,\alpha)\rangle=\sum_{\beta}B^*_{i,\alpha\beta}(k_x,k_z)m_x |\psi_i(k_x,k_z,\beta)\rangle\nonumber\\
\end{eqnarray}

The Abelian Berry connection can be obtained by taking the trace of the non-Abelian Berry connection, which is defined as $a_i^{\alpha\beta}(k_x,k_z)=-i\langle \psi(k_x,k_z,\alpha)|\partial_{k_i}|\psi(k_x,k_z,\beta)\rangle$.
The $x$ component of the non-Abelian Berry connection has the relation
\begin{eqnarray}
  &&a^{\alpha\beta}_x(-k_x,k_z)=-i\langle \psi(-k_x,k_z,\alpha)|\partial_{-k_x}|\psi(-k_x,k_z,\beta)\rangle \nonumber\\
  &&=-B_{\alpha\theta}(k_x,k_z)a^{\theta\gamma}_x(k_x,k_z) B^*_{\beta\gamma}(k_x,k_z)\nonumber\\&&-iB_{\alpha\theta}(k_x,k_z)\partial_{k_x} B^*_{\beta\theta}.
\end{eqnarray}
Then the $x$ component of the Abelian Berry connection has the relation
\begin{eqnarray}
  &&A_x(-k_x,k_z)=\mathrm{Tr}[a_x(-k_x,k_z)]\nonumber\\&&
  =-A_x(k_x,k_z)- i\mathrm{Tr}[B(k_x,k_z)\partial_{k_x}B^{\dag}(k_x,k_z)],
\end{eqnarray}
where the trace is taken for all the occupied bands in the considered subspace.
Similarly, the $z$ component of the Abelian Berry connection is
\begin{eqnarray}
  &&A_z(-k_x,k_z)=\mathrm{Tr}[a_z(-k_x,k_z)]\nonumber\\&&
  =A_z(k_x,k_z)+i\mathrm{Tr}[B(k_x,k_z)\partial_{k_z}B^{\dag}(k_x,k_z)].\label{Eq1}
\end{eqnarray}
Thus, the Berry curvature in the considered subspace has the property that
\begin{eqnarray}
  &&F(-k_x,k_z)\nonumber\\&&=\partial_{k_z}A_x(-k_x,k_z)-\partial_{-k_x} A_z(-k_x,k_z)\nonumber\\
  &&=-[\partial_{k_z}A_x(k_x,k_z)-\partial_{k_x} A_z(k_x,k_z)]\nonumber\\
  &&=-F(k_x,k_z).
\end{eqnarray}
The Berry curvature in each subspace is an odd function of $k_x$. Therefore, the Chern number, as an integral of the Berry curvature over the whole momentum space, in each subspace is zero.

In the spinful case, the mirror symmetry $m_x$ changes the mirror parity of the eigenfunctions under $m_y$. Therefore, two subspaces of $m_y$ are related to each other by $m_x$ in this case. As a consequence, one can follow the derivation above and obtain the result
\begin{eqnarray}
  F_{e}(-k_x,k_z)=-F_{o}(k_x,k_z).
\end{eqnarray}
Therefore, the Chern number in each subspace can be nonzero with the relation $C_e=-C_o$ and the MCN is $C_M=\frac{1}{2}(C_e-C_o)=C_e$. We conclude that the MCN for a system with the $pmm$ group must be zero for the spinless case but can be non-zero for the spinful case.

For the non-zero MCN in the spinful case of the $pmm$ group, we can further relate it to the HMC defined in the main text. This is because the Berry connection in the region $k_x\in(-\pi,0)$ can be determined by that in the region $k_x\in(0,\pi)$. Direct calculation gives
\begin{eqnarray}
  &&C_M=C_e=\frac{1}{2\pi}\int_{\mathrm{MIP}}dk_xdk_yF_e (k_x,k_z)\nonumber\\&&=\frac{1}{2\pi}\int_{-\pi}^{\pi} dk_z(\int_0^{\pi} dk_xF_e(k_x,k_z)+\int_{-\pi}^0dk_x F_e(k_x,k_z))\nonumber\\
  &&=\frac{1}{2\pi}\int_{-\pi}^{\pi} dk_z (\int_0^{\pi} dk_x F_e(k_x,k_z)+\int_{0}^{\pi}dk_x F_e(-k_x,k_z))\nonumber\\&&
  =\frac{1}{2\pi}\int_{-\pi}^{\pi} dk_z (\int_0^{\pi} dk_x F_e(k_x,k_z)-\int_{0}^{\pi}dk_x F_o(k_x,k_z))\nonumber\\
  &&=\frac{1}{2\pi}\int_{-\pi}^{\pi} dk_z \int_0^{\pi} dk_x(F_e(k_x,k_z)-F_o(k_x,k_z)) \nonumber\\&&
  =\chi_e(k_x,k_z)-\chi_o(k_x,k_z)
\end{eqnarray}
where $\chi_{e(o)}=\frac{1}{2\pi}\int_{-\pi}^{\pi} dk_z \int_0^{\pi} dk_xF_{e(o)}(k_x,k_z)$ is the integration of the Berry curvature on half of the mirror plane between two lines which project onto the high symmetry points in the surface BZ in even (odd) mirror parity subspace. The definition of $\chi_{e(o)}$ is actually the spinful version of the halved mirror chirality defined by Alexandradinata et al.\cite{alexandradinata2014spin} in the spinless case.

\section{Models of $p4m$, $p31m$ and $p6m$ in the spinful case}
\begin{figure}
\includegraphics[width=8.6cm]{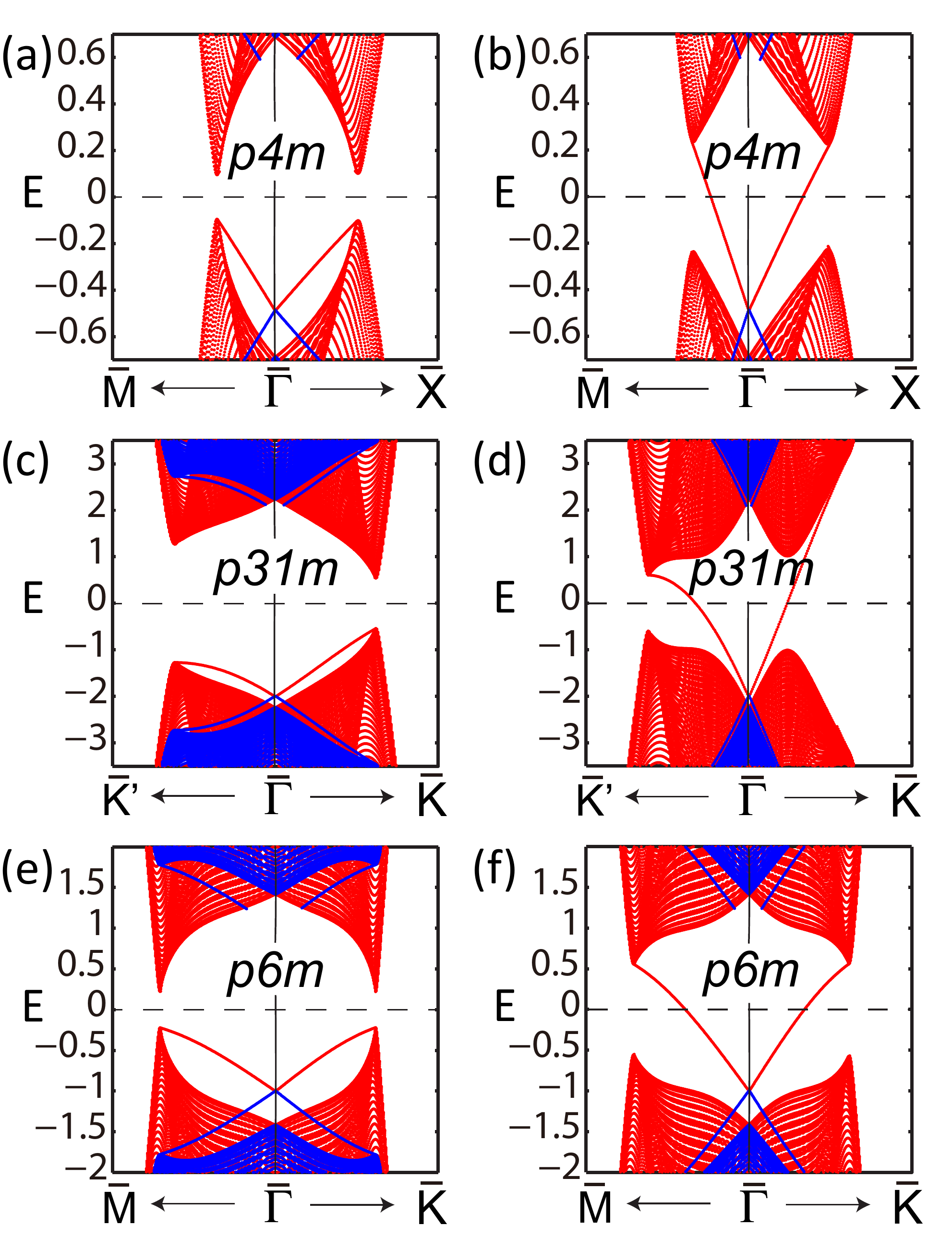}
\caption{(Color online) Band dispersions of the model of $p4m$, $p31m$, and $p6m$ group in the spinful case in a slab configuration. We only show the surface states on one surface of the slab. The red (blue) lines denote the bands with even (odd) mirror parity. (a) A trivial phase of the model of $p4m$ group, with $f_{AB}=0.8i$; (b) a TCI phase of the model of $p4m$ group with $f_{AB}=1.5i$; (c) a trivial phase of the model of $p31m$ group with $l_2=0.7$; (d) a TCI phase of the model of $p31m$ group with $l_2=2.5$; (e) a trivial phase of the model of $p6m$ group with $f_{AB1}=0.5e^{-i\pi/6}$; (f) a TCI phase of the model of $p6m$ group with $f_{AB1}=e^{-i\pi/6}$.}\label{Fig:11}
\end{figure}
\subsection{$p4m$ group in the spinful case}\label{App: p4m_model}
Here, we construct a model of a TCI in $p4m$ group in the spinful case. We consider a layered structure stacked along $z$ direction, of which each layer is a square lattice in $xy$ plane. There is one atom in each primitive cell. We consider the $s$ orbital (instead of the $p$ orbitals in the spinless model) of each atom in the simplest case. Assume there are two $s$ orbitals on each atom, denoted by $s_i~(i=A,B)$. The basis can be chosen as $\Psi_0=(|s_A,\uparrow\rangle, |s_A,\downarrow\rangle, |s_B,\uparrow\rangle, |s_B,\downarrow\rangle)^T$, where
\begin{eqnarray}
  |s_i,\sigma\rangle=\frac{1}{\sqrt{N}}\sum_{\mathbf{R}} e^{i\mathbf{k}\cdot\mathbf{R}} \phi_i(\mathbf{r}-\mathbf{R})|\sigma\rangle
\end{eqnarray}
with $i=A,B$ and $\sigma=\uparrow,\downarrow$.

The Hamiltonian has the form
\begin{eqnarray}
  H(\mathbf{k})=\left(
  \begin{array}{cc}
  H_A(\mathbf{k}) & H_{AB}(\mathbf{k})\\
  H_{AB}(\mathbf{k})^{\dag} & H_B(\mathbf{k})
  \end{array}\right),
\end{eqnarray}
where
\begin{eqnarray}
  &&H_{AB}=\{t_{j0}+t_{j1}[\cos(k_x)+\cos(k_y)]+t_{j2}\cos(k_z)\nonumber\\
  &&~~+t_{j3}\sin(k_z)\}\sigma_0+if_{j}[\sin(k_x)\sigma_y-\sin(k_y)\sigma_x],
\end{eqnarray}
where $j=A,B,AB$. In the basis $\Psi_0$, the matrix form of the mirror-symmetry operation $m_y$ is $D(m_y)=\sigma_0\otimes i\sigma_y$ and that of the fourfold-rotation symmetry operation $c_4$ is $D(c_4)=\sigma_0\otimes \frac{1}{\sqrt{2}}(\sigma_0+i\sigma_z)$. It can be checked that the Hamiltonian satisfies
\begin{eqnarray}
    H(\mathbf{k})=D(g)^{\dag}H(g\mathbf{k})D(g),
\end{eqnarray}
where $g=c_4,m_y$. Note that in the spinful case, although we only consider the $s$ orbital, the bands are doubly degenerate at $\bar{\Gamma}$, $\bar{\mathrm{X}}$, and $\bar{\mathrm{M}}$, and the topological classification is $\mathbb{Z}^3$. Set $t_{A0}=-t_{B0}=23$, $t_{A1}=-t_{B1}=-8$, $t_{A2}=-t_{B2}=-8$, $t_{A3}=t_{B3}=f_{A}=f_{B}=0$, $t_{AB0}=0.2$, $t_{AB1}=t_{AB2}=0.1$, $t_{AB3}=1i$. Tuning $f_{AB}$ from $0$ to $1.5i$, the system goes from a trivial insulating phase to a TCI phase with the HMCs $\chi(\bar{\Gamma}$-$\bar{\mathrm{X}})=\chi(\bar{\Gamma}$-$\bar{\mathrm{M}})=1$ and $\chi(\bar{\mathrm{X}}$-$\bar{\mathrm{M}})=0$. The band dispersions in a slab configuration with $f_{AB}=0.8i$ and $1.5i$ are shown in Fig.~\ref{Fig:11}(a) and (b), respectively.

\subsection{$p31m$ group in the spinful case}\label{App: p31m_model}
Reading the character table of $C_{3v}$ double group, we know that the basis $|\uparrow\rangle$ and $|\downarrow\rangle$ corresponds to the 2D Irrep in the spinful case. Thus, we consider a triangular lattice with $s$ orbital on each atom to build the model of the $p31m$ group. The bases are chosen as $\Psi_0=(|s_A,\uparrow\rangle, |s_A,\downarrow\rangle, |s_B,\uparrow\rangle, |s_B,\downarrow\rangle)^T$, where $A$ and $B$ denote two different $s$ orbitals on each atom. The Hamiltonian is
\begin{eqnarray}
  H(\mathbf{k})=\left(
  \begin{array}{cc}
  H_A(\mathbf{k}) & H_{AB}(\mathbf{k})\\
  H_{AB}(\mathbf{k})^{\dag} & H_B(\mathbf{k})
  \end{array}\right),
\end{eqnarray}
where
\begin{eqnarray}
  &&H_j=\{t_{j0}+t_{j1}[\cos(k_1)+\cos(k_2)+\cos(k_2-k_1)]\nonumber\\
  &&~~+t_{j2}\cos(k_z)+t_{j3}[\sin(k_1)-\sin(k_2)+\sin(k_2-k_1)]\nonumber\\
  &&~~+t_{j4}\sin(kz)\}\sigma_0+e^{i\pi/6}\cos\left(\frac{\pi}{6}\right)\{f_{j1}[\sin(k_2)\nonumber\\
  &&~~+\sin(k_2-k_1)]+f_{j2}[-\cos(k_2)+\cos(k_2-k_1)]\}\sigma_x\nonumber\\
  &&~~-e^{i\pi/6}\{f_{j1}[\sin(k_1)+(\sin(k_2)-\sin(k_2-k_1))\sin\left(\frac{\pi}{6}\right)]\nonumber\\
  &&~~+f_{j2}[\cos(k_1)-(\cos(k_2)+\cos(k_2-k_1))\sin\left(\frac{\pi}{6}\right)]\}\sigma_y\nonumber\\
\end{eqnarray}
where $j=A,B,AB$, and $k_1,k_2$ denote the momentum along the basis of the BZ $\mathbf{b}_1=(\sqrt{3},-1)\pi$ and $\mathbf{b}_2=(0,1)2\pi$, respectively. In the basis $\Psi_0$, the matrix form of the mirror symmetry operation $m_y$ is $D(m_y)=\sigma_0\otimes i\sigma_y$ and that of the three-fold rotation symmetry operation $c_3$ is $D(c_3)=\sigma_0\otimes \frac{1}{2}(\sigma_0+i\sqrt{3}\sigma_z)$. By direct calculation, we get $[m_y,c_3]\neq 0$, which indicates the double degeneracy at $\bar{\Gamma}$, $\bar{\mathrm{K}}$ and $\bar{\mathrm{K}'}$. Set $t_{A0}=-t_{B0}=23$, $t_{A1}=-t_{B1}=-8$, $t_{A2}=-t_{B2}=-8$, $t_{A3}=t_{B3}=0$, $f_{A1}=f_{A2}=f_{B1}=f_{B2}=0$, $t_{AB0}=2$, $t_{AB1}=t_{AB2}=t_{AB3}=0$, $t_{AB4}=i$, and define $l_1=(f_{AB1}+f_{AB2})e^{i\pi/6}/2$, $l_2=(f_{AB1}-f_{AB2})e^{i\pi/6}/2$. Fix $l_1=0$ and increase $l_2$ continuously. When $l_2=0$, it is the trivial insulating phase. When $l_2\sim 0.96$, gap closes on $\Gamma$-$\mathrm{K}$, and drive the system into to a semimetal phase. When $l_2\sim1.9$, the gap closes on $\Gamma$-$\mathrm{K}'$, and after that, the system becomes insulating again, with one surface Dirac cone at $\bar{\Gamma}$. The band dispersions in a slab configuration with $l_2=0.7$ and $2.5$ are shown in Figs.~\ref{Fig:11}(c) and (d), respectively.

\subsection{$p6m$ group in the spinful case}\label{App: p6m_model}
The Irreps of $C_{6v}$ (for $\bar{\Gamma}$) and $C_{2v}$ (for $\bar{\mathrm{M}}$) group are all 2D in the spinful case, thus we still can consider only the $s$ orbital of the atoms on a triangular lattice. Meanwhile, from the analysis of the $C_{3v}$ group, we know that with $s$ orbital the bands are doubly degenerate at $\bar{\mathrm{K}}$, too. The basis can be chosen as $\Psi_0=(|s_A,\uparrow\rangle, |s_A,\downarrow\rangle, |s_B,\uparrow\rangle, |s_B,\downarrow\rangle)^T$, where $A$ and $B$ denote two different $s$ orbitals on each atom. By setting the coefficients $t_{j3},f_{j2}$ $(j=A,B,AB)$ to zero in the Hamiltonian of $p31m$ group, we get the Hamiltonian of $p6m$ group. The matrix form of $c_{6}$ rotation operation is $D(c_6)=\sigma_0\otimes\frac{1}{2}(\sqrt{3}\sigma_0+i\sigma_z)$, and that of $m_y$ is still $\sigma_0\otimes i\sigma_y$. Set $t_{A0}=-t_{B0}=23$, $t_{A1}=-t_{B1}=-8$, $t_{A2}=-t_{B2}=-8$, $t_{A3}=t_{B3}=0$, $f_{A1}=f_{B1}=0$, $t_{AB0}=1$, $t_{AB1}=t_{AB2}=0$, $t_{AB3}=1i$. Tuning $f_{AB1}e^{i\pi/6}$ from 0 to 1, the system transforms from a trivial insulating phase to a TCI phase with one surface Dirac cone at $\bar{\Gamma}$. The band dispersions in a slab configuration with $f_{AB1}=0.5e^{-i\pi/6}$ and $e^{-i\pi/6}$ are shown in Fig.~\ref{Fig:11}(e) and (f), respectively.

\section{Model of $p4g$ group in the spinless case}\label{App: p4g_model}
\begin{figure}
\includegraphics[width=8.6cm]{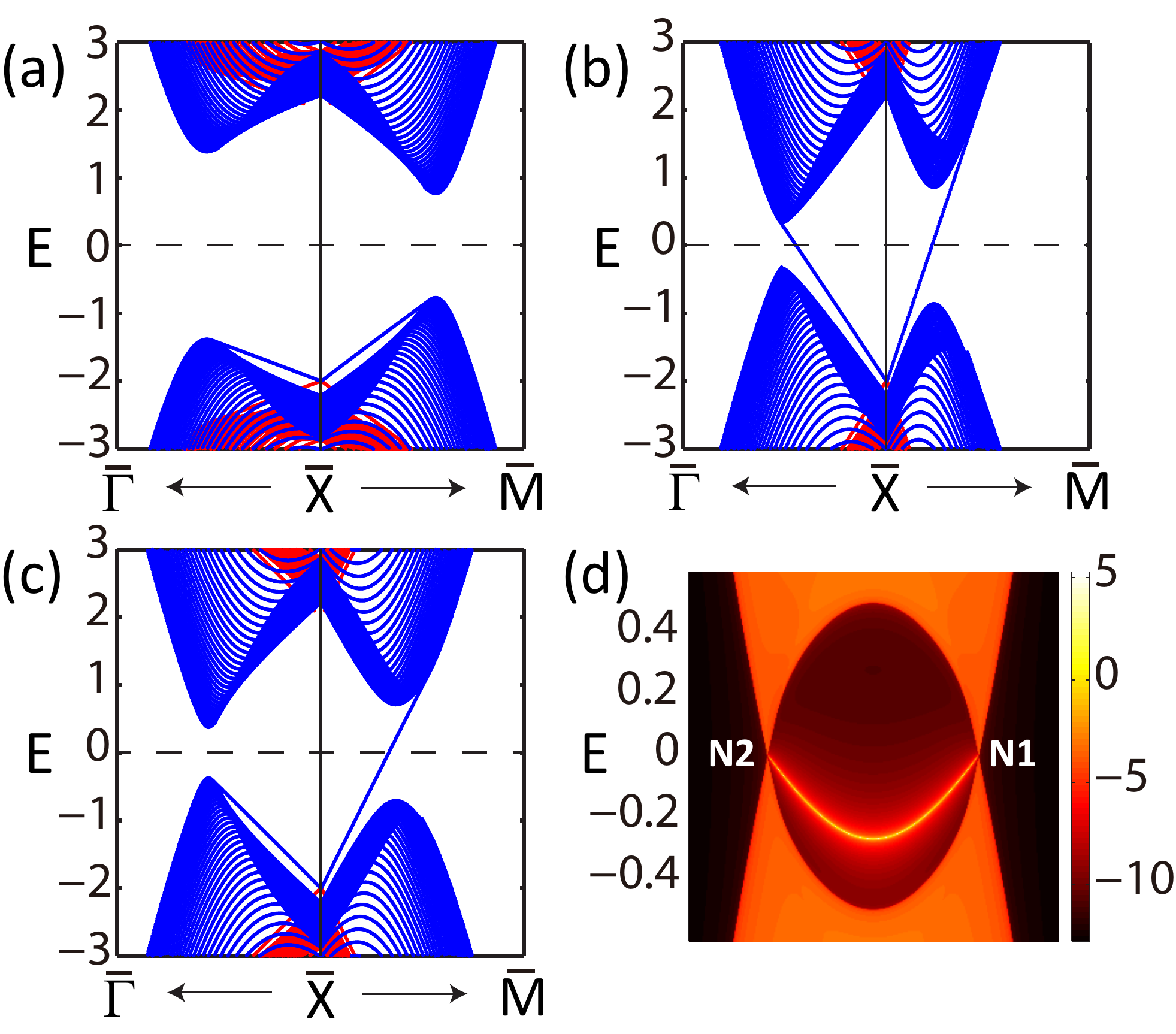}
\caption{(Color online) Band dispersions around $\bar{\mathrm{X}}$ of the model of $p4g$ group in the spinless case in a slab configuration. The parameter $l_1=0.3,1.2,0.8$ in (a), (b) and (c), respectively. We only show the surface states on one surface of the slab. The red (blue) lines denote the bands with even (odd) glide parity. (d) The density of states calculated by iterative Green functions on the line connecting two Weyl nodes (N1-N2) with $l_1=0.8$ in a semi-infinite configuration. The line N1-N2 parallel to $\bar{\Gamma}$-$\bar{\mathrm{X}}$ with the middle point on $\bar{\mathrm{X}}$-$\bar{\mathrm{M}}$.}\label{Fig:12}
\end{figure}

In this section, we give a model of $p4g$ group. An anti-ferromagnetic lattice can be constructed to realize the $p4g$ group. It is consisted of two interpenetrating square sublattices, which are connected by a translation $(\frac{1}{2},\frac{1}{2},0)$ to each other (the lattice constant in $xy$ plane is set as 1). The magnetic moments in each sublattice has a ferromagnetic order perpendicular to the $xy$ plane, while the magnetic directions of the two sublattices are opposite. Consider the $p_x$ and $p_y$ orbital on each atom, the basis can be chosen as $\Psi=(|A,1,p_x\rangle$, $|A,1,p_y\rangle$, $|A,2,p_x\rangle$, $|A,2,p_y\rangle$, $|B,1,p_x\rangle$, $|B,1,p_y\rangle$, $|B,2,p_x\rangle$, $|B,2,p_y\rangle)^T$, where $1$ and $2$ denote two sublattices, and $A$ and $B$ denote two sets of independent $p$ orbitals.
The Hamiltonian has the form
\begin{eqnarray}
  H(\mathbf{k})=\left(
  \begin{array}{cc}
  H_A(\mathbf{k}) & H_{AB}(\mathbf{k})\\
  H_{AB}(\mathbf{k})^{\dag} & H_B(\mathbf{k})
  \end{array}\right)
\end{eqnarray}
where
\begin{widetext}
\begin{eqnarray}
  H_{ja}&=&(t_{j0}+t_{j3} \cos(k_z)+t_{j4}\sin(k_z))\Gamma_{00}+(t_{j1}\cos(k_x) +t_{j2}\cos(k_y))\frac{(\Gamma_{00}+\Gamma_{03})}{2}\nonumber\\
  &&+(t_{j1}\cos(k_y)+t_{j2}\cos(k_x))\frac{(\Gamma_{00} -\Gamma_{03})}{2}+t_{j5}\sin(k_x)\sin(k_y)\Gamma_{33};\nonumber\\
  H_{jb}&=&(f_{j0}+f_{j3}\cos(k_z)+f_{j4}\sin(k_z))i\Gamma_{32} +(f_{j1}\cos(k_x)+f_{j2}\cos(k_y))\frac{(\Gamma_{31} +i\Gamma_{32})}{2}\nonumber\\
  &&+(f_{j1}\cos(k_y)+f_{j2}\cos(k_x)) \frac{(-\Gamma_{31}+i\Gamma_{32})}{2} +f_{j5}\sin(k_x)\sin(k_y)\Gamma_{01};\nonumber\\
  H_{jc}&=&[a_{j1}(1+\cos(k_x+k_y))+a_{j2}(\cos(k_x) +\cos(k_y))]\frac{(\Gamma_{10}+i\Gamma_{23})}{2}\nonumber\\
  &&+[a_{j1}\sin(k_x+k_y)+a_{j2}(\sin(k_x)+\sin(k_y))] \frac{(\Gamma_{20}-i\Gamma_{13})}{2}\nonumber\\
  &&+[a_{j2}(1+\cos(k_x+k_y))+a_{j1}(\cos(k_x)+\cos(k_y))] \frac{(\Gamma_{10}-i\Gamma_{23})}{2}\nonumber\\
  &&+[a_{j1}(\sin(k_x)+\sin(k_y))+a_{j2}\sin(k_x+k_y)] \frac{(\Gamma_{20}+i\Gamma_{13})}{2};\nonumber\\
  H_{jd}&=&[b_{j1}(1+\cos(k_x+k_y))+b_{j2}(\cos(k_x) +\cos(k_y))]\frac{(\Gamma_{11}-\Gamma_{22})}{2}\nonumber\\
  &&+[b_{j1}\sin(k_x+k_y)+b_{j2}(\sin(k_x)+\sin(k_y))] \frac{(\Gamma_{12}+\Gamma_{21})}{2}\nonumber\\
  &&-[b_{j2}(1+\cos(k_x+k_y))+b_{j1}(\cos(k_x)+\cos(k_y))] \frac{(\Gamma_{11}+\Gamma_{22})}{2}\nonumber\\
  &&+[b_{j2}\sin(k_x+k_y) +b_{j1}(\sin(k_x)+\sin(k_y))] \frac{(\Gamma_{12}-\Gamma_{21})}{2};\nonumber\\
  H_j&=&H_{ja}+H_{jb}+H_{jc}+H_{jd}.~~~~~~~~(j=A,B,AB)
\end{eqnarray}
\end{widetext}
The generator of the $p4g$ group can be chosen as glide-plane-symmetry operation $g_y=\{m_y|\bm{\tau} =(\frac{1}{2},\frac{1}{2})\}$ and four-fold rotation $c_4$. In the basis $\Psi$, the matrix forms of the symmetry operations are $D(g_y)=\sigma_0\otimes[e^{-ik_x/2} (\cos(k_x/2)\sigma_x+\sin(k_x/2)\sigma_y)\otimes\sigma_z]$, and $D(c_4)=\sigma_0\otimes[ ie^{-ik_x/2}(\cos(k_x/2)\sigma_0 +\sin(k_x/2)\sigma_z)\otimes\sigma_y]$. By direct calculation, it can be shown that at $\bar{\Gamma}$, $\{g_y,c_4\}=0$ and at $\bar{\mathrm{M}}$, $[g_y,c_4]=0$. Thus, in the basis $\Psi$, the bands are doubly degenerate at $\bar{\Gamma}$, while not degenerate at $\bar{\mathrm{M}}$. Since we have proved that with any basis the bands are doubly degenerate at $\bar{\mathrm{X}}$, one topological invariant HGC can be defined on $\bar{\Gamma}$-$\bar{\mathrm{X}}$. In our model, the representation of $c_4$ for the bands at $\bar{\mathrm{M}}$ depend on the value of the parameters of the Hamiltonian, thus in general, as discussed in the main text, the topological classification of our model is at least $\mathbb{Z}$. Set $t_{A0}=-t_{B0}=23$, $t_{A1}=-t_{B1}=8$, $t_{A2}=-t_{B2}=-8$, $t_{A3}=-t_{B3}=-8$, $t_{A4}=t_{B4}=0$, $t_{AB0}=1$, $t_{AB4}=i$. Define $l_1=(a_{AB1}+a_{AB2})/2$ and $l_2=(a_{AB1}-a_{AB2})/(2i)$ and fix $l_2=2l_1$. The other parameters can be chosen to be zero, or some arbitrary numbers with small absolute values, which will not change the topological properties. When $l_1<0.5$, the system is in a trivial insulating phase. Increase $l_1$ to 0.5, gap closes on $\mathrm{X}$-$\mathrm{M}$, and then drive the system into the semimetal phase with Weyl nodes. The band structure along two of the Weyl nodes are shown in Fig.~\ref{Fig:12}(d) with $l_1=0.8$. When $l_1\sim 1$, gap closes on $\Gamma$-$\mathrm{X}$. Increasing $l_1$ further, the system goes into a topological nontrivial phase with $\chi(\bar{\Gamma}$-$\bar{\mathrm{X}})=1$. The band dispersions in a slab configuration with $l_1=0.5,1.2,0.8$ are shown in Fig.~\ref{Fig:12}(a), (b) and (c).

\section{Useful character tables}\label{App:tab}
We have the following character tables (Tables III-V)
\begin{table}[h]
\centering
\caption{The character table of $C_{3v}$ single group. }
\begin{tabular}{c|ccc}
    \hline
    $C_{3v}$ & $E$ & $2C_3$ & $3m_v$ \\
    \hline
    $A_1$ & $1$ & $1$ & $1$ \\
    $A_2$ & $1$ & $1$ & $-1$  \\
    $E$ & $2$ & $-1$ & $0$ \\
    \hline
  \end{tabular}\label{tab:C3v}%
\end{table}

\begin{table}[h]
\centering
\caption{The character table of $C_{4v}$ single group. }
\begin{tabular}{c|ccccc}
    \hline
    $C_{4v}$ & $E$ & $C_2$ & $2C_4$ & $2m_v$ & $2m_d$ \\
    \hline
    $A_1$ & $1$ & $1$ & $1$ & $1$ & $1$ \\
    $A_2$ & $1$ & $1$ & $1$ & $-1$ & $-1$ \\
    $B_1$ & $1$ & $1$ & $-1$ & $1$ & $-1$ \\
    $B_2$ & $1$ & $1$ & $-1$ & $-1$ & $1$ \\
    $E$ & $2$ & $-2$ & $0$ & $0$ & $0$ \\
    \hline
  \end{tabular}\label{tab:C4v}%
\end{table}

\begin{table}[h]
\centering
\caption{The character table of $C_{6v}$ single group. }
\begin{tabular}{c|cccccc}
    \hline
    $C_{6v}$ & $E$ & $C_2$ & $2C_3$ & $2C_6$ & $2m_v$ & $2m_d$ \\
    \hline
    $A_1$ & $1$ & $1$ & $1$ & $1$ & $1$ & $1$ \\
    $A_2$ & $1$ & $1$ & $1$ & $1$ & $-1$ & $-1$\\
    $B_1$ & $1$ & $-1$ & $1$ & $-1$ & $-1$ & $1$\\
    $B_2$ & $1$ & $-1$ & $1$ & $-1$ & $1$ & $-1$\\
    $E_1$ & $2$ & $-2$ & $-1$ & $1$ & $0$ & $0$\\
    $E_2$ & $2$ & $2$ & $-1$ & $-1$ & $0$ & $0$\\
    \hline
  \end{tabular}\label{tab:C6v}%
\end{table}

\end{appendix}


\end{document}